\documentclass[aps,pra,twocolumn,superscriptaddress,dvipsnames]{revtex4}
\usepackage{graphics}
\usepackage[normalem]{ulem}
\usepackage{amsmath}
\usepackage{amssymb}
\usepackage[dvipsnames]{xcolor}
\usepackage{graphicx}
\usepackage{amsthm}
\usepackage{paralist}
\usepackage{verbatim}
\usepackage{chemarr}
\usepackage{braket}
\usepackage{enumitem}
\usepackage{tikzit}

\tikzstyle{dotted line}=[-, style=dotted, tikzit draw=brown]
\tikzstyle{dashed line}=[-, style=dashed, tikzit draw=cyan]
\tikzstyle{green fill line}=[-, fill={green!90!black}, tikzit draw=green]

\usetikzlibrary{circuits.ee.IEC}
\usetikzlibrary{patterns}
\usepackage[caption=false]{subfig}
\pgfdeclarelayer{nodelayer} 
\pgfdeclarelayer{edgelayer}   
\pgfsetlayers{nodelayer,edgelayer} 
\usepackage{commath}

\newtheorem{lemma}{Lemma}
\newtheorem{cond}{Condition}

\newcommand{\N}{\mathbb{N}}

\newcommand{\spc}[1]{\mathcal{#1}}

\def\>{\rangle}
\def\<{\langle}

\newcommand{\ketbra}[2]{\ket{#1} \!  \bra{#2}}

\newcommand{\map}[1]{\mathcal{#1}}
\newcommand{\Tr}{\operatorname{Tr}}

\newcommand{\St}{{\mathsf{St}}}

\newcommand{\Chan}{{\mathsf{Chan}}}

\newcommand{\Comb}{{\mathsf{Comb}}}

\newcommand{\Map}{{\mathsf{Map}}}
\newcommand{\TP}{{\mathsf{TP}}}
\newcommand{\CP}{{\mathsf{CP}}}

\newtheorem{prop}{Proposition}

\newtheorem{defi}{Definition}

\newtheorem{manualtheoreminner}{Condition}
\newenvironment{manualtheorem}[1]{%
	\renewcommand\themanualtheoreminner{#1}%
	\manualtheoreminner
}{\endmanualtheoreminner}
 
\newcommand{\Proof}{{\bf Proof. \,}}

\begin{document}

\title{Resource theories of communication}

\author{Hl\'er Kristj\'ansson}
\affiliation{Department of Computer Science, University of Oxford, Wolfson Building, Parks Road, Oxford, United Kingdom}
\affiliation{HKU-Oxford Joint Laboratory for Quantum Information and Computation}

\author{Giulio Chiribella}
\email{giulio.chiribella@cs.ox.ac.uk}
\affiliation{Department of Computer Science, The University of Hong Kong, Pokfulam Road, Hong Kong}
\affiliation{Department of Computer Science, University of Oxford, Wolfson Building, Parks Road, Oxford, United Kingdom}
\affiliation{HKU-Oxford Joint Laboratory for Quantum Information and Computation}
\affiliation{Perimeter Institute for Theoretical Physics, 31 Caroline Street North, Waterloo, Ontario, Canada}
\affiliation{HKU Shenzhen Institute of Research and Innovation, Kejizhong 2nd Road, Shenzhen, China}

\author{Sina Salek}
\affiliation{Center for Quantum Computing, Peng Cheng Laboratory, No.\ 2, Xingke 1st Street, Nanshan, Shenzhen, China}
\affiliation{Department of Computer Science, University of Oxford, Wolfson Building, Parks Road, Oxford, United Kingdom}

\author{Daniel Ebler}
\affiliation{Institute  for  Quantum  Science  and  Engineering,  Department  of  Physics, Southern  University  of  Science  and  Technology  (SUSTech),  Shenzhen,  China}
\affiliation{Wolfson College, University of Oxford, Linton Road, Oxford, United Kingdom}
\affiliation{Department of Computer Science, The University of Hong Kong, Pokfulam Road, Hong Kong}

\author{Matthew Wilson}
\affiliation{Department of Computer Science, University of Oxford, Wolfson Building, Parks Road, Oxford, United Kingdom}
\affiliation{HKU-Oxford Joint Laboratory for Quantum Information and Computation}

\begin{abstract}
	A series of recent works has shown that placing communication channels in a coherent  superposition of alternative configurations can boost their ability to transmit information. Instances of this phenomenon are the advantages arising from the use of communication devices in a superposition of alternative causal orders, and those arising from the transmission of information along a superposition of alternative trajectories. The relation among these advantages has been the subject of recent debate, with some authors claiming that the advantages of the superposition of orders could be reproduced, and even surpassed, by other forms of superpositions. To shed light on this debate, we develop a general framework of resource theories of communication. In this framework, the resources are communication devices, and the  allowed operations are (a) the placement of communication devices between the communicating parties, and (b) the connection of communication devices with local devices in the parties' laboratories. The allowed operations are required to satisfy the minimal condition that they do not enable communication independently of the devices representing the initial resources. The resource-theoretic analysis reveals that the aforementioned criticisms on the superposition of causal orders were based on an uneven comparison between different types of quantum superpositions, exhibiting different operational features.
\end{abstract}

\maketitle

\section{Introduction}
A fundamental task in information theory is to quantify the amount of information that a given set of communication devices can transmit.  
Claude E. Shannon addressed  this question for devices operating according to the laws of classical physics \cite{shannon1948mathematical},  laying down the foundations of our current communication technology. 
At the fundamental level, however, the classical laws are just an approximation of the laws of quantum physics. 
The ability to transmit quantum  data \cite{holevo1973bounds,Schumacher1997,Holevo1998,bennett2002entanglement} was shown to offer remarkable  advantages, such as the possibility of secure quantum key distribution \cite{BennettCh1984,ekert1991quantum}. Over time, the study of  communication protocols involving the exchange of quantum data
  led to  the establishment of the field of  \textit{quantum Shannon theory}
  \cite{wilde2013quantum}.

In a series of recent works, a further generalisation of quantum Shannon theory has been proposed where not only the transmitted data,  but also the configuration of the communication devices  can be  quantum   \cite{ebler2018enhanced, salek2018quantum,
chiribella2018indefinite,goswami2018communicating,procopio2019communication,procopio2020sending,wilson2020diagrammatic,gisin2005error,abbott2018communication,chiribella2019shannon2q,kristjansson2020latent}. 
This introduces a second level of quantisation of Shannon theory, generalising standard quantum Shannon theory where the transmitted data  are quantum but the configuration of the communication channels is classical.  In one of the new  frameworks, the available communication channels are combined in a superposition of different causal orders \cite{ebler2018enhanced,salek2018quantum,chiribella2018indefinite,goswami2018communicating,procopio2019communication,procopio2020sending,loizeau2020channel,wilson2020diagrammatic}, using an operation  known as the quantum {\tt SWITCH} \cite{chiribella2009beyond,Chiribella2013}.   In another framework,  information  can be sent along a superposition of trajectories \cite{gisin2005error,abbott2018communication,chiribella2019shannon2q,kristjansson2020latent}, leading to superpositions of alternative quantum evolutions \cite{Aharonov1990,oi2003interference,chiribella2019shannon2q}. In both frameworks, the superposition is generated by letting a quantum system control the configuration of the communication channels,  determining  either their order, or which of them is used to transmit  information.  Coherent control over the channels' configuration has  been shown to yield advantages in a wide range of communication scenarios, achieving rates beyond those that are possible 
 in standard quantum Shannon theory.  
Some of these  advantages stimulated experiments in quantum optics, both on the control of  orders \cite{goswami2018communicating,guo2020experimental} and on the  control  of trajectories \cite{lamoureux2005experimental}.

Recently, the works on the  coherent control of causal orders, in particular Refs.  \cite{ebler2018enhanced, salek2018quantum, chiribella2018indefinite},  have been criticised on the grounds that  similar advantages could be obtained with coherent control of  the choice of communication devices \cite{abbott2018communication}, or coherent control over different choices of encoding and decoding operations \cite{guerin2018shannon}.   Here we respond to these criticisms, by setting up a resource-theoretic framework that sheds light on the comparison between different extensions of quantum Shannon theory.

First, we point out that Refs.\  \cite{ebler2018enhanced, salek2018quantum, chiribella2018indefinite} only claimed that the superposition of causal orders offers an advantage with respect to {\em standard} quantum Shannon theory, where communication devices are composed in a definite order and no coherent control over their configuration is allowed.  The converse claim that every advantage over standard quantum Shannon theory must be due to  control over the causal  order was not  
made  in  \cite{ebler2018enhanced, salek2018quantum, chiribella2018indefinite}, and, in fact, was  known to be false, since Gisin {\em et al.}\  had  previously shown that  control over the choice of channels offers advantages over the standard model of quantum communication \cite{gisin2005error}.

Second, while it is clear that coherent control over devices  generically leads to communication advantages, it is important to distinguish between different types of control. 
Three distinct types of control have been considered so far: 
\begin{enumerate}
\item  control over the causal order of communication channels \cite{ebler2018enhanced, salek2018quantum, chiribella2018indefinite}
\item control over the choice of communication channels    \cite{gisin2005error,abbott2018communication,chiribella2019shannon2q}
\item   control over choices of encoding and decoding operations  \cite{guerin2018shannon}.
\end{enumerate} 
These three types of control are conceptually distinct and, as we will see,  have different operational features.

In this paper, we  construct a general framework for resource theories of communication, and use it to shed light on the different extensions of quantum Shannon theory that have been proposed so far. 
We formulate a minimal requirement of a resource theory of communication, namely that no allowed operation on the communication devices should bypass them, enabling communication independently of the communication devices available to the sender and receiver. 
Our framework captures the differences between the different types of  control 1--3, and helps clarify various comparisons that have been made   across protocols using them.

Applying our resource-theoretic framework, we argue that (a) the comparison between control  of causal orders and control of communication channels  proposed  in Ref.\ \cite{abbott2018communication} is uneven,  because the control  of communication channels requires (in principle) stronger initial resources than the control of causal orders,  and (b) the examples of communication  with  control over encoding and decoding proposed in Ref.\ \cite{guerin2018shannon}
do not satisfy the minimal requirement of a resource-theory of communication.

In \S \ref{sec:framework} of this work, we formalise standard quantum Shannon theory as a resource theory. In \S \ref{sec:genresc} we extend our framework to general resource theories of communication and formulate the minimal requirement that such a theory should satisfy.
\S \ref{sec:sups} presents the frameworks of superposition of causal orders \cite{ebler2018enhanced,salek2018quantum,chiribella2018indefinite} and superposition of trajectories \cite{chiribella2019shannon2q,abbott2018communication,gisin2005error}, showing that both are consistent with a resource-theoretic description.  \S \ref{sec:grenoblereply} comments on the comparisons made  in Ref.\ \cite{abbott2018communication} between the two frameworks, while \S \ref{sec:reply} argues that the communication protocols put forward in Ref.\ \cite{guerin2018shannon} do not admit a resource-theoretic formulation.

\section{Standard quantum Shannon theory as a resource theory of communication}\label{sec:framework}

We begin by reformulating   standard quantum Shannon theory as a  resource theory,  setting the scene for its extension to more general resource theories of communication.

\subsection{Quantum Shannon theory as a theory of resources \label{sec:resources}}

A central task in information theory  is  to quantify the amount of information that a given communication device  can transmit. In general, the amount of information can be classical or quantum, or of other types. In this paper, we will focus on classical and quantum information.    To make the quantification  unambiguous, it is essential to specify how the given device can be used.   The  device represents a resource, and the rules on the possible uses of this resource   can be formulated as a {\em resource theory} \cite{coecke2016,chitambar19resources}.  

A resource-theoretic approach to standard quantum Shannon theory  was initiated by Devetak, Harrow, and Winter \cite{devetak08resources}. Further resource-theoretic formalisations  have been put forward in Refs. \cite{ebler2018enhanced,liu-winter19resources,liu-yuan19resources,takagi2019resource} in a variety of communication scenarios. Related resource theories of quantum devices have been recently formulated in  Refs.\ \cite{theurer2019quantifying,dana2017resource,saxena2019dynamical} for purposes other than the theory of communication.

In this paper we will adopt the general  framework for resource theories proposed by  Coecke, Fritz, and Spekkens \cite{coecke2016}.  In this framework, the set of all possible resources is described by a set of objects, equipped with a set of operations acting on them. The set of operations is closed under sequential and parallel composition.  For example, the set of operations, hereafter denoted by $\mathsf M$, could be the set of all quantum channels (completely positive trace-preserving maps) acting on finite-dimensional quantum systems (the objects).     
The central idea of the resource-theoretic framework is to define a subset of  operations $\mathsf M_{\rm free} \subseteq \mathsf M$, which are regarded as {\em free}. The notion of resource is then defined relative to the set of free operations: a state or an operation is  a non-trivial resource if and only if it is not free, and a resource is more valuable than another if the former can be converted into the latter by means of free operations.  

Different choices of free operations generally define different resources. Intuitively, the set of free operations is meant to capture some operational restriction, which makes some operations ``easy to implement''.   In principle, however,  $\mathsf M_{\rm free}$ could be any subset of operations, as long as it is closed under sequential and parallel composition.  In this respect, the resource-theoretic approach is  a conceptual tool to understand the power of the  set   $\mathsf M_{\rm free}$, irrespectively of whether  implementing the operations in it is easy or not.

In quantum Shannon theory, the input resources are communication channels, or, more precisely,    {\em uses} of communication channels. For example, the ability to transfer a single qubit from a sender to a receiver is modelled as a single use of a single-qubit  identity  channel.  

To cast  quantum Shannon theory in the resource-theoretic framework of Ref.\ \cite{coecke2016}, one has to regard the various types of 
quantum channels as objects, and to defined the allowed  operations that transform input channels into output channels.  These operations are known as {\em quantum supermaps}
\cite{chiribella2008transforming,chiribella2009theoretical,Chiribella2013}.  In the following, we will define the sets of free supermaps $\mathsf M_{\rm free}$ for some of the basic scenarios in quantum Shannon theory,  setting the scene for the generalisations studied in the rest of the paper. 

\subsection{Notation}  

We will denote by $\spc H_A$ the Hilbert space associated to a given quantum system $A$, and by $L(\spc H_A)$ the space of all linear operators on $\spc H_A$. 
The set of all quantum states (positive semidefinite operators with unit trace) on  $\spc H_A$ will be  denoted by $\St (A) \subset L(H_A)$. For simplicity, we will restrict our attention to finite-dimensional systems, although this is not essential for our framework.    

The set of all linear maps from $L(\spc H_A)$  to $L(\spc H_B)$  will be denoted by $\Map (A,B)$.  
The set of quantum channels (completely positive trace-preserving maps) will be denoted by  $\Chan(A,B) \subset \Map (A,B)$. We will also use the shorthand $\Chan (A):  =  \Chan (A,A)$.  When the input and output are arbitrary, we will simply write $\Chan$. 
We will sometimes use the fact  that the action of a generic quantum channel $\map N\in \Chan (A,B)$ on an input state $\rho \in \St(A)$ can be written  in the Kraus representation, as $\map N (\rho) = \sum_i N_i \rho N_i^\dagger$, where $\{N_i\}$ is a  set of linear operators satisfying the normalisation condition $\sum_i N_i^\dagger N_i = I$ \cite{wilde2013quantum}.

We will denote by $A \otimes B$ the composite system consisting of subsystems $A$ and $B$.  
We  recall that 
 every map  $\map M  \in \Map  (A_1\otimes A_2,  B_1\otimes B_2)$ can be decomposed into a sum  of product maps, namely   $\map M   =  \sum_{j=1}^L  \,  \map M_{1,j} \otimes \map M_{2,j}$, with  $\map M_{1,j}  \in   \Map (A_1, B_1)$ and $\map M_{2,j} \in  \Map  (A_2,  B_2)$ for every $j\in  \{1,\dots,  L\}$.

A supermap is a linear transformation from $\Map  (A,B)$ to $\Map (A', B')$, where $A,A',B,B'$ are generic systems.   The tensor product  of two supermaps $\map S:  \Map (A_1,  B_1)  \to \Map (A_1',  B_1')$ and $\map T:  \Map (A_2,  B_2)  \to \Map (A_2',  B_2')$ is the supermap  $\map S\otimes \map T   :   \Map (A_1\otimes A_2, B_1\otimes B_2)  \to \Map (A_1'\otimes A_2',  B_1'\otimes B_2')$ defined by the condition  $(\map S\otimes \map T)  (  \map M_1  \otimes \map M_2)  :  =      \map S (\map M_1) \otimes \map T(\map M_2)$   for every $\map M_1 \in  \Map  (A_1,  B_1)$ and $\map M_2 \in \Map  (A_2,  B_2)$.  Since all the maps in $\Map (A_1\otimes A_2, B_1\otimes B_2)$  are linear combinations of product maps, this condition uniquely defines the supermap $\map S\otimes \map T$.

\subsection{Direct communication from a sender to a receiver through a single   channel}\label{subsect:direct_single}

Consider the basic  communication scenario where a sender (Alice) communicates directly to a receiver (Bob). At the fundamental level, the possibility of communication consists of two ingredients: the availability  of a piece of hardware that serves as a communication device, and  the placement of  that piece of hardware between the sender and the receiver. For example, the piece of hardware could be an optical fibre, and the placement could be provided by  a communication company that laid the fibre between the sender's and the receiver's locations.   In some situations, the placement is  implicit: for example, the sender and receiver could be communicating through a medium, such as the air between them, which has been placed there, as it were, by Nature itself.  

Mathematically, the communication device is described by a quantum channel $\map N  \in  \Chan (X, Y)$, which transforms systems of type $X$ into systems of type  $Y$.   For example, the systems  could be single qubits, encoded in the polarisation of single photons. At this level, the systems are not  assigned  a specific  location in spacetime. 
Accordingly, we will call the systems   $X$ and $Y$ {\em unplaced systems}, and   the  channel $\map N  \in  \Chan (X, Y)$ an {\em unplaced channel}.  
 
\begin{figure}
	\centering
	\begin{tikzpicture}[scale=1.2]
	\begin{pgfonlayer}{nodelayer}
	\draw[fill=red!90!white] (-3.75,0.25) -- (-3.75,1.75) -- (-1.75,1.75) -- (-1.75,0.25) -- (-3.75,0.25) -- (-3.75,1.75);
	\draw[fill=Turquoise!90!white] (-6.25,0.25) -- (-6.25,1.75) -- (-4.75,1.75) -- (-4.75,0.25) -- (-6.25,0.25) -- (-6.25,1.75);
	\draw[fill=Turquoise!90!white] (-0.75,0.25) -- (-0.75,1.75) -- (0.75,1.75) -- (0.75,0.25) -- (-0.75,0.25) -- (-0.75,1.75);
		\node [style=none] (0) at (-0.75, 1.75) {};
		\node [style=none] (1) at (-0.75, 0.25) {};
		\node [style=none] (2) at (0.75, 1.75) {};
		\node [style=none] (3) at (0.75, 0.25) {};
		\node [style=none] (4) at (-1.75, 1.75) {};
		\node [style=none] (5) at (-3.75, 1.75) {};
		\node [style=none] (6) at (-3.75, 0.25) {};
		\node [style=none] (7) at (-1.75, 0.25) {};
		\node [style=none] (8) at (-1.75, 1) {};
		\node [style=none] (9) at (-0.75, 1) {};
		\node [style=none] (10) at (0.75, 1) {};
		\node [style=none] (11) at (-3.75, 1) {};
		\node [style=none] (12) at (-4.75, 1) {};
		\node [style=none] (13) at (-1.25, 1.75) {$Y$};
		\node [style=none] (14) at (-4.25, 1.75) {$X$};
		\node [style=none, font={\large}] (15) at (-2.75, 1) {$\map{N}$};
		\node [style=none, font={\large}] (16) at (0, 1) {$\map{W}$};
		\node [style=none] (17) at (-7.25, 1) {};
		\node [style=none, font={\large}] (18) at (-5.5, 1) {$\map{V}$};
		\node [style=none] (19) at (-4.75, 1) {};
		\node [style=none] (20) at (-6.25, 1) {};
		\node [style=none] (21) at (-4.75, 0.25) {};
		\node [style=none] (22) at (-6.25, 0.25) {};
		\node [style=none] (23) at (-6.25, 1.75) {};
		\node [style=none] (24) at (-4.75, 1.75) {};
		\node [style=none] (25) at (-6.75, 1.75) {$A$};
		\node [style=none] (26) at (1.25, 1.75) {$B$};
		\node [style=none] (27) at (1.75, 1) {};
	\end{pgfonlayer}
	\begin{pgfonlayer}{edgelayer}
		\draw (5.center) to (4.center);
		\draw (4.center) to (7.center);
		\draw (7.center) to (6.center);
		\draw (6.center) to (5.center);
		\draw (0.center) to (2.center);
		\draw (2.center) to (3.center);
		\draw (3.center) to (1.center);
		\draw (1.center) to (0.center);
		\draw (11.center) to (12.center);
		\draw (23.center) to (24.center);
		\draw (24.center) to (21.center);
		\draw (21.center) to (22.center);
		\draw (22.center) to (23.center);
		\draw (20.center) to (17.center);
		\draw (10.center) to (27.center);
		\draw (8.center) to (9.center);
	\end{pgfonlayer}
\end{tikzpicture}
	\caption{\label{fig:place} {\em Basic placement supermap  $\map S^{A,B}_{\rm place}  (\map N)    : =  \map W^B \circ \map N \circ \map V^A
			$.} In this paper, the unplaced communication channels $\map N$ are drawn in red, while the placement supermaps (i.e.\ supermaps from unplaced channels to placed channels) are drawn in blue.} 
\end{figure}
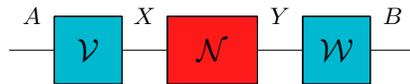

The placement of the device  can be described by introducing a {\em placement operation}, which corresponds to putting the input (output) system at the sender's (receiver's) location.   Mathematically, a placement operation is  a supermap  that  transforms channels in  $\Chan (X,Y) $ into channels in $\Chan (A, B)$, where system $A$  ($B$) is of the same type as system $X$  ($Y$), denoted as $A\simeq X$  ($B\simeq Y$), and is placed at the sender's  (receiver's) end, as  illustrated in Figure \ref{fig:place}.  
Explicitly, we define  the \textit{basic placement supermap} as: 
  \begin{align}\label{eq:place}
 \map S_{\rm place}^{A,B}  (\map N)    : =  \map W^B \circ \map N \circ \map V^A  \, , 
 \end{align} 
where $\map V^A \in  \Chan (A,X)$  and $\map W^B  \in  \Chan ( Y,B)$ are unitary channels implementing the isomorphisms $A\simeq X$ and $Y\simeq B$, respectively.  

We will call the systems $A$ and $B$ {\em placed systems},  and the channel $\map C:  = \map S^{A,B}_{\rm place} (\map N)$ a {\em placed channel}.    In the following, we will use the letters $\map N$ and $\map C$  for unplaced and placed channels, respectively.  In figures, we will represent  unplaced channels as red boxes, and placed channels as green boxes. This choice of colours reflects the fact that the placed channels are ready to be used by the communicating parties, while the unplaced channels have yet to be made available to them.

Once a device is in place, the sender and receiver can use it to communicate to one another. 
Typically, the communication is achieved by connecting the communication device  with other devices present at the sender's and receiver's locations. For example, one end of an optical fibre could be connected to  a computer, used by the sender to type  an email, and the other end of the fibre could be connected to another computer,  used by the receiver to read the email. The operations performed by the sender and receiver can be described by a supermap \cite{chiribella2008transforming,chiribella2009theoretical,Chiribella2013}  transforming placed channels in $\Chan (A, B)$ into  placed channels in  $\Chan (A', B')$, where $A'$ and $B'$ are two new input and output systems, also placed in the sender's and receiver's locations, respectively.

Ref.\ \cite{chiribella2008transforming} showed that the most general supermap $\map S$ transforming a generic input  channel $\map C  \in  \Chan (A, B)$ into an output channel $\map S (\map C)  \in  \Chan(A',B')$ has the form  
\begin{align}\label{allsupermaps}
\map S   (\map C)      =   \map D \circ (\map C\otimes \map I_{\rm Aux})  \circ   \map E \, ,
\end{align}
where $\rm Aux$ is an auxiliary quantum system, and $\map E  \in  \Chan (A',A\otimes {\rm Aux})$ and $\map D \in  \Chan (B \otimes {\rm Aux},B')$  are quantum channels.    These supermaps define the set of all possible operations on  input channels, and play the role of the set $\mathsf M$ in the general resource-theoretic framework described in Subsection \ref{sec:resources}.

\begin{figure}
	\centering
\begin{tikzpicture}[scale=1.2]
	\begin{pgfonlayer}{nodelayer}
	\draw[fill=green!90!white] (-3.75,0.25) -- (-3.75,1.75) -- (-1.75,1.75) -- (-1.75,0.25) -- (-3.75,0.25) -- (-3.75,1.75);
	\draw[fill=Orchid!90!white] (-6.25,0.25) -- (-6.25,1.75) -- (-4.75,1.75) -- (-4.75,0.25) -- (-6.25,0.25) -- (-6.25,1.75);
	\draw[fill=Orchid!90!white] (-0.75,0.25) -- (-0.75,1.75) -- (0.75,1.75) -- (0.75,0.25) -- (-0.75,0.25) -- (-0.75,1.75);
		\node [style=none] (0) at (-0.75, 1.75) {};
		\node [style=none] (1) at (-0.75, 0.25) {};
		\node [style=none] (2) at (0.75, 1.75) {};
		\node [style=none] (3) at (0.75, 0.25) {};
		\node [style=none] (4) at (-1.75, 1.75) {};
		\node [style=none] (5) at (-3.75, 1.75) {};
		\node [style=none] (6) at (-3.75, 0.25) {};
		\node [style=none] (7) at (-1.75, 0.25) {};
		\node [style=none] (8) at (-1.75, 1) {};
		\node [style=none] (9) at (-0.75, 1) {};
		\node [style=none] (10) at (0.75, 1) {};
		\node [style=none] (11) at (-3.75, 1) {};
		\node [style=none] (12) at (-4.75, 1) {};
		\node [style=none] (13) at (-1.25, 1.75) {$B$};
		\node [style=none] (14) at (-4.25, 1.75) {$A$};
		\node [style=none, font={\large}] (15) at (-2.75, 1) {$\map{C}$};
		\node [style=none, font={\large}] (16) at (0, 1) {$\map{D}$};
		\node [style=none] (17) at (-7.25, 1) {};
		\node [style=none, font={\large}] (18) at (-5.5, 1) {$\map{E}$};
		\node [style=none] (19) at (-4.75, 1) {};
		\node [style=none] (20) at (-6.25, 1) {};
		\node [style=none] (21) at (-4.75, 0.25) {};
		\node [style=none] (22) at (-6.25, 0.25) {};
		\node [style=none] (23) at (-6.25, 1.75) {};
		\node [style=none] (24) at (-4.75, 1.75) {};
		\node [style=none] (25) at (-6.75, 1.75) {$A'$};
		\node [style=none] (26) at (1.25, 1.75) {$B'$};
		\node [style=none] (27) at (1.75, 1) {};
	\end{pgfonlayer}
	\begin{pgfonlayer}{edgelayer}
		\draw (5.center) to (4.center);
		\draw (4.center) to (7.center);
		\draw (7.center) to (6.center);
		\draw (6.center) to (5.center);
		\draw (0.center) to (2.center);
		\draw (2.center) to (3.center);
		\draw (3.center) to (1.center);
		\draw (1.center) to (0.center);
		\draw (11.center) to (12.center);
		\draw (23.center) to (24.center);
		\draw (24.center) to (21.center);
		\draw (21.center) to (22.center);
		\draw (22.center) to (23.center);
		\draw (20.center) to (17.center);
		\draw (10.center) to (27.center);
		\draw (8.center) to (9.center);
	\end{pgfonlayer}
\end{tikzpicture}
\caption{\label{fig:encdec} {\em Encoding-decoding supermap $\map S_{\map E, \map D}   (\map C)   :=  \map D   \circ \map C  \circ \map E .$} In this paper, the placed quantum channels are drawn in green, while the encoding-decoding supermaps are drawn in violet.} 
\end{figure}
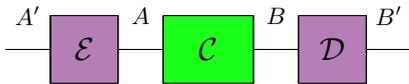

To specify the set of free operations, one has to specify a {\em subset} of the set of all supermaps. The standard choice  (in the absence of additional resources such as shared entanglement or shared randomness) is to require the  free operations to have  the form  
\begin{equation}\label{freesupermapsStandard}
\map S_{\map E,\map D}   (\map C)   :=  \map D   \circ \map C  \circ \map E   
\end{equation} 
(see  Figure \ref{fig:encdec} for an illustration).  Operationally, this choice of $\mathsf{M}_\textrm{free}$ is justified by the fact that the supermaps  \eqref{freesupermapsStandard} can be achieved by performing a local encoding operation $\map E$ at the sender's side and a local decoding operation $\map D$ at the receiver's side, without requiring the transmission of any  system   other than the system sent through the channel $\map C$.   

Note that while the supermaps \eqref{freesupermapsStandard} are the standard  choice, other choices could be made. For example, one could consider   quantum communication with the assistance of  classical communication \cite{bennett1996entanglement}, or classical communication with the assistance of shared entanglement  \cite{bennett1999entanglement}.  In these scenarios, the set of free supermaps is larger than the set of supermaps of the form (\ref{freesupermapsStandard}), and contains supermaps that can be achieved with the 
additional resources 
 under consideration. The characterisation of such supermaps is provided  in Appendix \ref{app:altsupermaps}. In the following, however, we  will stick to the simplest choice of free supermaps, namely the choice  in \eqref{freesupermapsStandard}. 

In general, we will refer to  supermaps from unplaced channels to placed channels as \textit{placement supermaps}, and we will interpret them as being performed either by a communication provider,  or by Nature itself. We will refer to supermaps from placed channels to placed channels as \textit{party supermaps}, and will  interpret them as being performed by the communicating parties.

\subsection{Direct communication from a sender to a receiver through multiple channels}\label{subsect:direct_mult}

So far, we have considered operations on a  single quantum channel. We now extend the resource-theoretic formulation to   scenarios where multiple communication channels  (or multiple uses of the same communication channel) are available.  

Consider a communication protocol that uses   $k$ communication devices, described by $k$ unplaced channels $\map N_1, \dots, \map N_k$, with $\map N_i \in \Chan(X_i,Y_i)$ for $i\in\{1,\dots, k\}$. We denote by $( \map N_1, \dots ,\map N_k)$  
 the resource corresponding to a single use of each device. Again, the list  $( \map N_1, \dots ,\map N_k)$ is interpreted as a description of the hardware before it is placed between the sender and receiver. For example, the hardware could be a list of optical fibres with some given specifications, {\em viz.}\ attenuation coefficient, bandwidth, and length. 
	
In Appendix \ref{sec:placed_unplaced} we show that  the list $( \map N_1, \dots ,\map N_k)$ can be interpreted as an equivalent notation for the product channel $ \map N_1\otimes \cdots \otimes \map N_k$, viewed as an element of a suitable set of channels (namely, $k$-partite no-signalling channels). In the following, we will use the list notation $( \map N_1, \dots ,\map N_k)$ as a visual reminder that the channels $ \map N_1, \dots ,\map N_k$ are unplaced.

In the direct communication scenario,  it is understood that all the input systems  are placed in Alice's laboratory, and all the output systems  are placed in Bob's laboratory.  Equivalently, this means that the communication devices are  placed  in parallel between the sender and the receiver.  The operation of placing the devices in parallel is described by the {\em parallel placement supermap}  $\map S_{\rm par}^{\bf A,  \bf B}$ defined by
 \begin{equation}\label{eq:place_par}
\map  S_{\rm par}^{\bf A, \bf B}  (\map N_1,\dots,  \map N_k) : =    \map S_{\rm place}^{A_1,B_1}   (\map N_1) \otimes \cdots \otimes   \map S_{\rm place}^{A_k,B_k}  (\map N_k) \, .
\end{equation}
where  ${\bf A} :  =  (A_1, \dots,  A_k )$   [${\bf B} :  =  (B_1, \dots,  B_k )$]  is a list of quantum systems placed in Alice's (Bob's) laboratory, with $A_i \simeq X_i$ and $B_i \simeq  Y_i$ for every $i\in  \{1,\dots, k\}$. 
The result of the supermap is a placed quantum channel in $\Chan (  A_1\otimes \cdots \otimes A_k ,  B_1\otimes \cdots \otimes B_k)$.

A large body of results in standard quantum Shannon theory  refers to  channels combined  in parallel as in Equation (\ref{eq:place_par}).  For example,  Smith and Yard    \cite{smith2008super} showed that, surprisingly,  the parallel composition of two channels with zero quantum capacity can give rise to a channel  with non-zero quantum capacity.  This phenomenon became known as {\em activation of the quantum capacity}.

\subsection{Network communication from a sender to a receiver}\label{subsec:network}

Let us now consider a communication scenario where the sender (Alice) and receiver (Bob) communicate through a network of communication devices. To begin with, we focus on the simple case where Alice and Bob  communicate through two devices, which are connected by an intermediate party (Ray), who 
  serves as a ``repeater'' passing to Bob the information received from Alice.   

The initial resource is described by a pair of unplaced channels $( \map N_1, \map N_2) \in \Chan (X_1,Y_1) \times  \Chan (X_2,  Y_2) $. 
The operation of placing channel $\map N_1$ between Alice and Ray, and channel $\map N_2$ between Ray and Bob is described by the {\em sequential placement supermap}  $\map S_{\rm seq}^{A,R,R',B} $ defined by
\begin{align}\label{eq:seq}
\map S_{\rm seq}^{A,R,R',B}  (\map N_1,  \map N_2)   :=   \map S_{\rm place}^{A, R}  (\map N_1)  \otimes  \map S_{\rm place}^{ R', B} ( \map N_2)   \,, 
\end{align}
  where system $A \simeq  X_1$ is placed in Alice's laboratory, systems $R\simeq Y_1$ and $R'\simeq X_2$ are placed in Ray's laboratory, and system $B\simeq Y_2$ is placed in Bob's laboratory. 
 
 Note that the sequential placement \eqref{eq:seq}  is formally identical to the parallel placement \eqref{eq:place_par}: in both cases, the placement of multiple channels is the tensor product of the placement of individual channels.  The difference between parallel and sequential placement arises  from the different spacetime locations in which the inputs and outputs of the channels are placed. In the parallel placement, all the input systems $\bf A$ are   at the sender's location, and  all  the output systems  $\bf B$ are  at the receiver's location.   In the sequential placement, the systems $A, R, R', B $ appear in  a strict sequential order: $A$ before $R$, $R$ before $R'$, $R'$ before $B$.    This difference is crucial when it comes to specifying how the output of the placement supermap is to be used: in the case of parallel placement, the output of the supermap can be connected with local operations at the sender's and receiver's ends. In the case of sequential placement, intermediate operations are possible.     
  
The difference between sequential and parallel placements is reflected by the different type of channels they generate. The output of the sequential placement supermap   \eqref{eq:seq} is a {\em two-step quantum process}, where the first step represents the transfer of information from $A$ to $R$, and the second step corresponds to the transfer of information from $R'$ to $B$. Mathematically, a two-step  process $\map C$ transforming system $S_1$ into system  $S_1'$ in the first step, and system $S_2$ into system $S_2'$ in the second step,  is described a quantum channel $\map C   \in  \Chan  (S_1\otimes S_2 ,  S_1'\otimes S_2')$ satisfying the condition   \cite{beckman2001causal,eggeling2002semicausal,chiribella2008quantum} 
\begin{equation}
\begin {split}
\label{comb1}  \Tr_{ S_2'}  [  \map C   (  \rho)]  &=   \Tr_{ S_2'}  \left[  \map C   \left(   \Tr_{S_2}  [\rho]\otimes \frac {I_{S_2}}{d_{S_2}}  \right)\right] \\
&\forall \rho  \in \St (S_1\otimes S_2) \, ,
\end {split}
\end{equation}
where $\Tr_S$, $I_S$, and $d_S$ denote the  trace over  $\spc H_S$, the identity on $\spc H_S$, and the dimension of $\spc H_S$, respectively.  The difference between a two-step process and a generic bipartite channel is that the two-step process has to satisfy the  additional condition \eqref{comb1}, which  ensures compatibility with the causal ordering of the systems $S_1,  S_1', S_2,$ and $S_2'$.   
  
  Two-step quantum processes are known in the literature as quantum combs \cite{chiribella2008quantum,chiribella2009theoretical},  quantum memory channels \cite{kretschmann2005quantum,caruso2014quantum}, and  non-Markovian quantum processes \cite{pollock2018nonmarkov}. Following Refs.\ \cite{chiribella2008quantum,chiribella2009theoretical}, we will refer to two-step quantum processes as {\em quantum $2$-combs}, and we will denote the corresponding set    as $ \Comb  [( S_1,S_1' )  \,,\,  (S_2,  S_2') ]$. 
  
 The sequential placement supermap (\ref{eq:seq}) transforms a pair of unplaced channels $(\map N_1,\map N_2)$ into a placed  2-comb $\map C_1\otimes \map C_2$, with $\map C_1 : =  \map S_{\rm place}^{A,R}  (\map N_1) $ and $\map C_2  :  =  \map S_{\rm place}^{R', B}  (\map N_2)$.     Note that, in general, the set of 2-combs  also  contains maps that are not of the product form  $\map C_1\otimes \map C_2$.  These maps correspond to two-step processes where a memory is passed from the first step to the second.  
 
 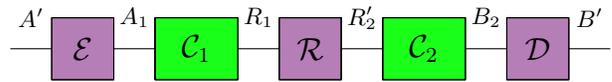
\begin{figure}
 	\centering
 	\begin{tikzpicture}[scale=1.1]
	\begin{pgfonlayer}{nodelayer}
	
	\draw[fill=green!90!white] (-3.75,0.25) -- (-3.75,1.75) -- (-1.75,1.75) -- (-1.75,0.25) -- (-3.75,0.25) -- (-3.75,1.75);
	\draw[fill=green!90!white] (1.75,0.25) -- (1.75,1.75) -- (3.75,1.75) -- (3.75,0.25) -- (1.75,0.25) -- (1.75,1.75);
	\draw[fill=Orchid!90!white] (0.75,0.25) -- (0.75,1.75) -- (-0.75,1.75) -- (-0.75,0.25) -- (0.75,0.25) -- (0.75,1.75);
	\draw[fill=Orchid!90!white] (4.75,0.25) -- (4.75,1.75) -- (6.25,1.75) -- (6.25,0.25) -- (4.75,0.25) -- (4.75,1.75);
	\draw[fill=Orchid!90!white] (-4.75,0.25) -- (-4.75,1.75) -- (-6.25,1.75) -- (-6.25,0.25) -- (-4.75,0.25) -- (-4.75,1.75);
		\node [style=none] (0) at (1.75, 0.25) {};
		\node [style=none] (1) at (1.75, 1.75) {};
		\node [style=none] (2) at (3.75, 1.75) {};
		\node [style=none] (3) at (3.75, 0.25) {};
		\node [style=none] (4) at (0.75, 1.75) {};
		\node [style=none] (5) at (0.75, 0.25) {};
		\node [style=none] (6) at (-0.75, 0.25) {};
		\node [style=none] (7) at (-0.75, 1.75) {};
		\node [style=none] (8) at (-1.75, 1.75) {};
		\node [style=none] (9) at (-3.75, 1.75) {};
		\node [style=none] (10) at (-3.75, 0.25) {};
		\node [style=none] (11) at (-1.75, 0.25) {};
		\node [style=none] (12) at (-1.75, 1) {};
		\node [style=none] (13) at (-0.75, 1) {};
		\node [style=none] (14) at (0.75, 1) {};
		\node [style=none] (15) at (1.75, 1) {};
		\node [style=none] (16) at (3.75, 1) {};
		\node [style=none] (18) at (-3.75, 1) {};
		\node [style=none] (20) at (-1.25, 1.75) {$R_1$};
		\node [style=none] (21) at (-4.25, 1.75) {$A_1$};
		\node [style=none, font={\large}] (22) at (-2.75, 1) {$\map{C}_1$};
		\node [style=none, font={\large}] (23) at (0, 1) {$\map{R}$};
		\node [style=none, font={\large}] (24) at (2.75, 1) {$\map{C}_2$};
		\node [style=none] (26) at (1.25, 1.75) {$R'_2$};
		\node [style=none] (27) at (4.25, 1.75) {$B_2$};
		\node [style=none] (28) at (6.25, 1.75) {};
		\node [style=none] (29) at (6.25, 0.25) {};
		\node [style=none] (30) at (4.75, 0.25) {};
		\node [style=none] (31) at (4.75, 1.75) {};
		\node [style=none] (32) at (3.75, 1) {};
		\node [style=none] (33) at (4.75, 1) {};
		\node [style=none] (34) at (6.25, 1) {};
		\node [style=none] (35) at (7.25, 1) {};
		\node [style=none, font={\large}] (36) at (5.5, 1) {$\map{D}$};
		\node [style=none] (37) at (-4.75, 1.75) {};
		\node [style=none] (38) at (-4.75, 0.25) {};
		\node [style=none] (39) at (-6.25, 0.25) {};
		\node [style=none] (40) at (-6.25, 1.75) {};
		\node [style=none] (41) at (-7.25, 1) {};
		\node [style=none] (42) at (-6.25, 1) {};
		\node [style=none] (43) at (-4.75, 1) {};
		\node [style=none] (44) at (-3.75, 1) {};
		\node [style=none, font={\large}] (45) at (-5.5, 1) {$\map{E}$};
		\node [style=none] (46) at (-6.75, 1.75) {$A'$};
		\node [style=none] (47) at (6.75, 1.75) {$B'$};
	\end{pgfonlayer}
	\begin{pgfonlayer}{edgelayer}
		\draw (9.center) to (8.center);
		\draw (8.center) to (11.center);
		\draw (11.center) to (10.center);
		\draw (10.center) to (9.center);
		\draw (7.center) to (4.center);
		\draw (4.center) to (5.center);
		\draw (5.center) to (6.center);
		\draw (6.center) to (7.center);
		\draw (1.center) to (2.center);
		\draw (2.center) to (3.center);
		\draw (3.center) to (0.center);
		\draw (0.center) to (1.center);
		\draw (15.center) to (14.center);
		\draw (13.center) to (12.center);
		\draw (31.center) to (28.center);
		\draw (28.center) to (29.center);
		\draw (29.center) to (30.center);
		\draw (30.center) to (31.center);
		\draw (35.center) to (34.center);
		\draw (33.center) to (32.center);
		\draw (40.center) to (37.center);
		\draw (37.center) to (38.center);
		\draw (38.center) to (39.center);
		\draw (39.center) to (40.center);
		\draw (44.center) to (43.center);
		\draw (42.center) to (41.center);
	\end{pgfonlayer}
\end{tikzpicture}
 	\caption{\label{fig:seq} {\em Encoding-repeater-decoding supermap $\map S_{\map E,\map R,\map D}  (\map C_1 \otimes \map C_2)  :=  \map D   \circ \map C_2 \circ \map R \circ \map C_1  \circ \map E $.} Party supermaps (i.e.\ supermaps from placed channels to placed channels) are drawn in violet.} 
 \end{figure}
  
  Once the devices have been placed,  the sender, repeater, and receiver  can  connect them with their local devices,  thus establishing a single channel that transfers information  directly from the sender to the receiver.   The most general supermaps  from quantum combs to quantum channels have been characterised in Ref.\ \cite{chiribella2009theoretical}. Their action on product combs $\map C_1\otimes \map C_2$ is given by 
 \begin{align}\label{3comb}  
 \map S  (  \map C_1 \otimes  \map C_2)    =      \map D \circ (  \map C_2\otimes \map I_{\rm {Aux_2}}) \circ  \map R \circ   (\map C_1\otimes \map I_{{\rm Aux}_1}) \circ \map E   \, ,
 \end{align}
 where ${\rm Aux}_1$ and ${\rm Aux}_2$ are auxiliary systems, and $\map E$, $\map R$, and $\map D$ are arbitrary channels in $\Chan  (A',  A\otimes {\rm Aux_1})$,  $\Chan  ( R \otimes {\rm Aux_1}  ,  R' \otimes {\rm Aux}_2 )$, and $\Chan ( B\otimes {\rm Aux}_2 ,  B')$, respectively.   
 
The standard choice of free supermaps is the supermaps that are achievable without the auxiliary systems $\rm Aux_1$ and $\rm Aux_2$, that is, the supermaps of the form 
\begin{equation}\label{eq:free_seq}
\map S_{\map E,\map R,\map D}  (\map C_1 \otimes \map C_2)  :=  \map D   \circ \map C_2 \circ \map R \circ \map C_1  \circ \map E \, , 
\end{equation} 
 illustrated in Figure \ref{fig:seq}.

Communication through a network of $k\ge 2$  devices is described by a direct generalisation of the above example. Consider the situation where a sender communicates to a receiver with the assistance of $k-1$ intermediate repeaters.  The communication devices are described by a  list of unplaced channels $( \map N_1, \dots ,\map N_{k}) \in \Chan (X_1,Y_1) \times \Chan (X_2, Y_2) \times \cdots \times \Chan (X_k,Y_k)$. The placement of the devices between the sender, repeaters, and receiver is described by the supermap  
\begin{align} 
\nonumber &\map S_{\rm seq}^{A,  R_1, R_1' ,\dots,  R_{k-1},  R_{k-1}'  ,B}  (\map N_1,\dots,  \map N_k)  \\
\label{eq:seq2}
 & :=   \map S_{\rm place}^{A, R_1}  (\map N_1)  \otimes \map S_{\rm place}^{R_1', R_2}  (\map N_2) \otimes \cdots \otimes  \map S_{\rm place}^{R_{k-1}', B}  (\map N_k)
 \, , 
\end{align} 
where  system $A\simeq X_1$ is placed in the sender's laboratory, system $B\simeq  Y_k$ is placed in the receiver's laboratory,  and  systems $R_i\simeq Y_i$  and $R_i'\simeq X_{i+1}$ are placed in the laboratory of the $i$-th repeater, for $i\in  \{1,\dots,  k-1\}$. 

The output of the supermap $\map S_{\rm seq}^{A,R_1,R_1',\dots,  R_{k-1}, R_{k-1}',B}$ is  a $k$-step quantum processes \cite{kretschmann2005quantum}, also known as a {\em quantum $k$-comb}  \cite{chiribella2009theoretical}. A quantum $k$-comb transforming system $S_i$ into system $S_i'$ at the $i$-th step is a quantum channel $\map C  \in  \Chan  ( S_1\otimes \cdots \otimes S_k  ,  S_1'\otimes \cdots \otimes S_k')$ satisfying a generalisation of condition \eqref{comb1} to $k$ steps (see Appendix \ref{sec:placed_unplaced} for the precise definition).
 The set of quantum $k$-combs  with the above input/output systems will be denoted by $\Comb  [  (S_1,S_1'), \dots, (S_k,  S_k')]$. 
 
Once the available devices have been placed, the communicating parties can connect their local devices to the placed communication channels. The corresponding supermap has the form 
\begin{equation}
\begin{split}
 &\map S_{\map E,  \map R_1, \dots, \map R_{k-1},\map D}  (\map C_1\otimes  \dots \otimes \map C_k) \\
\label{eq:enc-rep-dec}  & :  = \map D\circ \map C_k \circ \map R_{k-1}\circ \map C_{k-1} \circ \cdots \circ \map R_{1} \circ \map C_1\circ \map E \, ,
\end{split}
  \end{equation}
where $\map E \in \Chan  ( A',A)$ is the encoding operation performed by the sender, $\map R_i  \in \Chan  (R_i,R_i')$ is the repeater operation performed by the $i$-th intermediate party, and $\map D  \in \Chan  ( B,B' )$ is the decoding operation performed by the receiver.

\begin{figure}
	\centering
	\begin{tikzpicture}[scale=1.2]
	\begin{pgfonlayer}{nodelayer}
		\draw[fill=green!90!white] (-3.75,0.25) -- (-3.75,1.75) -- (-1.75,1.75) -- (-1.75,0.25) -- (-3.75,0.25) -- (-3.75,1.75);
		\draw[fill=green!90!white] (1.75,0.25) -- (1.75,1.75) -- (3.75,1.75) -- (3.75,0.25) -- (1.75,0.25) -- (1.75,1.75);
		\draw[fill=green!90!white] (1.75,-0.25) -- (1.75,-1.75) -- (3.75,-1.75) -- (3.75,-0.25) -- (1.75,-0.25) -- (1.75,-1.75);
		\draw[fill=Orchid!90!white] (0.75,-1.75) -- (0.75,1.75) -- (-0.75,1.75) -- (-0.75,-1.75) -- (0.75,-1.75) -- (0.75,1.75);
		\draw[fill=Orchid!90!white] (4.75,-1.75) -- (4.75,1.75) -- (6.25,1.75) -- (6.25,-1.75) -- (4.75,-1.75) -- (4.75,1.75);
		\draw[fill=Orchid!90!white] (-4.75,0.25) -- (-4.75,1.75) -- (-6.25,1.75) -- (-6.25,0.25) -- (-4.75,0.25) -- (-4.75,1.75);
		\node [style=none] (0) at (1.75, 0.25) {};
		\node [style=none] (1) at (1.75, 1.75) {};
		\node [style=none] (2) at (3.75, 1.75) {};
		\node [style=none] (3) at (3.75, 0.25) {};
		\node [style=none] (4) at (1.75, -0.25) {};
		\node [style=none] (5) at (1.75, -1.75) {};
		\node [style=none] (6) at (3.75, -1.75) {};
		\node [style=none] (7) at (3.75, -0.25) {};
		\node [style=none] (8) at (4.75, 1.75) {};
		\node [style=none] (9) at (4.75, -1.75) {};
		\node [style=none] (10) at (6.25, 1.75) {};
		\node [style=none] (11) at (6.25, -1.75) {};
		\node [style=none] (12) at (0.75, 1.75) {};
		\node [style=none] (13) at (0.75, -1.75) {};
		\node [style=none] (14) at (-0.75, -1.75) {};
		\node [style=none] (15) at (-0.75, 1.75) {};
		\node [style=none] (16) at (-1.75, 1.75) {};
		\node [style=none] (17) at (-3.75, 1.75) {};
		\node [style=none] (18) at (-3.75, 0.25) {};
		\node [style=none] (19) at (-1.75, 0.25) {};
		\node [style=none] (20) at (-1.75, 1) {};
		\node [style=none] (21) at (-0.75, 1) {};
		\node [style=none] (22) at (-0.75, -1) {};
		\node [style=none] (23) at (0.75, 1) {};
		\node [style=none] (24) at (1.75, 1) {};
		\node [style=none] (25) at (1.75, -1) {};
		\node [style=none] (26) at (0.75, -1) {};
		\node [style=none] (27) at (3.75, 1) {};
		\node [style=none] (28) at (4.75, 1) {};
		\node [style=none] (29) at (3.75, -1) {};
		\node [style=none] (30) at (4.75, -1) {};
		\node [style=none] (31) at (6.25, 1) {};
		\node [style=none] (32) at (6.25, -1) {};
		\node [style=none] (33) at (-3.75, 1) {};
		\node [style=none] (34) at (-4.75, 1) {};
		\node [style=none] (35) at (-1.25, 1.75) {$R_1$};
		\node [style=none] (36) at (-4.25, 1.75) {$A_1$};
		\node [style=none, font={\large}] (37) at (-2.75, 1) {$\map{C}_1$};
		\node [style=none, font={\large}] (38) at (0, 0) {$\map{R}$};
		\node [style=none, font={\large}] (39) at (2.75, 1) {$\map{C}_2$};
		\node [style=none, font={\large}] (40) at (2.75, -1) {$\map{C}_3$};
		\node [style=none, font={\large}] (41) at (5.5, 0) {$\map{D}$};
		\node [style=none] (42) at (-7.25, 1) {};
		\node [style=none, font={\large}] (43) at (-5.5, 1) {$\map{E}$};
		\node [style=none] (44) at (-4.75, 1) {};
		\node [style=none] (45) at (-6.25, 1) {};
		\node [style=none] (46) at (-4.75, 0.25) {};
		\node [style=none] (47) at (-6.25, 0.25) {};
		\node [style=none] (48) at (-6.25, 1.75) {};
		\node [style=none] (49) at (-4.75, 1.75) {};
		\node [style=none] (50) at (-6.75, 1.75) {$A'$};
		\node [style=none] (51) at (1.25, 1.75) {$R'_2$};
		\node [style=none] (52) at (4.25, 1.75) {$B_2$};
		\node [style=none] (53) at (1.25, -1.75) {$R'_3$};
		\node [style=none] (54) at (4.25, -1.75) {$B_3$};
		\node [style=none] (55) at (6.75, 1.75) {$B'$};
		\node [style=none] (56) at (7.25, 1) {};
	\end{pgfonlayer}
	\begin{pgfonlayer}{edgelayer}
		\draw (17.center) to (16.center);
		\draw (16.center) to (19.center);
		\draw (19.center) to (18.center);
		\draw (18.center) to (17.center);
		\draw (15.center) to (12.center);
		\draw (12.center) to (13.center);
		\draw (13.center) to (14.center);
		\draw (14.center) to (15.center);
		\draw (1.center) to (2.center);
		\draw (2.center) to (3.center);
		\draw (3.center) to (0.center);
		\draw (0.center) to (1.center);
		\draw (4.center) to (7.center);
		\draw (7.center) to (6.center);
		\draw (6.center) to (5.center);
		\draw (5.center) to (4.center);
		\draw (8.center) to (10.center);
		\draw (10.center) to (11.center);
		\draw (11.center) to (9.center);
		\draw (9.center) to (8.center);
		\draw (30.center) to (29.center);
		\draw (28.center) to (27.center);
		\draw (24.center) to (23.center);
		\draw (25.center) to (26.center);
		\draw (21.center) to (20.center);
		\draw (33.center) to (34.center);
		\draw (48.center) to (49.center);
		\draw (49.center) to (46.center);
		\draw (46.center) to (47.center);
		\draw (47.center) to (48.center);
		\draw (45.center) to (42.center);
		\draw (31.center) to (56.center);
	\end{pgfonlayer}
\end{tikzpicture}
	\caption{\label{fig:qst} {\em A composite supermap.} The figure shows the placement of $k=3$ channels between a sender, a single repeater ($r=1$), and a receiver. The placement  is then followed by an encoding-repeater-decoding supermap (in violet), representing the local operations performed by sender, repeater, and receiver. }
\end{figure}
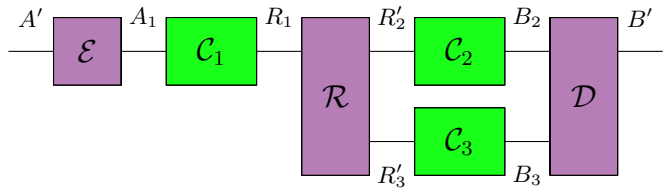

More generally, one can consider any placement of $k\ge 2$ devices with $r\le k-1$ intermediate repeaters. This includes placing some channels in parallel between two subsequent parties, in which case the placed channel is a quantum $(r+1)$-comb. An example of this situation is illustrated in Figure \ref{fig:qst}. 
The most general  placement supermaps corresponding to a definite causal structure of communicating parties are described in Appendix \ref{app:causal_structure}.

\subsection{Terminology}
In the rest of the paper,  the study of communication protocols involving only parallel placement between a sender and a receiver will be called \textit{standard quantum Shannon theory for direct communication}.  
The study of communication protocols involving both parallel and sequential placements between a sender, a receiver, and intermediate  parties will be called  \textit{standard quantum Shannon theory for network communication}, or simply, \textit{standard quantum Shannon theory}.  We will  not consider assisted scenarios, such as entanglement-assisted communication, which can nevertheless  be incorporated in our framework  as discussed in Appendix \ref{app:altsupermaps}.

\section{General resource theories of communication}\label{sec:genresc}  
  Here we  extend the framework of standard quantum Shannon theory to  general resource theories of communication, arguing that any such  theory must not include operations that enable communication independently of the communication devices initially available to the communicating parties.

\subsection{Basic structure}
The resource-theoretic formulation of standard quantum Shannon theory, discussed in the previous Section,  suggests a general scheme for constructing new resource theories   of communication. The basic scheme is as follows: 
\begin{enumerate}
	\item One use of each communication device is described by an \textit{unplaced quantum channel}, specifying how a given system type is transformed into another system type, but without assigning these system types to specific locations in spacetime.
	\item The (uses of the) available communication devices  are described by  a \textit{list of unplaced quantum channels}.
	\item  The sender, receiver, and possibly a set of intermediate parties are assigned spacetime regions, whose causal structure specifies who can send messages to whom.   The physical systems accessed by the communicating parties are {\em placed systems}, that is, systems assigned  specific locations in spacetime.
	\item The placement of the communication devices in between the communicating parties is described by  a \textit{placement supermap}, that is, a supermap transforming  lists of unplaced quantum channels into placed quantum channels.	A placed quantum channel has placed systems as inputs and outputs, and can in general be a multistep process, represented by a quantum comb.
	\item The operations performed by the sender, receiver, and intermediate parties are described by a  \textit{party supermap}, that is,  a supermap on the set of placed quantum channels. 
\end{enumerate}
In the above scheme, a resource theory of communication is formulated by specifying  which operations are considered as ``free''  in points 4 and 5 above. 

Free operations on placed channels (party supermaps) are interpreted as being implemented by the sender, the receiver, or  intermediate parties. Free operations from unplaced to placed channels  (placement supermaps) are interpreted as being performed by an external agent, e.g.\ a communication provider, or Nature itself. This is consistent with the intuitive idea that a communication infrastructure  has to be set up  before communication takes place.   Overall, a resource theory of communication  describes the  actions performed by  the communicating parties and by  an external agent  that places the communication devices between them.

In principle, one could also consider a third type of operations, from unplaced channels to unplaced channels. These operations would be performed by the third party {\em before the channels are placed between the sender and receiver}. For example, the third party could decide to discard one of the devices in the list $(\map N_1,\dots, \map N_k)$, and use only the remaining devices. 
For completeness, we will include the possibility of  these ``pre-placement operations'' in our general scheme.

 \subsection{Resource theories of communication}

 For a resource theory of communication, the broader set of operations $\mathsf M$ from which the free operations  $\mathsf M_{\rm free}$  are chosen  consists of (1) supermaps from unplaced channels to unplaced channels, (2) supermaps from unplaced channels to placed channels, and (3) supermaps from placed channels to placed channels. The mathematical classification of these three types of admissible supermaps is given in Appendix \ref{subsec:admissible}.

 A resource theory of communication is then specified by fixing the set of free operations: 
  
 \begin{defi}
	\emph{(Resource theory of communication.)}  A resource theory of communication is specified  by a set of free supermaps $ \mathsf{M}_{\rm free}\subset \mathsf M$, closed under sequential and parallel composition,  containing (1)  free supermaps from unplaced channels to unplaced channels, called {\em pre-placement supermaps},  (2) free supermaps from unplaced channels to placed channels, called {\em placement supermaps},   and (3) free supermaps from placed channels to placed channels, called {\em party supermaps}.  
\end{defi}

In pictures, we represent the placement supermaps by blue boxes, and the party supermaps by violet boxes.  

Mathematically, the different channel types are objects in a symmetric monoidal category, and the free operations
$\mathsf{M}_{\rm free}$ correspond to the morphisms between them. This scheme matches the general framework of Coecke, Fritz and Spekkens \cite{coecke2016}.

The set $\mathsf{M}_\textrm{free}$ can be specified by a generating set of operations \cite{coecke2016}.  For example, standard quantum Shannon theory is the resource theory of communication  where the free  operations $\mathsf M_{\rm standard}$ are generated from the following types of free operations:
\begin{enumerate}[label={(\roman*)}]
\item\label{op:place} \textit{Basic placement:}  For a single channel $\map N \in \Chan (X, Y)$, the map 
\begin{equation*}
\map S_{\rm place}^{A,B}  (\map N)    : =  \map W^B \circ \map N \circ \map V^A ,
\end{equation*}
where  $\map V \in \Chan  (A, X)$  [$\map W \in \Chan (Y,B)$] is  the unitary channel implementing the isomorphism between the unplaced system $X$  ($Y$) and the placed system $A$  ($B$).
\item\label{op:encdec} \textit{Insertion of  local devices:}   
For $l$ placed channels $\map C_1 \otimes \dots  \otimes \map C_l \in \Comb[(A,  R_1),  (R_1',  R_2), \dots,  (R_{l-1}',  B)] $, 
the {\em encoding map} 
\begin{equation}\label{eq:gen_enc}
\hspace{3em}\map S_{\map E}  (\map C_1 \otimes \dots  \otimes \map C_l)   :  = (\map C_1  \circ \map E)   \otimes  \map C_2\otimes \cdots  \otimes \map C_l \, , 
\end{equation}
the {\em repeater map}
\begin{equation}\label{eq:gen_rep}
\begin{split}
 \hspace{2.3em}& \map S_{\map R_m}  (\map C_1 \otimes \dots  \otimes \map C_l)  :  =  \map C_1  \otimes  \cdots \otimes \map C_{m-1} \\
    & \qquad   \otimes  (\map C_{m+1} \circ  \map R_m \circ \map C_{m}  )   \otimes  \map C_{m+2} \otimes \cdots  \otimes \map C_l \, , 
\end{split}
\end{equation}
and the {\em decoding map} 
\begin{equation}\label{eq:gen_dec}
\hspace{3em} \map S_{\map D}  (\map C_1 \otimes \dots  \otimes \map C_l)   :  = \map C_1  \otimes   \cdots  \otimes \map C_{l-1}   \otimes (\map D\circ \map C_l) \, , 
\end{equation}
where $  \map E  \in  \Chan  (  A',A)$, $\map R_m  \in  \Chan (  R_m, R_m')$, and $\map D \in \Chan  ( B,B')$ are quantum channels representing local devices at the sender's, $m$-th repeater's, and receiver's end, respectively. 
\end{enumerate} 
Note that we omitted pre-placement   supermaps,  because the set of such supermaps is  trivial  in standard quantum Shannon theory.

The other supermaps shown earlier in Section \ref{sec:framework} can be decomposed into the basic supermaps  \ref{op:place}--\ref{op:encdec}. For example, the parallel placement \eqref{eq:place_par} and sequential placement  \eqref{eq:seq2} are just the product of basic placement supermaps  \ref{op:place}, which place individual channels in the appropriate configuration.  
Similarly, the encoding-decoding supermap  \eqref{freesupermapsStandard} and the encoding-repeater-decoding-supermap \eqref{eq:free_seq} are just the result of multiple insertions of  local devices \ref{op:encdec}.

\subsection{Generalised channel capacities}
	
In standard quantum Shannon theory,
the classical (quantum) capacity of a quantum channel $\map N$ is defined as the maximum number bits (qubits) that can be transmitted over $n$ parallel uses of $\map N$,  per channel use and with vanishing error in the asymptotic limit $n \rightarrow \infty$. This is equivalent to the maximum number of classical (quantum) identity channels $\map I_\textrm{clas}$  ($\map I$) that the $n$ parallel uses of $\map N$ can simulate, per channel use and with vanishing error in the asymptotic limit $n \rightarrow \infty$, using arbitrary encoding/decoding channels \cite{wilde2013quantum} (the classical identity channel $\map I_\textrm{clas}$ being defined as the perfect dephasing channel with respect to a given orthonormal basis).

The standard definition of classical (quantum) capacity is appropriate for placed channels, which have already been arranged in between the sender and receiver, and therefore can only be used in parallel.   However,   unplaced channels  could be arranged  in more general configurations, generating a broader class of communication protocols.

In a general resource theory of communication,  
we define the {\em generalised classical (quantum) capacity} of $\map N$ as the maximum number of classical (quantum) identity channels $\map I_\textrm{clas}$ ($\map I$) that can be generated  by performing   free operations of $ \mathsf{M}_\textrm{free}$, per channel use and with vanishing error in the asymptotic limit of $n\to \infty$.   Other types of generalised capacities can be defined similarly, with respect to some given ideal reference channel.

The generalised capacity  is (trivially) a resource monotone \cite{chitambar19resources,coecke2016}, meaning that it cannot be increased by applying free operations. 
Moreover, the generalised capacity increases (or stays the same) whenever the set of free operations is enlarged. Example of this situation are   the  capacity enhancements observed in the presence of quantum control over the causal orders  \cite{ebler2018enhanced,salek2018quantum,chiribella2018indefinite,goswami2018communicating,procopio2019communication,procopio2020sending}: in these protocols,  the set of placements of standard quantum Shannon theory is enlarged to include placements in a superposition of alternative orders, and consequently various channel capacities have been shown to increase.

\subsection{A  minimal requirement for  any resource theory of communication}\label{subsec:minimal}

Formally, every set of free supermaps defines a resource theory of communication. However, such a resource theory may not be a meaningful one.   We argue that every meaningful resource theory of communication should at least satisfy a minimal requirement: 
the free operations should not 
allow the sender and receiver to communicate independently of the communication devices from which their communication protocol is built.

To illustrate this idea, consider the situation where two parties, Alice and Bob,  communicate through a noisy telephone line.   In the standard theory of communication, the key question is how to use this communication resource to transmit information reliably. 
Now, if Alice were to walk into Bob's room, he would clearly be able to hear her through the air, but this would not be a new way to use the telephone line. Rather, it would be a way to bypass it.  
 The air would act as a side-channel, allowing Alice and Bob to communicate to each other independently of how good or how bad their telephone line is.

The telephone line example has the following structure.  Initially, Alice and Bob have access to a noisy communication channel $\map N  \in \Chan (A,B)$.  
The operation of Alice moving into Bob's lab can be modelled as a {\em side-channel supermap} 
\begin{equation}\label{sidechannel}
\begin{split}
 \map S_{\rm side}^{(E,E')} : \quad  &  \Chan  (A,B)  \to \Chan  (A\otimes E,  B\otimes E')   \\
 &\map S (\map N)  :  \map N \mapsto \map N \otimes \map I_{E,E'} \, ,
 \end{split}
\end{equation}
 which juxtaposes the noisy channel $\map N$ with a side-channel  $\map I_{E,E'} \in \Chan(E,E')$ acting on some additional systems $E$ and $E'$ (the  air in the proximity of  Alice and  Bob, respectively). If the channel $\map I_{E,E'}$ is ideal,  then the supermap   $\map S^{(E,E')}_{\rm side}$ would let Alice communicate  perfectly to Bob. 
  This communication ``enhancement", however, is independent of the original channel $\map N$. 
 Every operation of the form \eqref{sidechannel} trivialises the notion of communication enhancement, and therefore should not be allowed in a resource theory of communication.

Building on the above example, we now propose a general notion of a side-channel generating operation: 
\begin{defi}\label{def:sc}
\emph{(Side-channel generating operations.)} 
A supermap $\map S   \in \mathsf M $  {\em  generates a classical (quantum) side-channel} if there exist two free supermaps $\map S_{1}  \in \mathsf{M}_{\rm free}$ and $\map S_{2} \in \mathsf{M}_{\rm free}$ such that,  for all choices of input  channels $( \map N_1, \dots, \map N_k)$ for supermap $\map S_1$, one has 
\begin{align}\label{eq:nsc}
(\map S_{2}  \circ \map S  \circ \map S_{1} )( \map N_1, \dots, \map N_k)  = \map C  \, ,
\end{align}
where $\map C$ is a placed  quantum channel with non-zero classical (quantum) capacity.   
\end{defi}

The above definition captures the idea that the supermap $\map S$ can be used to construct a communication protocol that works independently of the communication devices originally available to the communicating parties.    
In the telephone line example, the channel $\map C$ is the ideal channel $\map I_{E,E'}$ describing the transmission of a message through the air between Alice and Bob.

We demand that any sensible resource theory of communication should forbid side-channel generating operations:
\begin{cond} \label{nsc}
	\emph{(No Side-Channel Generation.)}
	In a resource theory of classical (quantum) communication, no free operation $\map S \in \mathsf M_{\rm free}$  should generate a classical (quantum) side-channel.
\end{cond}
We stress that Condition \ref{nsc} is a {\em minimal} requirement, and that, in particular cases, one may want to impose even stronger conditions on the allowed operations. 
In other words, we are not claiming that every resource theory of communication satisfying Condition  \ref{nsc}   is an interesting one. Rather, Condition  \ref{nsc}  is a bottom line that has to be satisfied when defining new resource theories of communication. 

It is immediate to verify that standard quantum Shannon theory  satisfies  Condition  \ref{nsc}. In the following, we will show that 
\begin{enumerate}
\item quantum Shannon theory with superpositions of causal orders satisfies Condition  \ref{nsc}
\item quantum Shannon theory with superpositions of trajectories satisfies Condition  \ref{nsc}
\item quantum Shannon theory with superpositions of encoding and decoding operations violates Condition  \ref{nsc}. 
\end{enumerate} 

In Appendix \ref{app:comparison}, we comment on the difference between our   framework and the  frameworks of Refs.  \cite{liu-yuan19resources,liu-winter19resources,takagi2019resource}, discussing an alternative to Condition \ref{nsc}, where the free supermaps are required to transform constant channels into constant channels.  In Appendix \ref{app:alternativesidechannel} we  discuss the difference between our definition of side-channels and another notion of side-channels proposed in Ref.\ \cite{guerin2018shannon}, assessing some of the claims made therein.

\section{Superposition of orders and superposition of trajectories \label{sec:sups}}

Here we formulate the resource theories of quantum Shannon theory with superpositions of causal orders and quantum Shannon theory with superpositions of trajectories, and we show that both theories satisfy the requirement of No Side-Channel Generation.

\subsection{Quantum Shannon theory with superpositions of causal orders}\label{subsec:switch}
The information-theoretic advantages of indefinite causal order in quantum computation were envisaged by Hardy \cite{hardy2009quantum}, and fleshed out  a few years later with the introduction of the quantum {\tt SWITCH} \cite{chiribella2009beyond,Chiribella2013}, a higher-order operation that places two quantum devices in a superposition of two alternative causal orders. Since then,  information-processing advantages of the quantum {\tt SWITCH} have been found in a variety of contexts, including  quantum query complexity \cite{Chiribella2012,araujo2014computational}, quantum communication complexity \cite{guerin2016exponential}, and quantum metrology \cite{zhao2019quantum}.   Other forms of indefinite causal order, and their advantages in non-local games, have been demonstrated in Ref.\ \cite{oreshkov2012quantum}.   In all the above works, the combination of quantum devices in an  indefinite causal order  was shown to offer performances that cannot be matched by any quantum protocol that uses the input devices in a definite order. 

A different category of advantages arises  in the context of quantum communication \cite{ebler2018enhanced, salek2018quantum, chiribella2018indefinite, goswami2018communicating,procopio2019communication,procopio2020sending,loizeau2020channel}. Here,  protocols that combine  communication channels through the quantum {\tt SWITCH} have been shown to offer advantages  with respect  to the protocols allowed in standard quantum Shannon theory, as defined earlier in this paper. These advantages are {\em not} advantages with respect to all possible protocols with definite causal order. They cannot be so, because the set of all  protocols with definite causal order   includes also trivial protocols where the original communication channels are juxtaposed with noiseless channels, as in the telephone line example of Equation (\ref{sidechannel}).     

The proper way to interpret the communication advantages shown in  Refs.\ \cite{ebler2018enhanced, salek2018quantum, chiribella2018indefinite, goswami2018communicating,procopio2019communication,procopio2020sending,loizeau2020channel} is to regard them as a comparison between two different resource theories of communication: standard quantum Shannon theory, and an extended resource theory that includes the quantum {\tt SWITCH} among its placements.

\begin{figure}
	\centering
	\begin{tikzpicture}[scale=1.2]
	\begin{pgfonlayer}{nodelayer}
	\draw[fill=red!90!white] (-1,1) -- (-1,2.5) -- (1,2.5) -- (1,1) -- (-1,1) -- (-1,2.5);
	\draw[fill=red!90!white] (-1,-1) -- (-1,-2.5) -- (1,-2.5) -- (1,-1) -- (-1,-1) -- (-1,-2.5);
	\draw[fill=Turquoise!90!white] (-3.5,2.5) -- (-2,2.5) -- (-2,0.5) -- (2,0.5) --  (2,2.5) -- (3.5,2.5) -- 
	(3.5,-2.5) -- (2,-2.5) -- (2,-0.5) -- (-2,-0.5) --  (-2,-2.5) -- (-3.5,-2.5)-- (-3.5,2.5) -- (-2,2.5);
	\draw[fill=Orchid!90!white] (-6,1) -- (-6,2.5) -- (-4.5,2.5) -- (-4.5,1) -- (-6,1) -- (-6,2.5);
	\draw[fill=Orchid!90!white] (6,2.5) -- (6,-2.5) -- (4.5,-2.5) -- (4.5,2.5) -- (6,2.5) -- (6,-2.5);
		\node [style=none] (0) at (-1, 1) {};
		\node [style=none] (1) at (-1, 2.5) {};
		\node [style=none] (2) at (1, 2.5) {};
		\node [style=none] (3) at (1, 1) {};
		\node [style=none] (4) at (-1, -1) {};
		\node [style=none] (5) at (-1, -2.5) {};
		\node [style=none] (6) at (1, -2.5) {};
		\node [style=none] (7) at (1, -1) {};
		\node [style=none] (8) at (2, 2.5) {};
		\node [style=none] (9) at (2, -2.5) {};
		\node [style=none] (10) at (3.5, 2.5) {};
		\node [style=none] (11) at (3.5, -2.5) {};
		\node [style=none] (12) at (4.5, -2.5) {};
		\node [style=none] (13) at (4.5, 2.5) {};
		\node [style=none] (14) at (6, 2.5) {};
		\node [style=none] (15) at (6, -2.5) {};
		\node [style=none] (16) at (-2, 2.5) {};
		\node [style=none] (17) at (-2, -2.5) {};
		\node [style=none] (18) at (-3.5, -2.5) {};
		\node [style=none] (19) at (-3.5, 2.5) {};
		\node [style=none] (20) at (-4.5, 2.5) {};
		\node [style=none] (21) at (-6, 2.5) {};
		\node [style=none] (22) at (-6, 1) {};
		\node [style=none] (23) at (-4.5, 1) {};
		\node [style=none] (24) at (-4.5, 1.75) {};
		\node [style=none] (25) at (-3.5, 1.75) {};
		\node [style=none] (26) at (-3.5, -1.75) {};
		\node [style=none] (27) at (-4.5, -1.75) {};
		\node [style=none] (28) at (-2, 1.75) {};
		\node [style=none] (29) at (-1, 1.75) {};
		\node [style=none] (30) at (-1, -1.75) {};
		\node [style=none] (31) at (1, 1.75) {};
		\node [style=none] (32) at (2, 1.75) {};
		\node [style=none] (33) at (1, -1.75) {};
		\node [style=none] (34) at (2, -1.75) {};
		\node [style=none] (35) at (3.5, 1.75) {};
		\node [style=none] (36) at (4.5, 1.75) {};
		\node [style=none] (37) at (4.5, -1.75) {};
		\node [style=none] (38) at (3.5, -1.75) {};
		\node [style=none] (39) at (-6, 1.75) {};
		\node [style=none] (40) at (-7, 1.75) {};
		\node [style=none] (41) at (6, 1.75) {};
		\node [style=none] (42) at (7, 1.75) {};
		\node [style=none] (43) at (-6.5, 2.5) {$A'$};
		\node [style=none] (44) at (-4, 2.5) {$A$};
		\node [style=none] (45) at (4, 2.5) {$B$};
		\node [style=none] (46) at (6.5, 2.5) {$B'$};
		\node [style=none] (47) at (-4, -2.5) {$O$};
		\node [style=none] (48) at (4, -2.5) {$O$};
		\node [style=none, font={\large}] (49) at (-5.25, 1.75) {$\map{E}$};
		\node [style=none, font={\large}] (50) at (0, 1.75) {$\map{N}_1$};
		\node [style=none, font={\large}] (51) at (0, -1.75) {$\map{N}_2$};
		\node [style=none, font={\large}] (52) at (5.25, 0) {$\map{D}$};
		\node [style=none] (53) at (-4.5, -1.25) {};
		\node [style=none] (54) at (-4.5, -2.25) {};
		\node [style=none] (56) at (-2, 0.5) {};
		\node [style=none] (57) at (-2, -0.5) {};
		\node [style=none] (58) at (2, 0.5) {};
		\node [style=none] (59) at (2, -0.5) {};
		\node [style=none] (60) at (-1.5, 2.5) {$X$};
		\node [style=none] (61) at (1.5, 2.5) {$X$};
		\node [style=none] (62) at (-1.5, -2.5) {$X$};
		\node [style=none] (63) at (1.5, -2.5) {$X$};
		\node [style=none, font={\large}] (64) at (0, 0) {};
		\node [style=none] (65) at (-2, -1.75) {};
		\node [style=none] (66) at (0, 0) {};
		\node [style=none] (67) at (2, -1.75) {};
		\node [style=none] (68) at (0, 0) {};
		\node [style=none] (69) at (2, 1.75) {};
		\draw [bend right=90, looseness=3.25,fill=Turquoise!90!white] (53.center) to (54.center);
	\end{pgfonlayer}
	\begin{pgfonlayer}{edgelayer}
	\node [style=none] (55) at (-5, -1.75) {$\, \, \omega$};
		\draw (21.center) to (20.center);
		\draw (20.center) to (23.center);
		\draw (23.center) to (22.center);
		\draw (22.center) to (21.center);
		\draw (19.center) to (16.center);
		\draw (17.center) to (18.center);
		\draw (18.center) to (19.center);
		\draw (1.center) to (2.center);
		\draw [in=90, out=-90, looseness=1.00] (2.center) to (3.center);
		\draw (3.center) to (0.center);
		\draw (0.center) to (1.center);
		\draw (4.center) to (7.center);
		\draw (7.center) to (6.center);
		\draw (6.center) to (5.center);
		\draw (5.center) to (4.center);
		\draw (8.center) to (10.center);
		\draw [in=90, out=-90, looseness=1.00] (10.center) to (11.center);
		\draw (11.center) to (9.center);
		\draw (13.center) to (14.center);
		\draw (14.center) to (15.center);
		\draw (15.center) to (12.center);
		\draw [in=-90, out=90, looseness=1.00] (12.center) to (13.center);
		\draw (42.center) to (41.center);
		\draw [in=0, out=180, looseness=1.00] (36.center) to (35.center);
		\draw (37.center) to (38.center);
		\draw (34.center) to (33.center);
		\draw (32.center) to (31.center);
		\draw (29.center) to (28.center);
		\draw (26.center) to (27.center);
		\draw (25.center) to (24.center);
		\draw (39.center) to (40.center);
		\draw (53.center) to (54.center);
		\draw (16.center) to (56.center);
		\draw (56.center) to (58.center);
		\draw [in=-90, out=90, looseness=1.00] (58.center) to (8.center);
		\draw (57.center) to (59.center);
		\draw (59.center) to (9.center);
		\draw (57.center) to (17.center);
		\draw [dashed, in=-30, out=0, looseness=1.75] (64.center) to (32.center);
		\draw [dashed, in=150, out=180, looseness=1.75] (66.center) to (65.center);
		\draw (65.center) to (30.center);
		\draw [dotted, in=30, out=0, looseness=1.75] (68.center) to (67.center);
		\draw [dotted, in=-150, out=180, looseness=1.75] (64.center) to (28.center);
		\draw [dashed] (25.center) to (28.center);
		\draw [dotted] (32.center) to (35.center);
		\draw [dashed, in=-135, out=0, looseness=1.00] (34.center) to (35.center);
		\draw [dotted, in=180, out=-45, looseness=1.00] (25.center) to (65.center);
	\end{pgfonlayer}
\end{tikzpicture}
	\vspace{0.5cm}
	\caption{\label{fig:switch} {\em Communication through the quantum {\tt SWITCH}.} The quantum {\tt SWITCH} placement $\map S^{A,B,\omega}_{\tt SWITCH}$ (in blue) places two quantum channels $(\map N_1$, $\map N_2)$ in a superposition of causal orders, determined by the fixed state $\omega \in \St(O)$, between a sender and receiver,  and is followed by the encoding-decoding supermap $\map S_{\rm \map E, \rm \map D}$ (in violet). The dashed and dotted lines illustrate the two alternative orders of applying $\map N_1$ and $\map N_2$, respectively.} 
\end{figure}
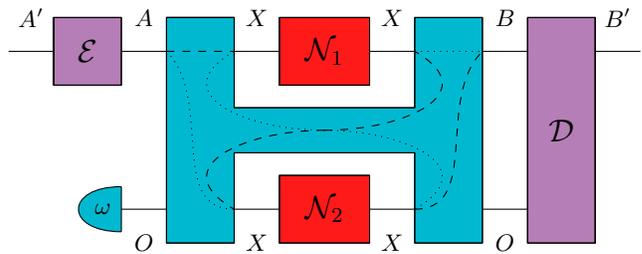

Here we explicitly define such a resource theory, which we call quantum Shannon theory with superpositions of causal orders (SCO). The corresponding set of free operations will be denoted by $\mathsf M_{\rm SCO}$. The generating free operations are operations  \ref{op:place}--\ref{op:encdec} of standard quantum Shannon theory, plus an additional placement supermap, based on   the quantum {\tt SWITCH}: 
\begin{enumerate}[label={(\roman*)}]
	\setcounter{enumi}{2}
	\item\label{op:switch} The {\em quantum {\tt SWITCH} placement} $\map S^{A,B,\omega}_{\tt SWITCH}$ maps a pair of  unplaced quantum channels $(\map N_1, \map N_2)  \in  \Chan  (X)\times \Chan (X)$ into a placed quantum channel $\map S^{A,B,\omega}_{\tt SWITCH} (\map N_1, \map N_2) \in  \Chan (A, B\otimes O)$, where $A\simeq  X$  ($B\simeq X$) is a quantum system placed at the sender's (receiver's) end, and $O$ is a qubit system, called the {\em order qubit},  placed at the receiver's end.   
	 Explicitly, the quantum channel $\map S^{A,B,\omega}_{\tt SWITCH} (\map N_1, \map N_2)$ is defined as 
	\begin{equation}\label{eq:switch}
\map S^{A,B,\omega}_{\tt SWITCH} (\map N_1, \map N_2)(\rho) = \sum_{i,j} S_{ij} (\rho  \otimes \omega) S^\dagger_{ij} \, ,
	\end{equation}
	where $\omega \in \St (O)$ is a state of the order qubit, and   
		\begin{equation}\label{controlorder}
\hspace{3em}S_{ij} := N^{(2)}_i N^{(1)}_j \! \otimes \! \ket{0}\bra{0} \! + N^{(1)}_j N^{(2)}_i \! \otimes \!  \ket{1}\bra{1} \, ,
	\end{equation} 
	$\{|0\>,|1\>\}$ being an orthonormal basis for the order qubit.    The quantum channel  $\map S^{A,B,\omega}_{\tt SWITCH} (\map N_1, \map N_2)$  is independent of the Kraus decomposition of the channels $\map N_1$ and $\map N_2$. 
	\end{enumerate}

A communication protocol using  the quantum {\tt SWITCH} placement is given in Figure \ref{fig:switch}. Note that the initial state of the order qubit 
is fixed as part of the placement,
and is thus inaccessible to the sender \cite{ebler2018enhanced,salek2018quantum}.

We stress that the quantum {\tt SWITCH} placement should be understood here as an abstract supermap from two quantum channels to a new quantum channel. Whether this supermap can be physically realised, and how it can be realised, is  entirely another matter. Various ways to reproduce the action of the quantum {\tt SWITCH}  have been proposed, using conventional physics \cite{procopio2015experimental,rubino2017experimental,goswami2018indefinite,goswami2018communicating,guo2020experimental,taddei2020experimental}, closed timelike curves \cite{Chiribella2013}, or quantum gravity scenarios \cite{zych2019bell,paunkovic2019quantum}. 
  However,  the resource theory $\mathsf M_{\rm SCO}$ should be considered as the abstract resource theory associated with the quantum {\tt SWITCH} transformation, without reference to a specific physical implementation. 
  
  The motivation  for including the quantum {\tt SWITCH} among the free operations  is to understand how the world \textit{could} be, if quantum devices could be combined in a superposition of alternative orders.  The study of  quantum Shannon theory with the addition of the quantum {\tt SWITCH} is similar in spirit the study of information tasks assisted by the Popescu-Rohrlich box \cite{popescu1994quantum}, a  fictional device that generates stronger than quantum correlations.   Like the Popescu-Rohrlich box,  the quantum {\tt SWITCH} serves as a conceptual device, used to better understand  standard quantum theory by comparing it to possible alternatives.

\subsection{Quantum Shannon theory with superpositions of trajectories}

The superposition of alternative evolutions was defined in Refs.\ \cite{Aharonov1990,oi2003interference}, and  applied to quantum communication  in Refs.\ \cite{gisin2005error,lamoureux2005experimental}, where the ability to send quantum particles along a superposition of different trajectories provided the working principle for a new technique called  error filtration.  Shannon-theoretic  advantages of the superposition of trajectories   were  demonstrated more recently in Refs.\ \cite{abbott2018communication,chiribella2019shannon2q,kristjansson2020latent}.

Here, we formulate the resource theory of quantum Shannon theory with superpositions of trajectories (ST) \cite{chiribella2019shannon2q}.
The set of free operations in this resource theory, denoted by $\mathsf M_{\rm ST}$, is generated by the standard free operations \ref{op:place}--\ref{op:encdec}, with the addition of a {\em superposition placement} \ref{op:superpos}, which creates a superposition of two alternative communication channels.

In order to define the superposition placement,  we need to  revise the way in which the communication hardware is modelled.   Normally, a quantum communication channel  $\map N  \in  \Chan (X)$ describes the action of a communication device {\em when a system is transmitted}.   However, the communication device  
 also exists when no system is sent through it.  The action of the device in the lack of an input  can be modelled by introducing a vacuum state, which can be sent to the device in alternative to states of system $X$.  Hence,  the overall action of the communication device  is described not by the original channel $\map N  \in \Chan (X)$, but  by another channel  $\widetilde{ \map N}$ that acts as $\map N$ when the input is restricted to $X$, and as the identity transformation $\map I _{\textrm{Vac}}$ when the input is in the vacuum state.  

The channel $\widetilde {\map N}$ is called a  \textit{vacuum extension} of the quantum channel  $\map N$ \cite{chiribella2019shannon2q}.   Mathematically, $\widetilde{\map N}$ is an element of  $\Chan (\widetilde X)$, where $\widetilde X:  =X\oplus {\rm Vac}$ is the quantum system with Hilbert space   $\spc H_{\widetilde X}  : =  \spc H_X  \oplus  \spc H_{\rm Vac}$,  $\spc H_{\rm Vac}$ being the Hilbert space of the vacuum.   Note that, in general,  a density matrix of system $\widetilde X$    can also have off-diagonal elements of the form $\ketbra{\psi}{\rm vac}$, with $\ket{\psi} \in \spc H_X$ and $\ket{\rm vac} \in \spc H_{\rm Vac}$, corresponding to the presence of quantum coherence between system $X$ and the vacuum.

In the following, we will assume for simplicity that the vacuum Hilbert space is one-dimensional, meaning that there exists a unique vacuum state $|{\rm vac}\>$, up to global phases.   With this assumption, the conditions for a channel $\widetilde{\map  N} \in  \Chan (\widetilde X)$ to be a vacuum extension of  channel $\map N  \in \Chan (X)$ are  
\begin{align}\label{vext1}
\widetilde  {\map N }   (  |{\rm vac}\>  \<  {\rm vac}|   )   =  |{\rm vac}\>  \<  {\rm vac}|  \, , 
\end{align}
and
\begin{align}\label{vext2}
\widetilde  {\map N }   (   P_X  \rho  P_X   )  = \map N (P_X  \rho  P_X )  \qquad \forall \rho  \in  \St (\widetilde X)  \, ,
  \end{align} 
where $P_X:  =  I-  |{\rm vac}\>  \<  {\rm vac}| $ is the projector on the subspace corresponding to system $X$.

Conditions  (\ref{vext1})  and (\ref{vext2}) imply that the Kraus operators of  channel $\widetilde{ \map N}$ are of the form 
\begin{equation}\label{vacext}
	\widetilde N_i  =  N_i  \oplus  \nu_i \,  |\rm vac \>\<\rm vac  |   \, ,
\end{equation}
where  $\{N_i\}$ are Kraus operators for  $\map N$, and $\{\nu_i\}$ are complex numbers  satisfying the condition  $\sum_i \, |   \nu_i|^2   =  1$.   In the following, the numbers $\{\nu_i\}$ will be  called  the \textit{vacuum amplitudes}  of channel $ \widetilde{ \map N}$.

The action of channel  $ \widetilde{ \map N}$ on a generic quantum state   $\rho  \in \St (\widetilde X)$ is 
\begin{equation}
\begin{split}
 \widetilde  {\map N }   (    \rho   )      =   \, &     {\map N }   (   P_X  \rho  P_X   )         +  \<{\rm vac}|    \rho  |{\rm vac}\>  \,      |\rm vac \>\<\rm vac  |     \\
& +   F  \, \rho \,   |\rm vac \>\<\rm vac  |     +   |\rm vac \>\<\rm vac  |   \,  \rho\,    {\mathit{F}}^{\dagger}  \, ,  \label{outputstate}
\end{split}
\end{equation}
where the operator
\begin{align}\label{vacint}
F   :=  \sum_i  \, \overline \nu_i\,  N_i \, ,
\end{align} is called the  {\em vacuum interference operator} \cite{chiribella2019shannon2q}.  Note that the operator $F$ depends only on the channel $\widetilde {\map N}$, and not on the choice of Kraus operators, as one can see by comparing the two sides of Equation (\ref{outputstate}).  

If the vacuum interference operator is zero, then the output state (\ref{outputstate}) is an incoherent mixture of a state of system $X$ and the vacuum.       
\begin{defi}\label{def:cohvac} For $F=  0$, we say that the vacuum extension  $\widetilde  {\map N }$ {\em has no coherence with the vacuum}, and we call it the {\em incoherent vacuum extension} of channel $\map N$. 
For  $F  \not =  0  $, we say that the vacuum extension $\widetilde  {\map N }$ {\em has coherence with the vacuum}.
\end{defi}

Mathematically, the vacuum extension of a given quantum channel is  highly non-unique: every channel has infinitely many vacuum extensions  \cite{chiribella2019shannon2q,abbott2018communication}.
 Physically, the choice of vacuum extension is part of the specification of the communication device, and can be determined through process tomography  \cite{oi2003interference}.

 Vacuum-extended channels represent communication devices that can act on the information carrier, or on the vacuum, or on any coherent superposition of the two. Using this feature, it is possible to coherently control the choice of channel through which the information carrier is sent. The result can be interpreted as   a placement of the given different channels in a superposition of being on the path of the information carrier:

 \begin{enumerate}[label={(\roman**)}]
\setcounter{enumi}{2}
\item\label{op:superpos}  The {\em superposition placement} $\map S^{A,B,\omega}_{\rm sup}$ maps a pair of unplaced vacuum-extended  channels $(\widetilde{\map N}_1,  \widetilde{ \map N}_2)  \in  \Chan  (\widetilde X)\times \Chan (\widetilde X)$  into a placed quantum channel $\map S^{A,B,\omega}_{\rm sup} (\map N_1, \map N_2) \in  \Chan (A, B\otimes P)$, where $A\simeq  X$  ($B\simeq X$) is a quantum system placed at the sender's (receiver's) end, and $P$ is a qubit system, called the {\em path qubit},  placed at the receiver's end.   
Explicitly, the quantum channel $\map S^{A,B,\omega}_{\rm sup} (\widetilde{\map N}_1, \widetilde{\map N}_2)$ is defined as  
\begin{equation}\label{supaction} 
\begin{split}
\hspace{2em}
\map S^{A,B,\omega}_{\rm sup} ( \widetilde {\map N}_1, \widetilde {\map N}_2 )  (\rho )   =    & \<1|  \omega |1\>    \,  \map N_1 (\rho)  \otimes |1\>\<1|    \\
     +  &\<  2|  \omega |2\>    \,  \map N_2 (\rho)  \otimes |2\>\<2|   \\
    +  & \<1|  \omega  |2\>    \,  F_1  \rho F_2^\dag \otimes |1\>\<2|    \\
    +  &\<2 | \omega|1\>    \,  F_2  \rho F_1^\dag \otimes |2\>\<1|    \, ,
 \end{split}
\end{equation}
where   $\omega \in  \St (P)$ is a state of the path qubit, $\{  |1\>  , |2\>\}$ is an orthonormal basis for the path qubit,  and $F_1$ and $F_2$ are the vacuum interference operators associated to channels $\widetilde  {\map N}_1$ and $\widetilde {\map N}_2$, respectively.   
\end{enumerate}

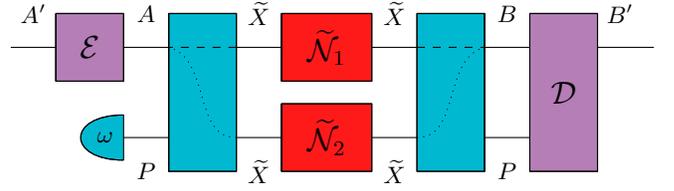
\begin{figure}
	\centering
	\begin{tikzpicture}[scale=1.2]
	\begin{pgfonlayer}{nodelayer}
	\draw[fill=red!90!white] (-1,0.25) -- (-1,1.75) -- (1,1.75) -- (1,0.25) -- (-1,0.25) -- (-1,1.75);
	\draw[fill=red!90!white] (-1,-0.25) -- (-1,-1.75) -- (1,-1.75) -- (1,-0.25) -- (-1,-0.25) -- (-1,-1.75);
	\draw[fill=Turquoise!90!white] (3.5,1.75) -- (3.5,-1.75) -- (2,-1.75) -- (2,1.75) -- (3.5,1.75) -- (3.5,-1.75);
	\draw[fill=Turquoise!90!white] (-3.5,1.75) -- (-3.5,-1.75) -- (-2,-1.75) -- (-2,1.75) -- (-3.5,1.75) -- (-3.5,-1.75);
	\draw[fill=Orchid!90!white] (-6,0.25) -- (-6,1.75) -- (-4.5,1.75) -- (-4.5,0.25) -- (-6,0.25) -- (-6,1.75);
	\draw[fill=Orchid!90!white] (6,1.75) -- (6,-1.75) -- (4.5,-1.75) -- (4.5,1.75) -- (6,1.75) -- (6,-1.75);
		\node [style=none] (0) at (-1, 0.25) {};
		\node [style=none] (1) at (-1, 1.75) {};
		\node [style=none] (2) at (1, 1.75) {};
		\node [style=none] (3) at (1, 0.25) {};
		\node [style=none] (4) at (-1, -0.25) {};
		\node [style=none] (5) at (-1, -1.75) {};
		\node [style=none] (6) at (1, -1.75) {};
		\node [style=none] (7) at (1, -0.25) {};
		\node [style=none] (8) at (2, 1.75) {};
		\node [style=none] (9) at (2, -1.75) {};
		\node [style=none] (10) at (3.5, 1.75) {};
		\node [style=none] (11) at (3.5, -1.75) {};
		\node [style=none] (12) at (4.5, -1.75) {};
		\node [style=none] (13) at (4.5, 1.75) {};
		\node [style=none] (14) at (6, 1.75) {};
		\node [style=none] (15) at (6, -1.75) {};
		\node [style=none] (16) at (-2, 1.75) {};
		\node [style=none] (17) at (-2, -1.75) {};
		\node [style=none] (18) at (-3.5, -1.75) {};
		\node [style=none] (19) at (-3.5, 1.75) {};
		\node [style=none] (20) at (-4.5, 1.75) {};
		\node [style=none] (21) at (-6, 1.75) {};
		\node [style=none] (22) at (-6, 0.25) {};
		\node [style=none] (23) at (-4.5, 0.25) {};
		\node [style=none] (24) at (-4.5, 1) {};
		\node [style=none] (25) at (-3.5, 1) {};
		\node [style=none] (26) at (-3.5, -1) {};
		\node [style=none] (27) at (-4.5, -1) {};
		\node [style=none] (28) at (-2, 1) {};
		\node [style=none] (29) at (-1, 1) {};
		\node [style=none] (30) at (-1, -1) {};
		\node [style=none] (31) at (-2, -1) {};
		\node [style=none] (32) at (1, 1) {};
		\node [style=none] (33) at (2, 1) {};
		\node [style=none] (34) at (1, -1) {};
		\node [style=none] (35) at (2, -1) {};
		\node [style=none] (36) at (3.5, 1) {};
		\node [style=none] (37) at (4.5, 1) {};
		\node [style=none] (38) at (4.5, -1) {};
		\node [style=none] (39) at (3.5, -1) {};
		\node [style=none] (40) at (-6, 1) {};
		\node [style=none] (41) at (-7, 1) {};
		\node [style=none] (42) at (6, 1) {};
		\node [style=none] (43) at (7.25, 1) {};
		\node [style=none] (44) at (-6.5, 1.75) {$A'$};
		\node [style=none] (45) at (-4, 1.75) {$A$};
		\node [style=none] (46) at (-1.5, 1.75) {$\widetilde{X}$};
		\node [style=none] (47) at (1.5, 1.75) {$\widetilde{X}$};
		\node [style=none] (48) at (4, 1.75) {$B$};
		\node [style=none] (49) at (6.5, 1.75) {$B'$};
		\node [style=none] (50) at (-4, -1.75) {$P$};
		\node [style=none] (51) at (-1.5, -1.75) {$\widetilde{X}$};
		\node [style=none] (52) at (1.5, -1.75) {$\widetilde{X}$};
		\node [style=none] (53) at (4, -1.75) {$P$};
		\node [style=none, font={\large}] (54) at (-5.25, 1) {$\map{E}$};
		\node [style=none, font={\large}] (56) at (0, 1) {$\map{\widetilde{N}}_1$};
		\node [style=none, font={\large}] (57) at (0, -1) {$\map{\widetilde{N}}_2$};
		\node [style=none, font={\large}] (59) at (5.25, 0) {$\map{D}$};
		\node [style=none] (60) at (-4.5, -0.5) {};
		\node [style=none] (61) at (-4.5, -1.5) {};
		\node [style=none] (63) at (3.5, 1) {};
		\node [style=none] (64) at (2, -1) {};
		\node [style=none] (65) at (-3.5, 1) {};
		\node [style=none] (66) at (-2, -1) {};
		\draw [bend right=90, looseness=3.25,fill=Turquoise!90!white] (60.center) to (61.center);
	\end{pgfonlayer}
	\begin{pgfonlayer}{edgelayer}
			\node [style=none] (62) at (-5, -1) {$\, \, \omega$};
		\draw (21.center) to (20.center);
		\draw (20.center) to (23.center);
		\draw (23.center) to (22.center);
		\draw (22.center) to (21.center);
		\draw (19.center) to (16.center);
		\draw (16.center) to (17.center);
		\draw (17.center) to (18.center);
		\draw [in=270, out=90] (18.center) to (19.center);
		\draw (1.center) to (2.center);
		\draw (2.center) to (3.center);
		\draw (3.center) to (0.center);
		\draw (0.center) to (1.center);
		\draw (4.center) to (7.center);
		\draw (7.center) to (6.center);
		\draw (6.center) to (5.center);
		\draw (5.center) to (4.center);
		\draw (8.center) to (10.center);
		\draw (10.center) to (11.center);
		\draw (11.center) to (9.center);
		\draw (9.center) to (8.center);
		\draw (13.center) to (14.center);
		\draw (14.center) to (15.center);
		\draw (15.center) to (12.center);
		\draw (12.center) to (13.center);
		\draw (43.center) to (42.center);
		\draw (37.center) to (36.center);
		\draw (38.center) to (39.center);
		\draw (26.center) to (27.center);
		\draw (25.center) to (24.center);
		\draw (40.center) to (41.center);
		\draw [bend right=90, looseness=3.25] (60.center) to (61.center);
		\draw (60.center) to (61.center);
		\draw [style=dashed line] (28.center) to (25.center);
		\draw [style=dashed line] (33.center) to (36.center);
		\draw [style=dotted line, in=-165, out=0] (64.center) to (63.center);
		\draw [style=dotted line, in=-15, out=180] (66.center) to (65.center);
		\draw (66.center) to (30.center);
		\draw (28.center) to (29.center);
		\draw (33.center) to (32.center);
		\draw (34.center) to (64.center);
	\end{pgfonlayer}
\end{tikzpicture}
	\vspace{0.5cm}
	\caption{\label{fig:sup} {\em Communication through a superposition of quantum channels.} The supermap $\map S^{A,B,\omega}_{\rm sup}$ (in blue) places two vacuum-extended channels $( \widetilde {\map N}_1, \widetilde {\map N}_2 )$ on two alternative paths, and lets the transmitted system travel along both paths (dashed line and dotted line, respectively) in a quantum superposition,  determined by the  state $\omega \in \St(P)$. The resulting channel then undergoes the encoding-decoding supermap $\map S_{\rm \map E, \rm \map D}$  (in violet), describing the local operations performed at the sender's and receiver's ends.} 
\end{figure}

An example of a communication protocol using the superposition placement is shown in Figure \ref{fig:sup}. The superposition placement  is physically implementable in photonic systems,  making the resource theory $\mathsf M_{\rm ST}$ interesting both from a purely information-theoretic point of view as well as a practical point of view.

For simplicity of presentation, here we considered only superpositions of two channels, both for the superposition of trajectories and for the superposition of orders.   Both the superposition placement  \ref{op:superpos} the quantum {\tt SWITCH} placement \ref{op:switch} can be straightforwardly generalised to $k$ channels. The corresponding definitions can be found in Refs.\  \cite{chiribella2019shannon2q} and \cite{colnaghi2012quantum}, respectively.

\subsection{Superpositions of causal orders and superpositions of trajectories do not generate side-channels}
 
We now show that  the supermaps \ref{op:switch} or \ref{op:superpos}, combined with  \ref{op:place}--\ref{op:encdec}, do not generate side-channels.
\begin{prop}\label{switch_nosc}
	No supermap composed from the quantum {\tt SWITCH} placement \emph{\ref{op:switch}}, basic placement \emph{\ref{op:place}}, and insertion of local devices \emph{\ref{op:encdec}} generates side-channels.   
\end{prop}
\Proof 
The supermaps \ref{op:place}--\ref{op:encdec} of standard quantum Shannon theory do not generate side channels. Hence,  it is sufficient to prove that  the quantum {\tt SWITCH} does not generate side-channels.

This is done by finding a choice of adversarial channels  $\map N_1$ and $\map N_2$ such that $\map S^{A,B,\omega}_{\tt SWITCH}  ({\map N}_1, {\map N}_2) $ is  a channel with zero classical capacity.     One such choice is to pick $\map N_1$ to be the identity channel $ \map I $ (with a single Kraus operator $N^{(1)}_i  =  I$,  $i=  1$) and $\map N_2$ to be the  constant channel $\map N_2 (\rho)  =  |\psi_0\>\<\psi_0| $  (with Kraus operators $N^{(2)}_j   = |\psi_0\>\<j|$, for some orthonormal basis $\{|j\>  \}$).  
 With this choice,  the Kraus operators (\ref{controlorder}) in the definition of the channel  $\map S^{A,B,\omega}_{\tt SWITCH}  ({\map N}_1, {\map N}_2) $  are 
 \begin{align}
 S_{ij}   =  |\psi_0\>\<j|  \otimes I \, ,
 \end{align} 
 and therefore one has
 \begin{equation}
 \begin{split}
 \map S^{A,B,\omega}_{\tt SWITCH}  ({\map N}_1,  {\map N}_2)  (\rho)     &:=  \sum_{i,j}  S_{ij}  (\rho \otimes \omega)  S_{ij}^\dag   \\ 
 &~=  |\psi_0\>\<\psi_0| \otimes \omega \quad \forall \rho  \in  \St (A) \, .
  \end{split}
 \end{equation}
Since the output of the channel    $\map S^{A,B,\omega}_{\tt SWITCH}  ({\map N}_1,  {\map N}_2) $  is independent of its input, the channel  has zero capacity (both classical and quantum), and no combination of it with the other supermaps \ref{op:place}--\ref{op:encdec} can generate a channel with non-zero capacity.    
\qed

\begin{prop}\label{sup_nosc}
	No supermap composed from the  superposition placement \emph{\ref{op:superpos}}, basic placement \emph{\ref{op:place}}, and insertion of local devices \emph{\ref{op:encdec}} generates side-channels.  
\end{prop}
\Proof 
As in the  proof of Proposition \ref{switch_nosc},  it is sufficient to prove that the superposition placement \ref{op:superpos} does not generate side-channels.
This is done by finding a choice of adversarial vacuum-extended channels  $\widetilde{\map N}_1$ and $\widetilde {\map N}_2$ such that $\map S^{A,B,\omega}_{\rm sup}  (\widetilde{\map N}_1, \widetilde {\map N}_2) $ is  a channel with zero classical capacity.     One such choice is to pick the vacuum-extended channels  $\widetilde{\map N}_1$ and $\widetilde {\map N}_2$ defined by
\begin{equation}
\begin{split}
 \widetilde{ \map N}_1 (\rho)   =\widetilde{ \map N}_2 (\rho)   
 &= \rho_0 \Tr[\rho\,  (I  -  |{\rm vac}\>\<  {\rm vac}|  ) ]   \\ &~~~+     |{\rm vac}\>\<  {\rm vac}|  \rho  |{\rm vac}\>\<  {\rm vac}| ~~~~ \forall \rho \in \St (\widetilde{X})   \, .
 \end{split}
\end{equation} 
In other words,  $\widetilde{\map N}_1  = \widetilde {\map N}_2$ is the incoherent vacuum-extension of the constant channel  that maps  every state into the fixed state $\rho_0$.   For the  vacuum-extended   channels $\widetilde{\map N}_1$ and $\widetilde {\map N}_2$,  the vacuum interference operators  are $F_1  =  F_2  =  0$, and  the superposition placement then yields the channel  
\begin{equation} \label{cleaner}
\map S^{A,B,\omega}_{\rm sup} ( \widetilde {\map N}_1, \widetilde{\map N}_2)  (\rho )    =  \rho_0  \otimes {\rm diag} (\omega) \, ,
\end{equation}  
with ${\rm diag}  (\omega) :  = \<1|  \omega |1\>  \, |1\>\<1|  + \<2 |\omega |2\> \, |2\>\<2|$, as one can verify from Equation (\ref{supaction}).      Since  the channel   $\map S^{A,B,\omega}_{\rm sup}  ( \widetilde {\map N}_1, \widetilde{\map N}_2)$ is constant, it has zero (classical and quantum) capacity.
\qed 

\vspace{1ex}

Propositions \ref{switch_nosc} and \ref{sup_nosc}  show that both quantum Shannon theory with superpositions of causal orders and  quantum Shannon theory with superpositions of trajectories satisfy the requirement of No Side-Channel Generation, as stated in Condition \ref{nsc}.

\section{Reply to  Abbott {\em et al.}\  }\label{sec:grenoblereply}

In Ref.\ \cite{abbott2018communication}, Abbott {\em et al.}\ give an example of a communication  protocol where two completely depolarising channels  are coherently superposed.  The authors quantify the transmission of information in terms of the  Holevo information (a lower bound for the classical capacity \cite{wilde2013quantum}), and show that the Holevo information achievable by superposing  the two  channels is  greater than the Holevo information achievable by putting them  in the quantum {\tt SWITCH}.

This observation is presented as a comparison  between two alternative ways to  turn two depolarising channels into  a new quantum channel with non-zero capacity. Based on this comparison, the authors argue that the communication advantages of the quantum {\tt SWITCH} ``\textit{should therefore rather be understood as resulting from coherent control of quantum communication channels},'' as opposed to being specifically due to indefinite causal order. 

The logic of this conclusion, however, does not seem to pass a careful scrutiny.   First, it is not clear how a  comparison between the values of the Holevo information for the quantum SWITCH and for the superposition of channels could be used to make any deduction  on the ``true origin'' of the respective advantages.  If anything, the comparison would show that the ability to control trajectories is  more powerful  than the ability to control causal orders. Second,
the  comparison made in \cite{abbott2018communication} is uneven,
because  
\begin{enumerate}
\item  it does not compare supermaps acting on the same input channels, and 
\item it does not compare superpositions where the depolarising channels act the same number of times.   
\end{enumerate}
A detailed analysis of these two points is provided in the following. 
\medskip 

{\em 1. Different input channels.} 
In quantum Shannon theory with superpositions of trajectories, the input  resources are  vacuum-extended channels,  while in  quantum Shannon theory with superpositions of causal orders the input resources are  ordinary (non-vacuum-extended)  channels.  

A vacuum-extended channel is  a stronger resource than the corresponding channel, because it can have coherence with the vacuum, in the sense of Definition \ref{def:cohvac}.   We now argue that coherence with the vacuum is indeed the  underlying resource implicit in the communication advantages of Ref.\ \cite{abbott2018communication}. 
Suppose that a particle is sent in a superposition of two paths, going through two communication devices, each of which  acts as  a completely depolarising channels on the internal degree of freedom of the particle.    The two devices are described by vacuum extensions of the completely depolarising channel, and act as  
\begin{align}
\nonumber \widetilde  {\map N }_{\rm dep}   (    \rho   )      =  &     (1- \<{\rm vac}|    \rho  |{\rm vac}\>  )  \, \frac I d       +  \<{\rm vac}|    \rho  |{\rm vac}\>  \,      |\rm vac \>\<\rm vac  |     \\
& +   F  \, \rho \,   |\rm vac \>\<\rm vac  |     +   |\rm vac \>\<\rm vac  |   \,  \rho\,   F^\dag  \, , 
\end{align}
where $F$ is the vacuum interference operator defined in Equation (\ref{vacint}). 
Now, if the channels have no coherence with the vacuum (that is, if $F=0$), then  their superposition yields the constant channel  
\begin{equation} 
\map S^{A,B,\omega}_{\rm sup} ( \widetilde {\map N}_{\rm dep}, \widetilde{\map N}_{\rm dep})  (\rho )    =  \frac I d    \otimes {\rm diag} (\omega) \, ,
\end{equation}  
following from Equation (\ref{cleaner}) with $\rho_0  = I/d$.  Since the output is independent of the input, the channel $\map S^{A,B,\omega}_{\rm sup}$ cannot be used to communicate.  

The above analysis shows that the presence of coherence with the vacuum is necessary for the advantages observed by Abbott {\em et al.}\ \cite{abbott2018communication}. In contrast, the presence of coherence with the vacuum is, in principle, unnecessary for the advantages of the quantum   {\tt SWITCH}.  For example, the implementation of the quantum {\tt SWITCH} via closed timelike curves \cite{chiribella2009beyond,Chiribella2013},  illustrated in Figure \ref{fig:ctc}, does not require any coherence with the vacuum.

\begin{figure}
	\centering
	\begin{tikzpicture}[scale=1.2, circuit ee IEC]
	\begin{pgfonlayer}{nodelayer}
		\draw[fill=red!90!white] (-1,0.25) -- (-1,1.75) -- (1,1.75) -- (1,0.25) -- (-1,0.25) -- (-1,1.75);
	\draw[fill=red!90!white] (-1,-0.25) -- (-1,-1.75) -- (1,-1.75) -- (1,-0.25) -- (-1,-0.25) -- (-1,-1.75);
	\draw[fill=Turquoise!90!white] (3.5,1.75) -- (3.5,-1.75) -- (2,-1.75) -- (2,1.75) -- (3.5,1.75) -- (3.5,-1.75);
	\draw[fill=Turquoise!90!white] (-3.5,1.75) -- (-3.5,-1.75) -- (-2,-1.75) -- (-2,1.75) -- (-3.5,1.75) -- (-3.5,-1.75);
		\node [style=none] (0) at (-1, 0.25) {};
		\node [style=none] (1) at (-1, 1.75) {};
		\node [style=none] (2) at (1, 1.75) {};
		\node [style=none] (3) at (1, 0.25) {};
		\node [style=none] (4) at (-1, -0.25) {};
		\node [style=none] (5) at (-1, -1.75) {};
		\node [style=none] (6) at (1, -1.75) {};
		\node [style=none] (7) at (1, -0.25) {};
		\node [style=none] (8) at (2, 1.75) {};
		\node [style=none] (9) at (2, -1.75) {};
		\node [style=none] (10) at (3.5, 1.75) {};
		\node [style=none] (11) at (3.5, -1.75) {};
		\node [style=none] (16) at (-2, 1.75) {};
		\node [style=none] (17) at (-2, -1.75) {};
		\node [style=none] (18) at (-3.5, -1.75) {};
		\node [style=none] (19) at (-3.5, 1.75) {};
		\node [style=none] (25) at (-3.5, 1) {};
		\node [style=none] (26) at (-3.5, -1) {};
		\node [style=none] (27) at (-4.5, 3) {};
		\node [style=none] (28) at (-2, 1) {};
		\node [style=none] (29) at (-1, 1) {};
		\node [style=none] (30) at (-1, -1) {};
		\node [style=none] (31) at (-2, -1) {};
		\node [style=none] (32) at (1, 1) {};
		\node [style=none] (33) at (2, 1) {};
		\node [style=none] (34) at (1, -1) {};
		\node [style=none] (35) at (2, -1) {};
		\node [style=none] (36) at (3.5, 1) {};
		\node [style=none] (39) at (3.5, -1) {};
		\node [style=none] (45) at (-4, 1.5) {$A$};
		\node [style=none] (48) at (4, 1.5) {$B$};
		\node [style=none] (50) at (-4, 3.5) {$O$};
		\node [style=none] (60) at (-4.5, 3.5) {};
		\node [style=none] (61) at (-4.5, 2.5) {};
		\node [style=none, font={\Large}] (63) at (-2.75, 3) {$\bullet$};
		\node [style=none] (64) at (-0.75, 3) {};
		\node [style=none] (65) at (0.75, 3) {};
		
		\node [style=none] (67) at (-2.75, 1.75) {};
		\node [style=none] (68) at (2.75, 1.75) {};
		\node [style=none] (69) at (-4.5, 3) {};
		\node [style=none] (70) at (5.5, 3) {};
		\node [style=none, font={\large}] (72) at (0, 1) {$\map N_1$};
		\node [style=none, font={\large}] (73) at (0, -1) {$\map N_2$};
		\node [style=none, font={\large}] (74) at (-2.75, 0) {${\tt SWAP}$};
		\node [style=none, font={\large}] (75) at (2.75, 0) {${\tt SWAP}$};
		\node [style=none] (76) at (-1.5, 1.5) {$X$};
		\node [style=none] (77) at (1.5, 1.5) {$X$};
		\node [style=none] (78) at (-1.5, -1.5) {$X$};
		\node [style=none] (79) at (1.5, -1.5) {$X$};
		\node [style=none] (80) at (5.5, 1) {};
		\node [style=none] (81) at (4.25, -1) {};
		\node [style=none] (82) at (-5.5, 1) {};
		\node [style=none] (83) at (-4.25, -1) {};
		\node [style=none] (84) at (-4.25, -3) {};
		\node [style=none] (85) at (4.25, -3) {};
		\node [style=none] (86) at (4.25, -1) {};
		\node [style=none] (87) at (4.25, -3) {};
		\node [style=none] (88) at (-4, -1.5) {$X$};
		\node [style=none] (89) at (4, -1.5) {$X$};
		\node [style=none] (91) at (4, 3.5) {$O$};
		\draw [bend right=90, looseness=3.25,fill=Turquoise!90!white] (60.center) to (61.center);
		\node [style=none, font={\Large}] (66) at (2.75, 3) {};
		\draw (66.center) to (68.center);
		\draw (66.center) to (70.center);
		\draw (65.center) to (66.center);
	\end{pgfonlayer}
	\begin{pgfonlayer}{edgelayer}
		\draw (64.center) to (65.center);
		\node [style=none] (62) at (-5, 3) {$\, \, \omega$};
		\draw (19.center) to (16.center);
		\draw (16.center) to (17.center);
		\draw (17.center) to (18.center);
		\draw (18.center) to (19.center);
		\draw (1.center) to (2.center);
		\draw (2.center) to (3.center);
		\draw (3.center) to (0.center);
		\draw (0.center) to (1.center);
		\draw (4.center) to (7.center);
		\draw (7.center) to (6.center);
		\draw (6.center) to (5.center);
		\draw (5.center) to (4.center);
		\draw (8.center) to (10.center);
		\draw (10.center) to (11.center);
		\draw (11.center) to (9.center);
		\draw (9.center) to (8.center);
		\draw (35.center) to (34.center);
		\draw (33.center) to (32.center);
		\draw (29.center) to (28.center);
		\draw (30.center) to (31.center);
		\draw [bend right=90, looseness=3.25] (60.center) to (61.center);
		\draw (60.center) to (61.center);
		\draw (67.center) to (63.center);
		\draw (63.center) to (64.center);

		\draw (69.center) to (63.center);
	
		\draw (82.center) to (25.center);
		\draw (36.center) to (80.center);
		\draw (83.center) to (26.center);
		\draw (81.center) to (39.center);
		\draw (84.center) to (85.center);
		\draw [in=-180, out=-180, looseness=2.00] (83.center) to (84.center);
		\draw [in=0, out=0, looseness=2.00] (86.center) to (87.center);
		\node [style=none, font={\Large},fill=white] (66) at (2.75, 3) {$\circ$};
	\end{pgfonlayer}
\end{tikzpicture}
	\vspace{0.5cm}
	\caption{\label{fig:ctc}{\em An implementation of the quantum {\tt SWITCH} placement (in blue) using closed timelike curves.} A quantum state $\rho \in \St(A)$ is  routed through one of the two channels $\map N_1$ or $\map N_2$ by a $ {\tt SWAP}$ gate controlled by the state of the order qubit $\omega \in \St(O)$. A second $ {\tt SWAP}$ gate (controlled in the opposite way) routes the state to a closed timelike curve, which transfers the incoming system back through the first  $ {\tt SWAP}$ gate, and through one of the two channels  $\map N_2$ or $\map N_1$.} 
\end{figure}

In summary, the advantages of Refs.\  \cite{ebler2018enhanced}   and \cite{abbott2018communication} arise from different input resources, with the resources used in   \cite{abbott2018communication} (vacuum-extended channels exhibiting coherence with the vacuum)   being strictly stronger  than the resources used in \cite{ebler2018enhanced} (ordinary,  non-vacuum-extended channels, possibly without coherence with the vacuum).  

\medskip

{\em 2.  Different numbers of uses of the depolarising channel.}    Refs. \cite{abbott2018communication} and \cite{ebler2018enhanced} 
 refer to two different communication scenarios: 
 \begin{enumerate}
 \item[ (a)] in  Ref.\ \cite{abbott2018communication}, the particle travels through {\em only one} depolarising channel ({\em either} $\map N_1$ \emph{or} $\map N_2$),
 \item[ (b)] in  Ref.\ \cite{ebler2018enhanced} the particle travels through {\em two} depolarising channels  ({\em both} $\map N_1$ {\em and}  $\map N_2$). 
\end{enumerate}	
From this point of view, there is little surprise that Scenario {(a)} allows more communication than Scenario {(b)}, given that in Scenario {(b)} the particle is exposed twice to depolarising noise, as acknowledged also by the authors of  Ref.\ \cite{abbott2018communication}.  

One may argue that the difference between Scenarios {(a)}  and  {(b)}    is irrelevant, because the completely  depolarising channel $\map N_{\rm dep}  (\cdot)  :  =  I/d \Tr[\cdot]$   satisfies  the equality   
\begin{align}\label{squaredep}
\map N_{\rm dep}  \circ \map N_{\rm dep}   = \map N_\textrm{dep} \,, 
\end{align} 
meaning that applying the channel twice in a row is the same as applying it once. 

 However, the input  resource for the superposition of two channels is not two depolarising channels themselves,  but rather  their vacuum extensions.   Crucially, algebraic identities like the one in Equation (\ref{squaredep}) do not carry over to the vacuum extensions: in general,  the relation $\map N_1 \circ  \map N_2   =  \map N_3$ {\em does not} imply the relation $\widetilde {\map N}_1   \circ \widetilde {\map N}_2  = \widetilde {\map N}_3$. In the particular case of depolarising channels, we have the following result: 
  \begin{prop}\label{prop:dep}
 	The condition   $\widetilde {\map N}_{\rm dep}  \circ \widetilde   {\map N}_{\rm dep}   = \widetilde   {\map N}_\textrm{dep}  $ is satisfied if and only if the vacuum extension  $\widetilde{ \map N}_{\rm dep}$ has no coherence with the vacuum.
 \end{prop}
\noindent The proof is given in Appendix \ref{app:dep_proof}.

In summary, the only case in which the equation $\widetilde{ \map N}_{\rm dep} \circ \widetilde{ \map N}_{\rm dep} = \widetilde{\map N}_{\rm dep}$ would justify a comparison between the Holevo information with a single depolarising channel and the Holevo information with two depolarising channels is exactly the case in which the vacuum extension $\widetilde {\map N}_\textrm{dep}$ has no coherence with the vacuum, and therefore  the protocol of Ref.\ \cite{abbott2018communication} provides no advantage. 

In order to make an even comparison with the quantum {\tt SWITCH}, one should analyse the scenario where information is sent along  a superposition of two paths, each visiting  {\em two} depolarising channels.  Mathematically, this superposition is described by the channel  $\map S^{A,B,\omega}_{\rm sup} ( \widetilde {\map N}_{\rm dep} \circ \widetilde {\map N}_{\rm dep}, \widetilde{\map N}_{\rm dep}\circ \widetilde {\map N}_{\rm dep})$, instead of the channel  $\map S^{A,B,\omega}_{\rm sup} ( \widetilde {\map N}_{\rm dep}, \widetilde{\map N}_{\rm dep})$ considered in Ref.\  \cite{abbott2018communication}.  

The  Holevo information of the channel $\map S^{A,B,\omega}_{\rm sup} ( \widetilde {\map N}_{\rm dep} \circ \widetilde {\map N}_{\rm dep}, \widetilde{\map N}_{\rm dep}\circ \widetilde {\map N}_{\rm dep})$ was recently evaluated in  Ref.\ \cite{kristjansson2020latent},  for a set of vacuum extensions constructed from the representation of the completely depolarising channel as a uniform  mixture of the four Pauli unitaries.  For this set of vacuum extensions,  the maximum Holevo information turned out to be 0.018, which is strictly  less than the value 0.049 of the Holevo information for the quantum {\tt SWITCH}.  While this comparison is limited to a specific set of vacuum extensions, it already shows that bringing the comparison to an even ground may actually change the conclusions of Ref.\ \cite{abbott2018communication}.

For the above reasons, we argue that the comparison between the quantum {\tt SWITCH} \cite{ebler2018enhanced} and superposition of independent communication channels presented in Ref.\ \cite{abbott2018communication} is uneven.  
We nevertheless acknowledge the importance of the initial question posed in Ref.\ \cite{abbott2018communication}, namely to what extent indefinite causal order {\em per se}, as opposed to the common element of coherent control, is responsible for the communication advantages of the quantum {\tt SWITCH}. 
With respect to this open question, we point out that there exist several  partial indications that indefinite causal order does indeed exhibit specific features that differentiate it from the coherent control of  communication channels.

First, Refs.\ \cite{salek2018quantum,chiribella2018indefinite} showed that the  quantum {\tt SWITCH} enables noiseless quantum communication through two noisy channels, a phenomenon that is impossible through the coherent control of the same channels, {\em even if access to  vacuum-extensions is granted}.  In other words, even if one overlooks the fact that the control over orders and the control over channels build on  different initial resources, there still exist phenomena that are specific to the control over orders.

Second, Ref.\ \cite{loizeau2020channel} presented numerical evidence that the underlying mechanism for activation of communication capacity through the quantum {\tt SWITCH} is  different from the activation arising from  the superposition of trajectories.  Specifically, Ref.\ \cite{loizeau2020channel}  observed that the superposition of channels generically increases their capacity, whereas the quantum {\tt SWITCH} of channels can either increase or decrease their capacity, with this behaviour appearing to depend on the amount of non-commutativity of the input channels, as measured by a certain function of their Kraus operators.  On average, the authors found that when the  quantum {\tt SWITCH} increases the Holevo information, it has higher probability to increase it  by larger amounts compared to the superposition of channels.  

These findings provide  a numerical indication that the communication advantages of the quantum {\tt SWITCH} arise from an interplay between the coherent control of channels \textit{and} the non-commutativity of the Kraus operators,  a feature not present in the superposition of independent channels.

\section{Reply to Gu\'erin {\em et al.}}\label{sec:reply}
A recent paper by Gu\'erin, Rubino, and Brukner \cite{guerin2018shannon} argues that, in order to claim meaningful communication advantages,  the quantum {\tt SWITCH} should be compared to a general class of operations termed ``superpositions of direct pure processes''. In this Section we analyse their arguments and examples, concluding that they rest  on a communication model that violates  the basic resource-theoretic framework.  Before analysing that communication model, we also reply to two criticisms directed at the papers \cite{ebler2018enhanced, salek2018quantum, chiribella2018indefinite}.

\subsection{Reply to criticisms}

Refs.\ \cite{ebler2018enhanced, salek2018quantum, chiribella2018indefinite} proved that a quantum Shannon theory enriched with the quantum {\tt SWITCH} offers advantages over  standard quantum Shannon theory.   Ref.\   \cite{guerin2018shannon} criticises the fact that the comparison is restricted to standard  quantum Shannon theory, writing {\em ``[\dots] it is also important to keep a relatively large class of causally ordered processes against which the process under consideration can be compared; otherwise any advantage would be empty of practical significance.''}

The last comment on the ``practical significance" appears to be misplaced, given that  standard quantum Shannon theory (and not other models of communication with  causally ordered processes) underpins all implementations of quantum communication currently considered  in practice. This said, we stress that the motivation for studying  the communication advantages of the quantum {\tt SWITCH} is not directly a practical one: the motivation  is to explore how the theory of quantum communication as we currently know it would be affected by the possibility to combine quantum channels in a superposition of orders.   As mentioned  in Subsection \ref{subsec:switch}, the interest in  quantum communication assisted by the quantum {\tt SWITCH} is similar to the interest in communication and computation assisted by  Popescu-Rohrlich boxes, namely to better understand standard quantum theory by comparing it to possible alternatives.

We agree with the  authors of Ref.\ \cite{guerin2018shannon}  that it may be interesting to contrast the superposition of causal orders with other extensions of quantum Shannon theory.   However,  the particular extension proposed in Ref.\   \cite{guerin2018shannon}  appears to be problematic, in that it does not satisfy the minimal requirement for a resource theory of communication: as we will see in the following Subsections,   the operations proposed  in   Ref.\   \cite{guerin2018shannon}  generally create side-channels.

Ref.\   \cite{guerin2018shannon} also criticises the use of the term {\em causal activation} to describe the phenomenon in Refs.\ \cite{ebler2018enhanced, salek2018quantum, chiribella2018indefinite} of achieving a non-zero capacity when combining two zero-capacity channels in a superposition of alternative orders. The reason for the criticism is that {\em `` [\dots] there are causally ordered processes that offer  the  same  advantages  and  can  be  considered  as equivalent  resources.''}

Again, the criticism appears to be misplaced.   The term ``causal activation''  was  not meant to be a statement about the origin of the advantage. Instead, it was meant to be a way to distinguish the new type of activation  from the already known ``activation of the quantum capacity", introduced in the   seminal work of Smith and Yard    \cite{smith2008super}. Since the activation phenomenon observed in Refs.\ \cite{ebler2018enhanced, salek2018quantum, chiribella2018indefinite} was radically different from the standard activation of the quantum capacity, the authors  added the attribute ``causal'' to stress the different context of the activation observed in their work.   

Besides the choice of terminology, the claim that ``there are causally ordered processes that offer  the  same  advantages  and  can  be  considered  as equivalent  resources" appears to be unclear, as the authors of Ref.\   \cite{guerin2018shannon} did not provide a  resource-theoretic analysis.  In the following, we will assess their  claim within the resource-theoretic framework developed in this paper.

\subsection{The framework of SDPPs}

The authors of Ref.\ \cite{guerin2018shannon} argue that quantum Shannon theory with superpositions of causal orders should be considered within a general framework of ``superpositions of direct pure processes'' (SDPPs), which includes the quantum {\tt SWITCH}: ``\textit{It seems that any reasonable resource theory
that contains the quantum switch---a superposition of direct pure processes with different causal orders---should
also allow superpositions of direct pure processes with the same causal order}'' \cite{guerin2018shannon}. It is claimed, therefore, that the advantages of the quantum {\tt SWITCH} should be compared to SDPPs with a definite causal order. 
  In the following, we analyse the above claim, showing that, while the quantum {\tt SWITCH} and the SDPPs considered in Ref.\ \cite{guerin2018shannon} share a similar mathematical structure, they have different operational features: in particular, the specific SDPPs compared with the quantum {\tt SWITCH} in Ref.\ \cite{guerin2018shannon} generate side-channels, making the proposed advantages trivial from the resource-theoretic point of view. 
   
In the language of this paper, the SDPPs of Ref.\ \cite{guerin2018shannon} are  supermaps  that take two channels $(\map N_1, \map N_2)$, and return a  superposition  of $k$ channels which are individually of the form $\map D_j \circ \map N_2  \circ \map R_j \circ \map N_1  \circ \map E_j$ or $\map D_j' \circ \map N_1  \circ \map R_j' \circ \map N_2  \circ \map E_j' \, , j \in \{1,2,\dots,k\}$, for some encoding, repeater and decoding operations $\map E_j,\map E_j',  \map R_j, \map R_j',  \map D_j$ and $\map D_j'$. Here, a superposition of  $N$ channels  $\map A_j \in \Chan(A_j)$, for  $j\in  \{1,\dots, N\}$, 
 is defined in the most general way  as any channel $\map S \in \Chan(A_1 \oplus \cdots \oplus  A_N)$ which acts as $\map A_j$   when the input is restricted to  state in sector $A_j$ \cite{chiribella2019shannon2q}. SDPPs with a definite causal order are  defined as SDPPs which are superpositions of terms where  the input channels $\map N_1$ and $\map N_2$ occur in the same, fixed order. 
 
In the resource-theoretic scheme of our paper, the SDPPs should be regarded as a set of free operations.  One could, for example, consider them as a broader set of party supermaps, or alternatively,  as a set of placement supermaps with internal encoding,  decoding, and repeater operations, which are in a quantum superposition controlled by some quantum degree of freedom that is part of  the placement.

The resource theory based on SDPPs is different from the resource theory of  quantum Shannon theory with superpositions of trajectories. 
An important difference is that SDPPs are supermaps acting directly on the original channels, rather than their vacuum extensions.  In this respect, SDPPs and the quantum {\tt SWITCH} operate on the same kind of input resource, and a comparison between them would indeed be even.   However, in the following we will show that building a resource theory of communication where all SDPPs are taken as free operations is problematic, because it  violates the requirement of No Side-Channel Generation.  This fact will be illustrated by analysing the specific examples of SDPPs proposed in Ref.\ \cite{guerin2018shannon}.

\subsection{Some SDPPs generate classical  side-channels}

\begin{figure}
	\centering
	\begin{tikzpicture}[scale=1.2]
	\begin{pgfonlayer}{nodelayer}
		\draw[fill=red!90!white] (1,0) -- (1,1.5) -- (3,1.5) -- (3,0) -- (1,0) -- (1,1.5);
		\draw[fill=red!90!white] (4,0) -- (4,1.5) -- (6,1.5) -- (6,0) -- (4,0) -- (4,1.5);
		\draw[fill=Turquoise!90!white] (-1.5,0) -- (-1.5,1.5) -- (0,1.5) -- (0,0) -- (-1.5,0) -- (-1.5,1.5);
		\node [style=none] (0) at (1, 0) {};
		\node [style=none] (1) at (1, 1.5) {};
		\node [style=none] (2) at (3, 1.5) {};
		\node [style=none] (3) at (3, 0) {};
		\node [style=none] (4) at (4, 1.5) {};
		\node [style=none] (5) at (4, 0) {};
		\node [style=none] (6) at (6, 0) {};
		\node [style=none] (7) at (6, 1.5) {};
		\node [style=none] (8) at (0, 1.5) {};
		\node [style=none] (9) at (-1.5, 1.5) {};
		\node [style=none] (10) at (-1.5, 0) {};
		\node [style=none] (11) at (0, 0) {};
		\node [style=none] (12) at (0, 0.75) {};
		\node [style=none] (13) at (1, 0.75) {};
		\node [style=none] (14) at (4, 0.75) {};
		\node [style=none] (15) at (3, 0.75) {};
		\node [style=none] (16) at (6, 0.75) {};
		\node [style=none] (17) at (-1.5, 0.75) {};
		\node [style=none] (18) at (0.5, 1.5) {$M$};
		\node [style=none] (19) at (-3, 1.5) {$M$};
		\node [style=none, font={\large}] (20) at (-0.75, 0.75) {$\map{X}$};
		\node [style=none, font={\large}] (21) at (2, 0.75) {$\map{N}_1$};
		\node [style=none, font={\large}] (22) at (5, 0.75) {$\map{N}_2$};
		\node [style=none] (23) at (3.5, 1.5) {$M$};
		\node [style=none] (24) at (6.5, 1.5) {$M$};
		\node [style=none] (25) at (7, 0.75) {};
		\node [style=none] (26) at (-2, -1.75) {$C$};
		\node [style=none] (27) at (-3.5, 0.75) {};
		\node [style=none, font={\Large}] (28) at (-0.75, -1) {$\bullet$};
		\node [style=none] (29) at (-0.75, 0) {};
		\node [style=none] (31) at (-2.5, -1.5) {};
		\node [style=none] (32) at (-2.5, -1) {};
		\node [style=none] (33) at (7, -1) {};
		\node [style=none] (34) at (-2.5, -0.5) {};
		\draw [bend right=90, looseness=3.25,fill=Turquoise!90!white] (34.center) to (31.center);
	\end{pgfonlayer}
	\begin{pgfonlayer}{edgelayer}
		\node [style=none] (30) at (-3, -1) {$\, \, +$};
		\draw (9.center) to (8.center);
		\draw (8.center) to (11.center);
		\draw (11.center) to (10.center);
		\draw (10.center) to (9.center);
		\draw (1.center) to (2.center);
		\draw (2.center) to (3.center);
		\draw (3.center) to (0.center);
		\draw (0.center) to (1.center);
		\draw (4.center) to (7.center);
		\draw (7.center) to (6.center);
		\draw (6.center) to (5.center);
		\draw (5.center) to (4.center);
		\draw (16.center) to (25.center);
		\draw (17.center) to (27.center);
		\draw (12.center) to (13.center);
		\draw (15.center) to (14.center);
		\draw (29.center) to (28.center);
		\draw (33.center) to (32.center);
		\draw [bend right=90, looseness=3.25] (34.center) to (31.center);
		\draw (34.center) to (31.center);
	\end{pgfonlayer}
\end{tikzpicture}
	\vspace{0.5cm}
	\caption{\label{fig:cx} {\em An SDPP that  transfers classical information through a side-channel, bypassing the original communication devices.}  A control qubit $C$ is prepared in the state $\ket{+}$ and is sent together with the message $M$ through a {\tt CNOT} gate. If the message is initialised in either of the states $\ket{\pm}$, then its  interaction with the control through the {\tt CNOT} will output the state $\ket{\pm}$ in $C$. The receiver is thus able to decode the original message by measuring the control qubit, irrespectively of the  channels $\map N_1$ and $\map N_2$.  Overall, this SDPP fails to satisfy the requirement of No Side-Channel Generation, which we regard as a minimal requirement for a sensible resource theory of communication.} 
\end{figure}
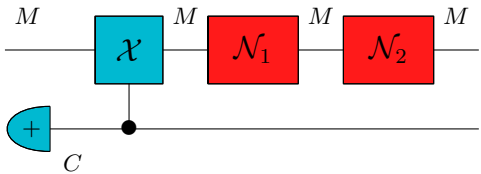

One  of the SDPPs proposed in Ref.\ \cite{guerin2018shannon} is the supermap depicted in Figure \ref{fig:cx}. This supermap corresponds to a protocol where two qubit channels $\map  N_1$ and $\map N_2$ are applied after a {\tt CNOT} gate, acting on the message qubit, and on an additional control qubit $C$.  The supermap, here denoted as $\map F : (\map N_1, \map N_2) \mapsto \map F (\map N_1, \map N_2)$, produces the  output channel  defined by
\begin{equation}\label{eq:sig_enc}
\map F (\map N_1, \map N_2) (\rho ) = [(\map N_2 \circ  \map N_1) \otimes \map I_C] \circ \map U^{\tt CNOT} (\rho \otimes |+\>\<+|),
\end{equation}
where   $\map U^{\tt CNOT} := {U^{\tt CNOT}}^\dagger (\cdot) U^{\tt CNOT}$ is the unitary channel corresponding to the {\tt CNOT} gate 
\begin{equation}\label{eq:cnot}
U^{\tt CNOT} := I_M \otimes \ketbra{0}{0}_C + X_M \otimes 
\ketbra{1}{1}_C \, ,
\end{equation}
 $X$ being the {\tt NOT} gate.
To avoid  overloading the  notation, here we have omitted  the isomorphisms between unplaced and placed systems, and simply denoted the (placed and unplaced) message system by $M$.

Now,  the map $\map F$ enables perfect classical communication of one bit independently of the communication channels $\map N_1$ and $\map N_2$ \cite{chiribella2019shannon2q,guerin2018shannon}. Using the phase kickback mechanism of the {\tt CNOT} gate \cite{cleve1998quantum}, information encoded in the states $|\pm\>: =  (|0\>  \pm  |1\>)/\sqrt 2$ is transferred from the message $M$ to the control $C$ \textit{before}  the noisy channels $\map N_1$ and $\map N_2$ are applied. Then, the information is safely carried by the control system to the receiver, completely bypassing the communication channels $\map N_1$ and $\map N_2$, and avoiding the resulting noise.    In other words, this example of an SDPP is analogous to the example of the noisy telephone line discussed in Subsection \ref{subsec:minimal}: it achieves communication by completely bypassing the original channels.

More formally, one can see that  the operation $\map F$ generates a classical side-channel in the sense of Definition \ref{def:sc}.  Indeed, one can consider the party supermap corresponding to the  encoding channel $\map E   =  \map I_M $
and the decoding channel $\map D  =  \Tr_M$, which discards the message qubit.  
The result is the channel  
\begin{equation}
\begin{split}
	 \map C (\cdot)  &=   \map D \circ \map F (\map N_1, \map N_2)  \circ \map E   (\cdot) \\
	 &=   \ketbra{+}{+}   \cdot  \ketbra{+}{+}    +    \ketbra{-}{-}     \cdot \ketbra{-}{-}  \, , 
\end{split}
\end{equation}
which is independent of $\map N_1$ and $\map N_2$ and provides a perfect transmission  line for classical communication.  In conclusion, the ``communication enhancement"  of the SDPP (\ref{eq:sig_enc}) arises from a classical side-channel, which completely bypasses the original communication devices. 

The authors of Ref.\ \cite{guerin2018shannon} also consider a specific interferometric implementation of the operation $\map F$, which they claim   avoids the criticism above. In Appendix \ref{app:int}, we analyse the arguments provided in  Ref.\ \cite{guerin2018shannon}, and conclude that, in fact, the above criticism still applies.

\subsection{Some SDPPs generate quantum  side-channels}  
In Appendix B of Ref.\  \cite{guerin2018shannon}, the authors  present an SDPP, stating that it {\em ``allows us to perfectly  transmit  one  qubit  of quantum information, for all channels [\dots]''}.  This statement is an explicit acknowledgement that the SDPP model permits the strongest possible kind of side-channels: perfect side-channels for quantum communication. 

The example in Appendix B of Ref.\  \cite{guerin2018shannon} is presented as one that  {\em ``generalises,  and  improves  upon, the observations made in the main text,''} the improvement being that  only four control qubits are used instead of eight, which is the number of qubits used by  the protocol in the main text.  
Here we review the example, showing that, in fact, one can improve it even further: the same perfect qubit side-channel can be generated by an SDPP that uses  only two control qubits, instead of four. 

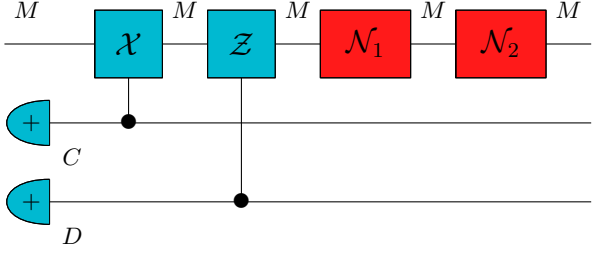
\begin{figure}
	\centering
	\begin{tikzpicture}[scale=1.2]
	\begin{pgfonlayer}{nodelayer}
	\draw[fill=red!90!white] (1,0) -- (1,1.5) -- (3,1.5) -- (3,0) -- (1,0) -- (1,1.5);
	\draw[fill=red!90!white] (4,0) -- (4,1.5) -- (6,1.5) -- (6,0) -- (4,0) -- (4,1.5);
	\draw[fill=Turquoise!90!white] (-1.5,0) -- (-1.5,1.5) -- (0,1.5) -- (0,0) -- (-1.5,0) -- (-1.5,1.5);
	\draw[fill=Turquoise!90!white] (-4,0) -- (-4,1.5) -- (-2.5,1.5) -- (-2.5,0) -- (-4,0) -- (-4,1.5);
		\node [style=none] (0) at (1, 0) {};
		\node [style=none] (1) at (1, 1.5) {};
		\node [style=none] (2) at (3, 1.5) {};
		\node [style=none] (3) at (3, 0) {};
		\node [style=none] (4) at (4, 1.5) {};
		\node [style=none] (5) at (4, 0) {};
		\node [style=none] (6) at (6, 0) {};
		\node [style=none] (7) at (6, 1.5) {};
		\node [style=none] (8) at (-2.5, 1.5) {};
		\node [style=none] (9) at (-4, 1.5) {};
		\node [style=none] (10) at (-4, 0) {};
		\node [style=none] (11) at (-2.5, 0) {};
		\node [style=none] (12) at (-2.5, 0.75) {};
		\node [style=none] (13) at (1, 0.75) {};
		\node [style=none] (14) at (4, 0.75) {};
		\node [style=none] (15) at (3, 0.75) {};
		\node [style=none] (16) at (6, 0.75) {};
		\node [style=none] (17) at (-4, 0.75) {};
		\node [style=none] (18) at (-5.5, 1.5) {$M$};
		\node [style=none, font={\large}] (19) at (-3.25, 0.75) {$\map{X}$};
		\node [style=none, font={\large}] (20) at (2, 0.75) {$\map{N}_1$};
		\node [style=none, font={\large}] (21) at (5, 0.75) {$\map{N}_2$};
		\node [style=none] (22) at (3.5, 1.5) {$M$};
		\node [style=none] (23) at (6.5, 1.5) {$M$};
		\node [style=none] (24) at (7, 0.75) {};
		\node [style=none] (25) at (-4.5, -1.75) {$C$};
		\node [style=none] (26) at (-6, 0.75) {};
		\node [style=none, font={\Large}] (27) at (-3.25, -1) {$\bullet$};
		\node [style=none] (28) at (-3.25, 0) {};
		\node [style=none] (29) at (-4.5, -3.5) {$D$};
		\node [style=none] (30) at (-1.5, 1.5) {};
		\node [style=none, font={\Large}] (31) at (-0.75, -2.75) {$\bullet$};
		\node [style=none] (32) at (0, 0) {};
		\node [style=none] (33) at (-2, 1.5) {$M$};
		\node [style=none] (34) at (0.5, 1.5) {$M$};
		\node [style=none] (35) at (-1.5, 0.75) {};
		\node [style=none] (36) at (-1.5, 0) {};
		\node [style=none] (37) at (0, 1.5) {};
		\node [style=none] (38) at (0, 0.75) {};
		\node [style=none] (39) at (-0.75, 0) {};
		\node [style=none, font={\large}] (40) at (-0.75, 0.75) {$\map{Z}$};
		\node [style=none] (42) at (-5, -1.5) {};
		\node [style=none] (43) at (-5, -1) {};
		\node [style=none] (44) at (7, -1) {};
		\node [style=none] (45) at (-5, -0.5) {};
		\node [style=none] (47) at (-5, -3.25) {};
		\node [style=none] (48) at (-5, -2.75) {};
		\node [style=none] (49) at (7, -2.75) {};
		\node [style=none] (50) at (-5, -2.25) {};
		\draw [bend right=90, looseness=3.25,fill=Turquoise!90!white] (45.center) to (42.center);
		\draw [bend right=90, looseness=3.25,fill=Turquoise!90!white] (50.center) to (47.center);
	\end{pgfonlayer}
	\begin{pgfonlayer}{edgelayer}
		\node [style=none] (46) at (-5.5, -2.75) {$\, \, +$};
		\node [style=none] (41) at (-5.5, -1) {$\, \, +$};
		\draw (9.center) to (8.center);
		\draw (8.center) to (11.center);
		\draw (11.center) to (10.center);
		\draw (10.center) to (9.center);
		\draw (1.center) to (2.center);
		\draw (2.center) to (3.center);
		\draw (3.center) to (0.center);
		\draw (0.center) to (1.center);
		\draw (4.center) to (7.center);
		\draw (7.center) to (6.center);
		\draw (6.center) to (5.center);
		\draw (5.center) to (4.center);
		\draw (16.center) to (24.center);
		\draw (17.center) to (26.center);
		\draw (15.center) to (14.center);
		\draw (28.center) to (27.center);
		\draw (30.center) to (37.center);
		\draw (37.center) to (32.center);
		\draw (32.center) to (36.center);
		\draw (36.center) to (30.center);
		\draw (39.center) to (31.center);
		\draw (12.center) to (35.center);
		\draw (38.center) to (13.center);
		\draw (44.center) to (43.center);
		\draw [bend right=90, looseness=3.25] (45.center) to (42.center);
		\draw (45.center) to (42.center);
		\draw (49.center) to (48.center);
		\draw [bend right=90, looseness=3.25] (50.center) to (47.center);
		\draw (50.center) to (47.center);
	\end{pgfonlayer}
\end{tikzpicture}
	\vspace{0.5cm}
	\caption{\label{fig:cxcz}{\em An SDPP that transfers quantum information through a side-channel, for any noisy channels acting on the message.} Two control qubits $C$ and $D$ are both prepared in the state $\ket{+}$. A message $M$ in state $\rho$ is to be communicated. The composite system $M \otimes C$ is sent through a {\tt CNOT} gate, followed by the composite system $M \otimes D$ going through a {\tt CPHASE} gate. As shown in Equation \eqref{eq:perfect}, the receiver is able to recover the original input by measuring the control qubit $D$ and performing a conditional correction on $C$, independently of the choice of noisy channels $\map N_1$ and $\map N_2$ that act on the message itself.} 
\end{figure}

 Our improved version of the SDPP in Ref.\  \cite{guerin2018shannon} is depicted in Figure \ref{fig:cxcz}. It uses two control qubits $C$ and $D$, in addition to the message qubit $M$.  The  corresponding supermap  is given by
\begin{equation}\label{eq:perfect_enc}
\begin{split}
	&\map G^{\omega,\xi} : (\map N_1, \map N_2) \mapsto \map G^{\omega,\xi} (\map N_1, \map N_2) \\
	 	&\map G^{\omega,\xi} (\map N_1, \map N_2) (\rho_M ) \\ 
	&~~~~~~= [(\map N_2 \circ \map N_1) \otimes \map I_{C} \otimes \map I_{D}] \circ (\map U^{\tt CPHASE}_{MD} \otimes \map I_C)   \\  
 &~~~~~~~~~~~	\circ     (\map U^{\tt CNOT}_{MC} \otimes \map I_D)  (\rho_M \otimes \omega_C \otimes \xi_D) \, ,
\end{split}
\end{equation}
where  $\omega$ and $\xi$ are quantum states of the control qubits $C$ and $D$, respectively,  and $\map U^{\tt CNOT}$ and $\map U^{\tt CPHASE}$ are the {\tt CNOT} and {\tt CPHASE} gates, respectively, defined by 
\begin{equation}\label{eq:cxcz}
\begin{split}
U^{\tt CNOT}_{MC} &:= I_M \otimes \ketbra{0}{0}_C + X_M \otimes 
\ketbra{1}{1}_C\\
U^{\tt CPHASE}_{MD} &:= I_M \otimes \ketbra{0}{0}_D + Z_M \otimes 
\ketbra{1}{1}_D \ , 
\end{split}
\end{equation}
$X$ and $Z$ being Pauli gates.

Explicit calculation reveals that for both control qubits initialised in the $\ket{+}$ state, one obtains
\begin{equation}\label{eq:perfect}
\begin{split}
	&\Tr_M  \left[\map G^{|+\>\<+|,  \,  |+\>\<+|} (\map N_1, \map N_2) (\rho_M )   \right]\\ 
	&~~~~~=  \rho_C \otimes  \ketbra{+}{+}_D +
	 X\rho X_C \otimes  \ketbra{-}{-}_D \, .
\end{split}
\end{equation}
Therefore, the initial state $\rho$ can be perfectly recovered independently of the noisy channels $\map N_1$ and $\map N_2$, by measuring $D$ in the Fourier basis and then applying a {\tt NOT} gate on $C$ if the outcome is $\ket{-}$.  The  supermap defined by Equation  \eqref{eq:perfect_enc} is an example of an SDPP that generates a perfect quantum side-channel, as it can perfectly transmit one qubit of quantum information for any choice of noisy channels.

The authors of Ref.\ \cite{guerin2018shannon} conclude with regard to their protocol:  ``\emph{This example shows that SDPPs [...] can be used to perfectly send one qubit of information, essentially trivialising the problem of enhancing quantum and classical channel capacity if one were to take the set of all SDPPs as a resource''.}   We agree, and argue that this is  the reason why the set of all SDPPs does not define a sensible  resource theory of communication.     

A possible direction of future research would be to compare the quantum {\tt SWITCH} with the subset of SDPPs that have definite causal order and do not generate side-channels. This may shed light on the mechanism that leads to enhancements in the quantum {\tt SWITCH}, and on whether or not the characteristics of this mechanism can be reproduced by SDPPs  with definite causal order.   

More interestingly, it would be important to compare the side-channel non-generating SDPPs with definite causal order with {\em all} the side-channel non-generating SDPPs with indefinite causal order, rather than just restricting the comparison to the quantum {\tt SWITCH}.  For a given pair of channels, the maximum communication capacity achievable with indefinite  causal order is---by definition---always larger than or equal to  the maximum communication capacity  achievable with definite causal order. The interesting question is whether there is a gap between the two, meaning that there exist communication advantages that can be achieved {\em only} with indefinite causal order.

\section{Summary and outlook}

We established  a general framework of resource theories of communication.  In our framework, the input resources are communication devices, which can be placed between the communicating parties, and combined with local operations performed by the communicating parties. A resource theory is specified by a choice of placement operations, describing how the communication devices are arranged, and by a choice of party operations, describing the action of the communicating parties.    

We formulated a minimal requirement that every resource theory of communication should satisfy: no combination of the allowed operations should be able to  bypass the communication devices initially available to the communicating parties. 
We have shown that  quantum Shannon theory with superpositions of causal order of communication channels \cite{ebler2018enhanced} and quantum Shannon theory with superpositions of trajectories of  information carriers \cite{chiribella2019shannon2q} satisfy this  requirement, while quantum Shannon theory with  superpositions of encoding and decoding operations   \cite{guerin2018shannon} does not.  

We pointed out the importance of distinguishing between different forms of coherent control, rather than conflating them into a generic label.  Specifically, we distinguished between three different types of superpositions: the superposition of causal orders of communication channels, the superposition of trajectories through independent communication channels, and the superposition of encoding/decoding operations. We observed that the superposition of causal orders is in principle different from the superposition of trajectories, because these two different superpositions use different input resources (the original channels in the former case, and an extension of the original channels in the latter case).       In turn, the superposition of orders and the superposition of  trajectories are strikingly different from the superposition of encoding/decoding operations, in that the former do not generate side-channels, while the latter does. 

Our definition of resource theories of communication can be extended straightforwardly to allow  correlated quantum channels as input resources, where the noisy processes occurring in the application of a device at time $t+1$ may be affected by the application of the same device at time $t$. Such correlated channels are known as quantum memory channels \cite{macchiavello02correlations,kretschmann2005quantum},  quantum combs \cite{chiribella2008transforming,chiribella2009theoretical} and non-Markovian quantum processes \cite{pollock2018nonmarkov,giarmatzi2018nonmarkov}, and have been shown to provide interesting communication advantages over uncorrelated channels \cite{macchiavello02correlations,chiribella2019shannon2q,kristjansson2020latent}.

Overall, the resource-theoretic framework proposed in this paper allows for  rigorous comparisons between different resource theories of communication, and can be used for the exploration of new models of quantum communication, with both foundational and  practical implications.

\begin{acknowledgments}
We acknowledge fruitful discussions with Alastair Abbott, {\v C}aslav Brukner, Bob Coecke,  Fabio Costa, Mile Gu, Philippe Allard Gu\'erin, Wenxu Mao, Ognyan Oreshkov, Giulia Rubino,  David Schmid, Carlo Sparaciari, Robert Spekkens, Philip Walther and Elie Wolfe. We are grateful to Alastair Abbott, {\v C}aslav Brukner,  Philippe Allard Gu\'erin, Giulia Rubino, Philip Walther,  and an anonymous referee for valuable comments on the manuscript. This work was supported by the National Natural Science Foundation of China through grant 11675136, the Hong Kong Research Grant Council through grant 17307719, the Croucher Foundation, the UK Engineering and Physical Sciences Research Council (EPSRC), and the John Templeton Foundation through grants 60609, Quantum Causal Structures, and  61466, The Quantum Information Structure of Spacetime  (qiss.fr).
Research at the Perimeter Institute is supported by the Government of Canada through the Department of Innovation, Science and Economic Development Canada and by the Province of Ontario through the Ministry of Research, Innovation and Science. The opinions expressed in this publication are those of the authors and do not necessarily reflect the views of the John Templeton Foundation. 
\end{acknowledgments}

\appendix

\section{Free supermaps in assisted communication scenarios}\label{app:altsupermaps}

 \begin{figure}[b]
	\centering
	\begin{tikzpicture}[scale=1.2]
	\begin{pgfonlayer}{nodelayer}
		\draw[fill=green!90!white] (-1,0.25) -- (-1,1.75) -- (1,1.75) -- (1,0.25) -- (-1,0.25) -- (-1,1.75);
	\draw[fill=Orchid!90!white] (-1,-0.25) -- (-1,-1.75) -- (1,-1.75) -- (1,-0.25) -- (-1,-0.25) -- (-1,-1.75);
	\draw[fill=Orchid!90!white] (3.5,1.75) -- (3.5,-1.75) -- (2,-1.75) -- (2,1.75) -- (3.5,1.75) -- (3.5,-1.75);
	\draw[fill=Orchid!90!white] (-3.5,1.75) -- (-3.5,-1.75) -- (-2,-1.75) -- (-2,1.75) -- (-3.5,1.75) -- (-3.5,-1.75);
		\node [style=none] (0) at (2, 1.75) {};
		\node [style=none] (1) at (2, -1.75) {};
		\node [style=none] (2) at (3.5, 1.75) {};
		\node [style=none] (3) at (3.5, -1.75) {};
		\node [style=none] (4) at (1, 1.75) {};
		\node [style=none] (5) at (-1, 1.75) {};
		\node [style=none] (6) at (-1, 0.25) {};
		\node [style=none] (7) at (1, 0.25) {};
		\node [style=none] (8) at (1, 1) {};
		\node [style=none] (9) at (2, 1) {};
		\node [style=none] (10) at (3.5, 1) {};
		\node [style=none] (11) at (-1, 1) {};
		\node [style=none] (12) at (-2, 1) {};
		\node [style=none] (13) at (1.5, 1.75) {$B$};
		\node [style=none] (14) at (-1.5, 1.75) {$A$};
		\node [style=none, font={\large}] (15) at (0, 1) {$\map{C}$};
		\node [style=none, font={\large}] (16) at (2.75, 0) {$\map{D}$};
		\node [style=none] (17) at (-4.75, 1) {};
		\node [style=none, font={\large}] (18) at (-2.75, 0) {$\map{E}$};
		\node [style=none] (19) at (-2, 1) {};
		\node [style=none] (20) at (-3.5, 1) {};
		\node [style=none] (21) at (-2, -1.75) {};
		\node [style=none] (22) at (-3.5, -1.75) {};
		\node [style=none] (23) at (-3.5, 1.75) {};
		\node [style=none] (24) at (-2, 1.75) {};
		\node [style=none] (25) at (-4, 1.75) {$A'$};
		\node [style=none] (26) at (4, 1.75) {$B'$};
		\node [style=none] (27) at (4.75, 1) {};
		\node [style=none] (28) at (1, -0.25) {};
		\node [style=none] (29) at (-1, -0.25) {};
		\node [style=none] (30) at (-1, -1.75) {};
		\node [style=none] (31) at (1, -1.75) {};
		\node [style=none] (32) at (1, -1) {};
		\node [style=none] (33) at (-1, -1) {};
		\node [style=none, font={\large}] (34) at (0, -1) {$\map{I}_{\rm clas}$};
		\node [style=none] (35) at (-2, -1) {};
		\node [style=none] (36) at (2, -1) {};
		\node [style=none, font={\footnotesize}] (37) at (-1.5, -1.75) {$\rm Aux$};
		\node [style=none, font={\footnotesize}] (38) at (1.5, -1.75) {$\rm Aux$};
	\end{pgfonlayer}
	\begin{pgfonlayer}{edgelayer}
		\draw (5.center) to (4.center);
		\draw (4.center) to (7.center);
		\draw (7.center) to (6.center);
		\draw (6.center) to (5.center);
		\draw (0.center) to (2.center);
		\draw (2.center) to (3.center);
		\draw (3.center) to (1.center);
		\draw (1.center) to (0.center);
		\draw (11.center) to (12.center);
		\draw (23.center) to (24.center);
		\draw (24.center) to (21.center);
		\draw (21.center) to (22.center);
		\draw (22.center) to (23.center);
		\draw (20.center) to (17.center);
		\draw (10.center) to (27.center);
		\draw (8.center) to (9.center);
		\draw (29.center) to (28.center);
		\draw (28.center) to (31.center);
		\draw (31.center) to (30.center);
		\draw (30.center) to (29.center);
		\draw (35.center) to (33.center);
		\draw (32.center) to (36.center);
	\end{pgfonlayer}
\end{tikzpicture}
	\caption{\label{fig:classical} {\em Supermap describing encoding and decoding operations assisted by free classical communication.}} 
\end{figure}
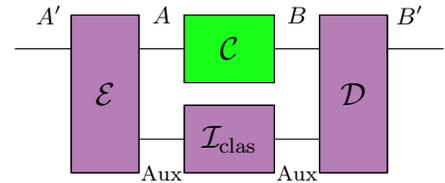

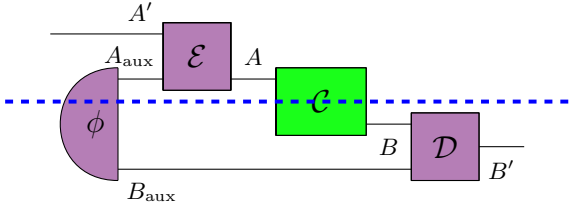
\begin{figure}[htb]
	\centering
	\begin{tikzpicture}[scale=1.2]
	\begin{pgfonlayer}{nodelayer}
	\draw[fill=green!90!white] (-3.75,0) -- (-3.75,1.5) -- (-1.75,1.5) -- (-1.75,0) -- (-3.75,0) -- (-3.75,1.5);
	\draw[fill=Orchid!90!white] (-0.75,-1) -- (-0.75,0.5) -- (0.75,0.5) -- (0.75,-1) -- (-0.75,-1) -- (-0.75,0.5);
	\draw[fill=Orchid!90!white] (-6.25,1) -- (-6.25,2.5) -- (-4.75,2.5) -- (-4.75,1) --  (-6.25,1) -- (-6.25,2.5) -- (-4.75,2.5) -- (-4.75,1);
		\node [style=none] (0) at (-0.75, 0.5) {};
		\node [style=none] (1) at (-0.75, -1) {};
		\node [style=none] (2) at (0.75, 0.5) {};
		\node [style=none] (3) at (0.75, -1) {};
		\node [style=none] (4) at (-1.75, 1.5) {};
		\node [style=none] (5) at (-3.75, 1.5) {};
		\node [style=none] (6) at (-3.75, 0) {};
		\node [style=none] (7) at (-1.75, 0) {};
		\node [style=none] (8) at (-1.75, 0.25) {};
		\node [style=none] (9) at (-0.75, 0.25) {};
		\node [style=none] (10) at (0.75, -0.25) {};
		\node [style=none] (11) at (-3.75, 1.25) {};
		\node [style=none] (12) at (-4.75, 1.25) {};
		\node [style=none] (13) at (-1.25, -0.25) {$B$};
		\node [style=none] (14) at (-4.25, 1.75) {$A$};
		\node [style=none, font={\large}] (15) at (-2.75, 0.75) {$\map{C}$};
		\node [style=none, font={\large}] (16) at (0, -0.25) {$\map{D}$};
		\node [style=none] (17) at (-8.75, 2.25) {};
		\node [style=none, font={\large}] (18) at (-5.5, 1.75) {$\map{E}$};
		\node [style=none] (20) at (-6.25, 2.25) {};
		\node [style=none] (21) at (-4.75, 1) {};
		\node [style=none] (22) at (-6.25, 1) {};
		\node [style=none] (23) at (-6.25, 2.5) {};
		\node [style=none] (24) at (-4.75, 2.5) {};
		\node [style=none] (25) at (-6.75, 2.75) {$A'$};
		\node [style=none] (26) at (1.25, -0.75) {$B'$};
		\node [style=none] (27) at (1.75, -0.25) {};
		\node [style=none] (28) at (-7.25, 1.5) {};
		\node [style=none] (29) at (-7.25, -1) {};
		\node [style=none] (30) at (-6.25, 1.25) {};
		\node [style=none] (31) at (-7.25, 1.25) {};
		\node [style=none] (32) at (-7.25, -0.75) {};
		\node [style=none] (33) at (-0.75, -0.75) {};
		\node [style=none] (37) at (-6.5, -1.25) {$B_{\rm aux}$};
		\node [style=none] (38) at (-7, 1.75) {$A_{\rm aux}$};
		\node [style=none] (40) at (-9.75, 0.75) {};
		\node [style=none] (41) at (2.75, 0.75) {};
			\draw [fill=Orchid!90!white,bend right=90, looseness=1.75] (28.center) to (29.center);
	\end{pgfonlayer}
	\begin{pgfonlayer}{edgelayer}
			\node [style=none] (35) at (-7.75, 0.25) {\large{$\phi$}};
		\draw (5.center) to (4.center);
		\draw (4.center) to (7.center);
		\draw [in=360, out=180] (7.center) to (6.center);
		\draw (6.center) to (5.center);
		\draw (0.center) to (2.center);
		\draw (2.center) to (3.center);
		\draw (3.center) to (1.center);
		\draw (1.center) to (0.center);
		\draw (11.center) to (12.center);
		\draw (23.center) to (24.center);
		\draw (24.center) to (21.center);
		\draw (21.center) to (22.center);
		\draw (22.center) to (23.center);
		\draw (20.center) to (17.center);
		\draw (10.center) to (27.center);
		\draw (8.center) to (9.center);
		\draw [in=90, out=-90] (28.center) to (29.center);
		\draw (32.center) to (33.center);
		\draw (31.center) to (30.center);
		\draw [style=dashed, color=blue, line width=1.5pt] (40.center) to (41.center);
	\end{pgfonlayer}
\end{tikzpicture}
	\caption{\label{fig:entanglement} {\em Supermap describing encoding and decoding operations assisted by shared entanglement.} The blue dashed line denotes the partition between Alice (top) and Bob (bottom).} 
\end{figure}

Here we provide examples of supermaps that arise in the presence of assistance from classical communication and entanglement.  

Let us consider first  the assistance of free classical communication \cite{bennett1996entanglement}, as illustrated in Figure \ref{fig:classical}. In this case, the free supermaps on placed channels have the form  
\begin{align}
\map S_{ \map E, \map D, \rm clas}   (\map C)   :=  \map D   \circ  (\map C  \otimes \map I^{\rm clas}_{\rm Aux})  \circ \map E \, ,
\end{align} 
where $\map I^{\rm clas}_{\rm Aux}$  is the classical identity channel, defined as  $  \map I^{\rm clas}_{\rm Aux}  (\rho)  =  \sum_j  \,  |j\>\<j| \,  \<  j|  \rho  |j  \>$ for some orthonormal basis $\{  |j\>\}$,  and $\map E  \in  \Chan (A',A\otimes {\rm Aux})$ and $\map D \in  \Chan (B \otimes {\rm Aux},B')$  are quantum channels. 
	
Let us consider now classical communication with the assistance of shared entanglement \cite{bennett1999entanglement}.  In  this case, the free operations on placed channels are those that can be achieved by performing encoding and decoding operations that act on a shared entangled state, as shown in  Figure \ref{fig:entanglement}.  Mathematically, these operations correspond to  free supermaps of the form 
\begin{align}
\nonumber \map S_{\map E, \map D, \rm ent}  (\map C)   &:=     \map D_{BB_{\rm aux}} \circ  (\map C_A \circ \map E_{A'A_{\rm aux}}  \otimes \map I_{B_{\rm aux}})   \\
&  \quad \quad  \circ  (\map I_{A'}\otimes \phi_{A_{\rm aux}B_{\rm aux}}) \,,
\end{align}
 where $\phi_{A_{\rm aux}B_{\rm aux}}$ is an entangled state on system $A_{\rm aux} \otimes B_{\rm aux}$, $\map E  \in  \Chan  (A' \otimes A_{\rm aux},  A)$  and $\map D   \in   \Chan  (B\otimes B_{\rm aux},  B')$ are encoding and decoding channels, respectively, and the subscripts indicate the input systems of all  channels.

\section{Mathematical structure of unplaced channels, placed channels and admissible supermaps}

\subsection{Placed and unplaced channels }\label{sec:placed_unplaced}

Here we  specify  the mathematical structure of   placed and unplaced channels.
In the following, the subset of completely positive (CP) maps will be denoted by $\CP (A,B) \subset \Map (A,B)$, and the subset of trace-preserving (TP) maps will be denoted by $\TP (A,B) \subset \Map (A,B)$.

For a single use of a single device, the distinction between  placed and unplaced channels  concerns only the type of inputs and outputs.  Mathematically, placed and unplaced channels are both described by completely positive trace-preserving maps.

When multiple devices are involved, the distinction is more substantial. As described in Subsections \ref{subsect:direct_mult}--\ref{subsec:network},  $k$ devices  can be placed in parallel,  giving rise to a multipartite quantum channel,
or in sequence,  giving rise to a  quantum $k$-comb, or, more generally,  in any combination of parallel and sequence,  giving rise to a quantum $l$-comb, with any $l$ from $1$ to $k$.  

Mathematically, a quantum $k$-comb transforming system $S_i$ into system $S_i'$ at the $i$-th step is a completely positive map $\map C  \in  \CP  ( S_1\otimes \cdots \otimes S_k  ,  S_1'\otimes \cdots \otimes S_k')$ satisfying the conditions  
\begin{equation}
\begin{split}
\Tr_{S_k'}  [ \map C(\rho)  ]   &=  \Tr_{S_k'}  \left[ \map C\left( \Tr_{S_k}[\rho] \otimes  \frac{ I_{S_k}}{d_k}  \right)  \right] \\
\Tr_{  S_{k\!-\!1}'  S_k'}  \! [ \map C(\rho)  ]  & =  \Tr_{ S_{k\!-\!1}'  S_k'}  \left[ \map C\left(   \Tr_{S_{k\!-\!1} S_k}[\rho] \! \otimes \! \frac{ I_{S_{k\!-\!1}}}{d_{k\!-\!1}}  \! \otimes \! \frac{ I_{S_k}}{d_k}  \right)  \right]\\
&~\,\vdots  \\
\Tr_{S_1' \cdots  S_k'}   [ \map C(\rho)  ]    & = 1  \qquad \forall \rho  \in \St (S_1\otimes \cdots\otimes S_k) \, .  
\label{kcomb}
\end{split}
\end{equation}
The set of quantum $k$-combs with the input/output systems above is denoted by $\Comb  [  (S_1,S_1'), \dots, (S_k,  S_k')]$.

In contrast, unplaced channels belong to a different set of completely positive maps. We will now specify this set explicitly. So far, we represented unplaced channels by lists, such as $( \map N_1, \dots ,\map N_k)$.
However, the set of lists is not closed under probabilistic mixtures,  which arise quite naturally when some of the parameters of the communication devices are subject to random fluctuations.

Probabilistic mixtures can be described by  convex combinations of the form $ \sum_{i=1}^L \, p_i\, ( \map N_{1,i}, \dots ,\map N_{k,i})$, where $(p_i)_{i=1}^L$  is a probability distribution, and, for every $i \in  \{1,\dots, L\}$, $( \map N_{1,i}, \dots ,\map N_{k,i})  \in  \Chan (X_1,Y_1) \times \dots \times \Chan  (X_k, Y_k)$ is a list of  channels.   Note that the convex combination $ \sum_{i=1}^L \, p_i\, ( \map N_{1,i}, \dots ,\map N_{k,i})$ must satisfy a basic consistency requirement: if    a channel in the list $( \map N_1, \dots ,\map N_k)$ is  a convex combination of channels, say $\map N_1= \sum_{i}\, p_i  \, \map N_{1,i}$, then the list  $( \map N_1, \dots ,\map N_k)$ should be equal to the corresponding  convex combination  $\sum_i  p_i ( \map N_{1,i}, \map N_2  , \dots ,\map N_k)$.   Requiring this consistency property to hold for every entry of the list implies that the convex combinations $ \sum_{i=1}^L \, p_i\, ( \map N_{1,i}, \dots ,\map N_{k,i})$ can be represented as elements of the tensor product  space $\TP (X_1,Y_1) \otimes \TP (X_2,Y_2) \otimes \cdots \otimes \TP (X_k,Y_k)$, which consists of all linear combinations of the form  
\begin{align}\label{generalnosig} 
\map M  =  \sum_{i=1}^L  \,  c_i  \,   \map N_{1,i} \otimes  \map N_{2,i} \otimes \cdots \otimes \map N_{k,i} \, ,
\end{align} where $(c_i)_{i=1}^L$ are real coefficients, and each $\map N_{j,k}$  is a trace-preserving map in  $\TP(  X_j,  Y_j)$.   

In summary,  the unplaced channels can be regarded as elements of the tensor product space 
$\TP (X_1,Y_1) \otimes \TP (X_2,Y_2) \otimes \cdots \otimes \TP (X_k,Y_k)$.  Precisely, the list $( \map N_1, \dots ,\map N_k)$ can be regarded as the product channel $ \map N_1\otimes \cdots \otimes \map N_k$.   In this paper, we use the list notation $( \map N_1, \dots ,\map N_k)$ and the tensor product notation $ \map N_1\otimes \cdots \otimes \map N_k$, interchangeably, depending on which representation is more convenient.  

As can be seen from Equation \eqref{generalnosig},  the tensor product space  $\TP (X_1,Y_1) \otimes \TP (X_2,Y_2) \otimes \cdots \otimes \TP (X_k,Y_k)$ also contains  channels that are not of the product form. The set of all such channels is in one-to-one correspondence with  the set of $k$-partite {\em no-signalling channels}, i.e.\ channels    $\map N \in  \Chan(  X_1\otimes \cdots \otimes X_k,  Y_1\otimes \cdots \otimes Y_k)$  with the additional property that the reduced state of any subset of the outputs depends only on the reduced state of the corresponding subset of inputs \cite{Chiribella2013}.   

In the following, we  use the notation $  {\sf   NSChan}   [(X_1,  Y_1 ),   \dots  ,  (X_k,  Y_k)]$  to denote the set of all quantum channels in  $\TP (X_1,Y_1) \otimes \TP (X_2,Y_2) \otimes \cdots \otimes \TP (X_k,Y_k)$, possibly including channels of the non-product form.  When the inputs and outputs are arbitrary, we will use the notation  $  {\sf   NSChan} $.

\subsection{Admissible supermaps}\label{subsec:admissible}

In order to specify the free operations in a resource theory, one has  to first specify  the broader set of operations from which the free operations are chosen. For a resource theory of communication, the operations are (1) supermaps from unplaced channels to unplaced channels, (2) supermaps from unplaced channels to placed channels, and (3) supermaps from placed channels to placed channels.   This specification will be provided in the following.

{\em (3)  Supermaps from placed channels to placed channels.}   An admissible supermap transforming channels in $\Chan (A,B)$ into channels in $\Chan (A', B')$ is a linear transformation $\map S$ from the set $\Map  (A,B)$ to the set $\Map (A', B')$, where $A,B,A',B'$ are placed systems \cite{chiribella2008transforming,chiribella2009theoretical}. 

Admissibility is the requirement that $\map S$ should transform unplaced channels into channels, even when acting locally on parts of larger devices.   Mathematically, the admissibility condition can be formulated by introducing an additional system describing the environment:   a supermap $\map  S$ is admissible if, for every environment system $E$, and for every channel $\map C \in  \Chan  (A , B  \otimes E)$, the map $(\map S\otimes \map I_E)  (\map C)$ belongs to $\Chan  (A' ,  B'\otimes E)$. 

The admissible supermaps from $\Chan (A,B)$ to  $\Chan (A', B')$ have been characterised in Ref.\ \cite{chiribella2008transforming}, which showed that any supermap from placed channels to placed channels can be obtained by sandwiching the input channel between a pre-processing channel and a post-processing channel, as in Equation \eqref{allsupermaps}. 

For channels placed in a sequence, the admissible supermaps transform quantum $k$-combs  in $\Comb[ (A_1,B_1), \dots,  (A_k,  B_k)]$ into quantum $l$-combs in $\Comb[ (A_1',B_1'), \dots,  (A_l', B_l')]$. They are defined as linear    transformations  $\map S$ from the set $\Map  ( A_1\otimes \cdots \otimes A_k,  B_1\otimes \cdots \otimes B_k)$ to the set $\Map  (A_1' \otimes \cdots \otimes A_l',  B_1'\otimes \cdots \otimes B_l')$ satisfying the property that, for every environment system $E$, and for every $k$-comb  $\map C \in  \Comb  [(A_1 , B_1 )  , \dots  (A_k , B_k \otimes  E)]$, the map $(\map S \otimes \map I_E)  (\map C)$ belongs to $\Comb  [(A_1', B_1'),\dots, (A_l' , B_l' \otimes  E)]$. The admissible supermaps with $l=1$ have been characterised in Ref.  \cite{chiribella2009theoretical}, and correspond to $(k+1)$-combs.   The general form of  such supermaps is shown in Equation \eqref{3comb} in the special case of $k=2$.

{\em (2)  Supermaps from unplaced channels to placed channels.} These supermaps represent the possible placements of quantum channels.  An admissible supermap transforming $k$ unplaced channels into a single placed channel is a linear supermap  $\map S$ from the set $\Map (X_1\otimes \cdots \otimes X_k,  Y_1\otimes \cdots \otimes Y_k )$ to the set  $\Map  (A,  B)$ where  $A$ and $B$ are placed systems, and $X_1,  \dots  ,  X_k,  Y_1,\dots,  Y_k$ are unplaced systems.  An admissible supermap $\map S$ should satisfy   the condition that, for every environment system $E$, and for every channel $\map C \in  \Chan  (X_1  \otimes \cdots \otimes X_k, Y_1  \otimes \cdots \otimes Y_k \otimes E)$ satisfying the condition  $\Tr_E   \map C  \in {\sf NSChan}  [(X_1,Y_1) ,  \dots, (X_k,Y_k)]$, the map $(\map S\otimes \map I_E)  (\map C)$ belongs to $\Chan  (A ,  B\otimes E)$.

Examples of  admissible supermaps are the basic placement supermap of Equation \eqref{eq:place},  the parallel placement supermap of Equation \eqref{eq:place_par}, and the sequential placement supermaps of Equations \eqref{eq:seq} and \eqref{eq:seq2}.   Note that, in fact, the parallel and sequential placements are just the products of many basic placements, and that the parallel or sequential nature of a given placement just depends on the causal structure of the spacetime points in which the inputs and outputs of the channels are placed. 

Other examples of admissible placement supermaps are  the quantum {\tt SWITCH} placement \ref{op:switch} of Ref.\ \cite{Chiribella2013}, and the superposition placement \ref{op:superpos} of Ref.\ \cite{chiribella2019shannon2q}.

{\em (1)  Supermaps from unplaced channels to unplaced channels.}  An admissible supermap from unplaced channels  in  ${\sf NSChan}  [(X_1,Y_1) ,  \dots, (X_k,Y_k)]$  to unplaced channels in ${\sf NSChan}  [(X_1',Y_1') ,  \dots, (X_{l}',Y_{l}')]$  is a linear map $\map S$ from the set $\Map (X_1\otimes \cdots \otimes X_k,  Y_1\otimes \cdots \otimes Y_k)$ to the set  $\Map (X_1'\otimes \cdots \otimes X_{l}',  Y_1'\otimes \cdots \otimes Y_{l}')$, where $(X_i)_{i=1}^k,  (Y_i)_{i=1}^k, (X'_j)_{j=1}^{l}, (Y'_j)_{j=1}^{l}$ are unplaced systems. Admissibility is the condition that, for every environment system $E$, and for every channel $\map C \in  \Chan  (X_1  \otimes \cdots \otimes X_k , Y_1  \otimes \cdots \otimes Y_k \otimes E)$ satisfying the condition  $\Tr_E   \map C  \in {\sf NSChan}  [(X_1,Y_1) ,  \dots, (X_k,Y_k)]$, the map $\map C'  := (\map S\otimes \map I_E)  (\map C)$ satisfies the condition     $\Tr_E   \map C'  \in {\sf NSChan}  [(X_1',Y_1') ,  \dots, (X_l',Y_l')]$.

An example of a supermap from unplaced channels to unplaced channels is the {\em discarding supermap} 
\begin{align}\label{discard}
\map S^m_{\rm discard}  (  \map N_1  , \dots,  \map N_k)  :  = (  \map N_1  , \dots, \map N_{m-1}, \map N_{m+1}, \dots,  \map N_k)  \,,
\end{align} 
which discards the $m$-th channel from a list of $k$ channels. Intuitively, the discarding supermap should always be included in the  set of free  supermaps, as the communication provider can always decide to discard a communication device in the construction of a communication network.      We say that the set of  free supermaps from unplaced channels to unplaced channels is {\em trivial} if it consists only of discarding supermaps and identity supermaps.

\section{Placement of channels in an arbitrary (definite) causal structure}\label{app:causal_structure}

\begin{figure*}
	\centering
	\subfloat[\label{fig:net_placed}]{%
		\input{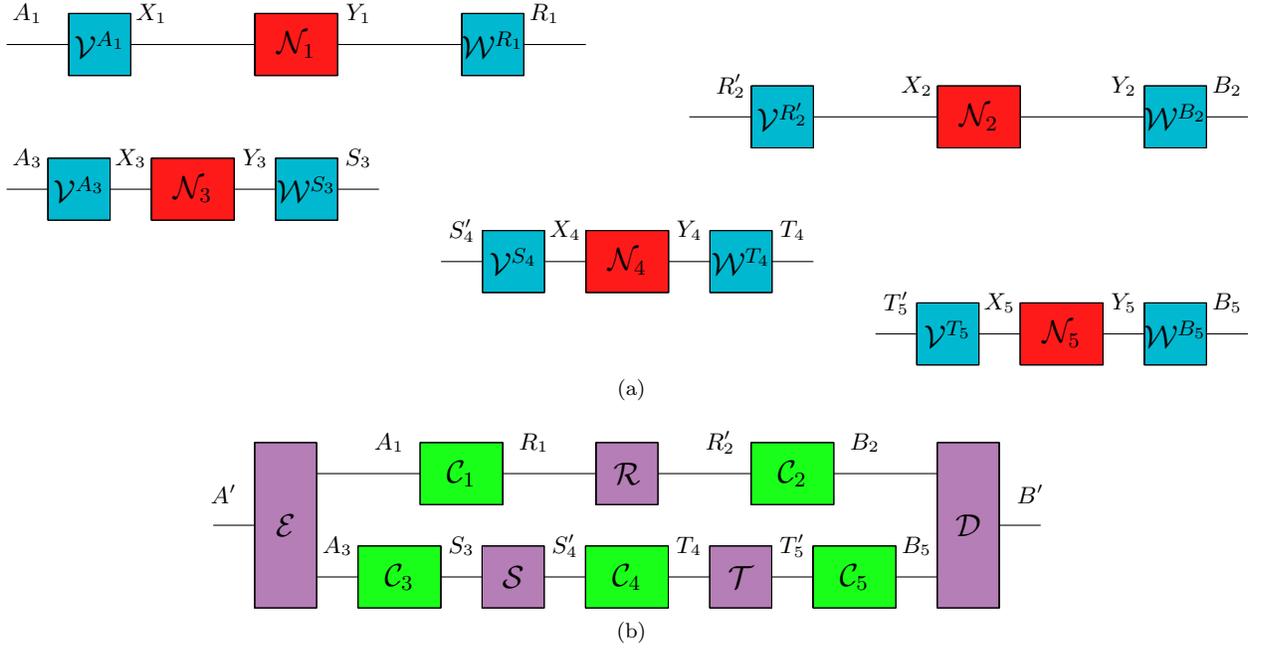}%
	}
	
	\subfloat[\label{fig:net_connected}]{%
		\begin{tikzpicture}[scale=1.1]
	\begin{pgfonlayer}{nodelayer}
	\draw[fill=Orchid!90!white] (-0.75,0.5) -- (-0.75,2) -- (0.75,2) -- (0.75,0.5) -- (-0.75,0.5) -- (-0.75,2);
	\draw[fill=green!90!white] (-3,0.5) -- (-3,2) -- (-5,2) -- (-5,0.5) -- (-3,0.5) -- (-3,2);
	\draw[fill=green!90!white] (3,0.5) -- (3,2) -- (5,2) -- (5,0.5) -- (3,0.5) -- (3,2);
		\draw[fill=green!90!white] (-1,-0.5) -- (-1,-2) -- (1,-2) -- (1,-0.5) -- (-1,-0.5) -- (-1,-2);
		\draw[fill=green!90!white] (4.5,-0.5) -- (4.5,-2) -- (6.5,-2) -- (6.5,-0.5) -- (4.5,-0.5) -- (4.5,-2);
		\draw[fill=green!90!white] (-4.5,-0.5) -- (-4.5,-2) -- (-6.5,-2) -- (-6.5,-0.5) -- (-4.5,-0.5) -- (-4.5,-2);
	\draw[fill=Orchid!90!white] (-3.5,-0.5) -- (-3.5,-2) -- (-2,-2) -- (-2,-0.5) -- (-3.5,-0.5) -- (-3.5,-2);
	\draw[fill=Orchid!90!white] (3.5,-0.5) -- (3.5,-2) -- (2,-2) -- (2,-0.5) -- (3.5,-0.5) -- (3.5,-2);
	\draw[fill=Orchid!90!white] (-9,2) -- (-9,-2) -- (-7.5,-2) -- (-7.5,2) -- (-9,2) -- (-9,-2);
	\draw[fill=Orchid!90!white] (9,2) -- (9,-2) -- (7.5,-2) -- (7.5,2) -- (9,2) -- (9,-2);
	
		\node [style=none] (4) at (-3, 2) {};
		\node [style=none] (5) at (-5, 2) {};
		\node [style=none] (6) at (-5, 0.5) {};
		\node [style=none] (7) at (-3, 0.5) {};
		\node [style=none] (8) at (-3, 1.25) {};
		\node [style=none] (11) at (-5, 1.25) {};
		\node [style=none, font={\large}] (15) at (-4, 1.25) {$\map{C}_1$};
		\node [style=none] (17) at (-7.5, 1.25) {};
		\node [style=none] (25) at (-5.75, 2) {$A_1$};
		\node [style=none] (26) at (-2.25, 2) {$R_1$};
		\node [style=none] (32) at (-4.5, -0.5) {};
		\node [style=none] (33) at (-6.5, -0.5) {};
		\node [style=none] (34) at (-6.5, -2) {};
		\node [style=none] (35) at (-4.5, -2) {};
		\node [style=none] (36) at (-4.5, -1.25) {};
		\node [style=none] (39) at (-6.5, -1.25) {};
		\node [style=none, font={\large}] (43) at (-5.5, -1.25) {$\map{C}_3$};
		\node [style=none] (45) at (-7.5, -1.25) {};
		\node [style=none] (53) at (-7, -0.5) {$A_3$};
		\node [style=none] (54) at (-4, -0.5) {$S_3$};
		\node [style=none] (55) at (-3.5, -1.25) {};
		\node [style=none] (60) at (5, 2) {};
		\node [style=none] (61) at (3, 2) {};
		\node [style=none] (62) at (3, 0.5) {};
		\node [style=none] (63) at (5, 0.5) {};
		\node [style=none] (64) at (5, 1.25) {};
		\node [style=none] (67) at (3, 1.25) {};
		\node [style=none, font={\large}] (71) at (4, 1.25) {$\map{C}_2$};
		\node [style=none] (73) at (0.75, 1.25) {};
		\node [style=none] (81) at (2.25, 2) {$R'_2$};
		\node [style=none] (82) at (5.75, 2) {$B_2$};
		\node [style=none] (83) at (7.5, 1.25) {};
		\node [style=none] (89) at (1, -0.5) {};
		\node [style=none] (90) at (-1, -0.5) {};
		\node [style=none] (91) at (-1, -2) {};
		\node [style=none] (92) at (1, -2) {};
		\node [style=none] (93) at (1, -1.25) {};
		\node [style=none] (96) at (-1, -1.25) {};
		\node [style=none, font={\large}] (100) at (0, -1.25) {$\map{C}_4$};
		\node [style=none] (102) at (-2, -1.25) {};
		\node [style=none] (110) at (-1.5, -0.5) {$S'_4$};
		\node [style=none] (111) at (1.5, -0.5) {$T_4$};
		\node [style=none] (112) at (2, -1.25) {};
		\node [style=none] (117) at (6.5, -0.5) {};
		\node [style=none] (118) at (4.5, -0.5) {};
		\node [style=none] (119) at (4.5, -2) {};
		\node [style=none] (120) at (6.5, -2) {};
		\node [style=none] (121) at (6.5, -1.25) {};
		\node [style=none] (124) at (4.5, -1.25) {};
		\node [style=none, font={\large}] (128) at (5.5, -1.25) {$\map{C}_5$};
		\node [style=none] (130) at (3.5, -1.25) {};
		\node [style=none] (138) at (4, -0.5) {$T'_5$};
		\node [style=none] (139) at (7, -0.5) {$B_5$};
		\node [style=none] (140) at (7.5, -1.25) {};
		\node [style=none] (141) at (-0.75, 1.25) {};
		\node [style=none] (142) at (-0.75, 2) {};
		\node [style=none] (143) at (-0.75, 0.5) {};
		\node [style=none] (144) at (0.75, 2) {};
		\node [style=none] (145) at (0.75, 0.5) {};
		\node [style=none] (146) at (-0.75, 1.25) {};
		\node [style=none] (147) at (0.75, 1.25) {};
		\node [style=none, font={\large}] (148) at (0, 1.25) {$\map{R}$};
		\node [style=none] (149) at (-2, -1.25) {};
		\node [style=none] (150) at (-3.5, -1.25) {};
		\node [style=none] (151) at (-3.5, -0.5) {};
		\node [style=none] (152) at (-3.5, -2) {};
		\node [style=none] (153) at (-2, -0.5) {};
		\node [style=none] (154) at (-2, -2) {};
		\node [style=none] (155) at (-3.5, -1.25) {};
		\node [style=none] (156) at (-2, -1.25) {};
		\node [style=none, font={\large}] (157) at (-2.75, -1.25) {$\map{S}$};
		\node [style=none] (158) at (2, -1.25) {};
		\node [style=none] (159) at (3.5, -1.25) {};
		\node [style=none] (160) at (3.5, -1.25) {};
		\node [style=none] (161) at (2, -1.25) {};
		\node [style=none] (162) at (2, -0.5) {};
		\node [style=none] (163) at (2, -2) {};
		\node [style=none] (164) at (3.5, -0.5) {};
		\node [style=none] (165) at (3.5, -2) {};
		\node [style=none] (166) at (2, -1.25) {};
		\node [style=none] (167) at (3.5, -1.25) {};
		\node [style=none, font={\large}] (168) at (2.75, -1.25) {$\map{T}$};
		\node [style=none] (169) at (-7.5, -1.25) {};
		\node [style=none] (171) at (-9, 2) {};
		\node [style=none] (172) at (-9, -2) {};
		\node [style=none] (173) at (-7.5, 2) {};
		\node [style=none] (174) at (-7.5, -2) {};
		\node [style=none] (176) at (-7.5, 1.25) {};
		\node [style=none, font={\large}] (177) at (-8.25, 0) {$\map{E}$};
		\node [style=none] (182) at (7.5, 2) {};
		\node [style=none] (183) at (7.5, -2) {};
		\node [style=none] (184) at (9, 2) {};
		\node [style=none] (185) at (9, -2) {};
		\node [style=none, font={\large}] (188) at (8.25, 0) {$\map{D}$};
		\node [style=none] (189) at (-10, 0) {};
		\node [style=none] (190) at (-9, 0) {};
		\node [style=none] (191) at (9, 0) {};
		\node [style=none] (192) at (10, 0) {};
		\node [style=none] (193) at (9.75, 0.75) {$B'$};
		\node [style=none] (194) at (-9.75, 0.75) {$A'$};
	\end{pgfonlayer}
	\begin{pgfonlayer}{edgelayer}
		\draw (5.center) to (4.center);
		\draw (4.center) to (7.center);
		\draw (7.center) to (6.center);
		\draw (6.center) to (5.center);
		\draw (33.center) to (32.center);
		\draw (32.center) to (35.center);
		\draw (35.center) to (34.center);
		\draw (34.center) to (33.center);
		\draw (61.center) to (60.center);
		\draw (60.center) to (63.center);
		\draw (63.center) to (62.center);
		\draw (62.center) to (61.center);
		\draw (90.center) to (89.center);
		\draw (89.center) to (92.center);
		\draw (92.center) to (91.center);
		\draw (91.center) to (90.center);
		\draw (118.center) to (117.center);
		\draw (117.center) to (120.center);
		\draw (120.center) to (119.center);
		\draw (119.center) to (118.center);
		\draw (142.center) to (144.center);
		\draw (144.center) to (145.center);
		\draw (145.center) to (143.center);
		\draw (143.center) to (142.center);
		\draw (151.center) to (153.center);
		\draw (153.center) to (154.center);
		\draw (154.center) to (152.center);
		\draw (152.center) to (151.center);
		\draw (162.center) to (164.center);
		\draw [in=90, out=-90] (164.center) to (165.center);
		\draw (165.center) to (163.center);
		\draw (163.center) to (162.center);
		\draw (171.center) to (173.center);
		\draw (173.center) to (174.center);
		\draw (174.center) to (172.center);
		\draw (172.center) to (171.center);
		\draw (182.center) to (184.center);
		\draw (184.center) to (185.center);
		\draw (185.center) to (183.center);
		\draw (183.center) to (182.center);
		\draw (190.center) to (189.center);
		\draw (192.center) to (191.center);
		\draw (176.center) to (11.center);
		\draw (8.center) to (146.center);
		\draw (147.center) to (67.center);
		\draw (64.center) to (83.center);
		\draw (121.center) to (140.center);
		\draw (124.center) to (167.center);
		\draw (166.center) to (93.center);
		\draw (96.center) to (156.center);
		\draw (155.center) to (36.center);
		\draw (39.center) to (169.center);
	\end{pgfonlayer}
\end{tikzpicture}%
	}
	\caption{\label{fig:network} (a) \textit{An illustration of the placement supermap} $\map S_{\rm network}^{A_1 R_1 R'_2 B_2 A_3 S_3 S'_4 T_4 T'_5 B_5 }$, \textit{described by Eq.\ \eqref{eq:net5}, acting on a list of five unplaced channels} $( \map N_1   , \dots ,\map N_5)$. (b) \textit{An illustration of the encoding-repeater-decoding supermap} $\map S_{ \map E, \map R, \map S,\map T, \map D}$, \textit{described by Eq.\ \eqref{eq:free_net5}, acting on the resulting placed channel of (a).}}
\end{figure*}

 Communication through a network of $k\ge 2$  devices, connected via $r \le k-1$ intermediate parties, is described by specifying 
the causal structure of the communicating parties, and by considering supermaps that are compatible with that causal structure.  In the case of a sender $\mathrm{\bf A}$, receiver $\mathrm{\bf B}$ and a single repeater $\mathrm{\bf R}$ (where a boldface letter $\mathrm{\bf R}$ is identified with the list of input/output systems $(R_1, \dots, R_k,R'_1, \dots, R'_l)$ accessible to a given communicating party), 
the causal structure is implicitly given by the totally ordered set  $\{\mathrm{\bf A} \preceq \mathrm{\bf R}, \mathrm{\bf R} \preceq \mathrm{\bf B} \}$, where $\mathrm{\bf A} \preceq \mathrm{\bf B}$ denotes that {\bf B} is in the future light cone of {\bf A}. In this case, it is clear that only placements between {\bf A} and {\bf R}, {\bf A} and {\bf B}, or {\bf R} and {\bf B} are allowed.    

 In the case of $r \ge 2$ repeaters, a general causal structure is described by a partially ordered set (poset), with a choice of possible relations between the intermediate parties \{{\bf R},{\bf S}, \dots, {\bf T}\}. Formally, a poset is a set endowed with a binary relation, which is reflexive, antisymmetric and transitive. The latter two properties ensure that loops in the causal structure are not allowed, i.e.\ if {\bf A} precedes {\bf R} and {\bf R} precedes {\bf S}, then {\bf A} precedes {\bf S}, and therefore {\bf S} cannot also precede {\bf A} (unless {\bf A = R = S}). Physically, the description of causal structure as a poset is motivated by the structure of spacetime as described by special relativity \cite{dowker2006causal}. 

Overall, the placement of communication devices between the communicating parties is described by a tensor product of basic placement supermaps, with the constraint that a placement from $\Chan(X_i,Y_i)$ to $\Chan(S'_i,T_i)$ is only possible if $\textrm{\bf S} \preceq \textrm{\bf T}$ in the causal structure.

We illustrate the scheme for a network of multiple repeaters with an example.
Consider the communication scenario with a sender, a receiver, and $r=3$ intermediate parties \{{\bf R},{\bf S},{\bf T}\}, arranged in a causal structure described by the poset $\{\textrm{\bf A} \preceq \textrm{\bf R}, \textrm{\bf R} \preceq \textrm{\bf B}, \textrm{\bf A} \preceq \textrm{\bf S}, \textrm{\bf S} \preceq \textrm{\bf T}, \textrm{\bf T} \preceq \textrm{\bf B}\}$. Suppose that the communicating parties have access to $k=5$ devices, described by the list of unplaced channels $( \map N_1, \dots ,\map N_{5}) \in \Chan (X_1,Y_1) \times \cdots \times \Chan (X_5,Y_5)$.  The use of the devices is specified by placing them in a particular configuration between the sender, receiver, and repeaters. 
One possible placement is given by
\begin{equation}\label{eq:net5}
\begin{split}
&\map S_{\rm network}^{A_1 R_1 R'_2 B_2 A_3 S_3 S'_4 T_4 T'_5 B_5 } ( \map N_1   , \dots ,\map N_5) \\
&~~= 
\map  S_{\rm place}^{A_1 R_1} (\map N_1) \otimes \map S_{\rm place}^{R'_2 B_2} (\map N_2) \\
&~~~~~~\otimes  \map S_{\rm place}^{A_3 S_3} (\map N_3) \otimes \map  S_{\rm place}^{S'_4 T_4} (\map N_4)  \otimes \map  S_{\rm place}^{T'_5 B_5} (\map N_5)  \\
&~~= 
\map W^{R_1} \!\circ\! \map N_1 \!\circ\! \map V^{A_1} 
\otimes 
\map W^{B_2} \!\circ\! \map N_2 \!\circ\!  \map V^{R'_2} 
\\
&~~~~\otimes \map W^{S_3} \!\circ\! \map N_{3} \!\circ\! \map V^{A_3}
\otimes
\map W^{T_4}\!\circ\! \map N_4 \!\circ\! \map V^{S'_4} 
\otimes
\map W^{B_5} \!\circ\! \map N_5 \!\circ\! \map V^{T'_5} \, ,
\end{split}
\end{equation}
essentially consisting of two sequences through repeaters, {\bf R} and ({\bf S},{\bf T}), placed in parallel between the sender and receiver, as illustrated in Figure \ref{fig:net_placed}. 

Note, that here the different intermediate parties are labelled $\textrm{\bf R}, \textrm{\bf S}, \dots, \textrm{\bf T}$. The subscript $i$ ($j$) of the placed system $R_i$ ($R_j'$) at the communicating party {\bf R} labels which input system $X_i$ (output system $Y_j$) it corresponds to. In contrast, in the main text where each party only had access to a single system, $R_i$ ($R_i'$) denoted the single input (output) system of the $i$-th repeater party.

With the devices placed within the network of communicating parties, we once again consider the free operations on the placed channels.
Consider a subset of $l \leq k$ placed channels $\map C_1 \otimes \dots  \otimes \map C_l \in \Chan(\cdot, R_1) \times \cdots \times \Chan(\cdot, R_m) \times  \Chan(R'_{m+1},\cdot) \times \cdots \times \Chan(R'_l,\cdot)
$, where the first $m \leq l$ channels have output systems at {\bf R} (and any arbitrary placed input systems), and the remaining $l-m$ channels have input systems at {\bf R} (and any arbitrary placed output systems). The final $k-l$ channels have neither input nor output systems at {\bf R}. The free operations that can be performed at {\bf R} are taken to be those of the form
\begin{equation}\label{eq:free_net}
\begin{split}
\map S_{\map R}  (\map C_{1}  \otimes  &\dots  \otimes  \map C_k)  
:= [ ( \map C_{m+1}  \otimes  \dots \otimes \map C_l) \circ \map R \\ \circ  &(\map C_{1}  \otimes \dots \otimes \map C_m ) ] 
\otimes ( \map C_{l+1}  \otimes  \dots \otimes \map C_{k}) \, , 
\end{split}
\end{equation} 
where $\map R \in \Chan(R_1 \otimes \cdots \otimes R_m,R'_{m+1} \otimes \cdots \otimes R'_l)$.
This includes as a special case the free operations $\map S_{\map E}$ and $\map S_{\map D}$ that can be performed by the sender and receiver, respectively, in which case $m=0$ or $m=l$. 
Overall, the choice of free operations on placed channels is taken to be any sequential or parallel composition of (local) party supermaps of the form of Equation \eqref{eq:free_net}.  When two supermaps $\map S_{\map R} $ and $\map S_{\map T}$ commute, we use the shorthand $\map S_{\map R, \map T} := \map S_{\map T} \circ  \map S_{\map R}  = \map S_{\map R} \circ  \map S_{\map T} $.

As an example, consider the placed channels given in Equation \eqref{eq:net5} and let $\map C_i = \map W \circ \map N_i \circ \map V$. Then the action of the most general free supermap on the placed channels $\map C_1 \otimes \dots  \otimes \map C_5$ is given by
\begin{equation}\label{eq:free_net5}
\begin{split}
& \map S_{ \map E, \map R, \map S,\map T, \map D} (\map C_1  \otimes \dots  \otimes \map C_5) \\ 
&~~=  \map D  \circ \left[ ( \map C_2 \circ \map R \circ \map C_1 ) \otimes ( \map C_5 \circ \map T \circ  \map C_4 \circ \map S \circ \map C_3)  \right] \circ \map E \, ,
\end{split}
\end{equation} 
and is illustrated in Figure \ref{fig:net_connected}.

\section{Comparison with other frameworks}\label{app:comparison}

Our framework is based on the approach of Coecke, Fritz, and Spekkens \cite{coecke2016}, where the set of free operations is taken as the starting point from which the notion of resource is defined.   
  An alternative approach is to start from a set of {\em ``zero resources''} and to define the free operations as those that preserve this set.   For resource theories of quantum channels, this approach was adopted in Refs.\ \cite{liu-yuan19resources,liu-winter19resources,takagi2019resource}, where free channels were specified first, and free operations were defined as those supermaps that transform free channels into free channels. 
  
In  standard quantum Shannon theory, a natural choice for the set of free channels is the set of constant channels: no communication protocol in standard quantum Shannon theory can achieve communication using only constant channels.    The set of supermaps that transform constant channels into constant channels was characterised in Ref.\ \cite{takagi2019resource}, where the authors showed that a supermap preserves the set of constant channels if and only if it is of the form $\map S  (\map N)   =  \sum_i   \,  c_i   \,  \map D_i   \circ \map N  \circ  \map E_i$, where $\map E_i$ and $\map D_i$ are suitable channels, and $(c_i)$ are real (possibly negative) coefficients, such that the map $\sum_i  \,  c_i \, \map E_i\otimes \map D_i$ is a quantum channel.  Physically, these supermaps correspond to the transformations that can be achieved with the assistance of free no-signalling channels between the sender's and receiver's locations.

Going from standard quantum Shannon theory to its extensions, it is not clear whether constant channels should still be regarded as free. Clearly, a {\em placed} constant channel is useless for communication, because it does not transfer any information from the sender's laboratory to the receiver's laboratory.   Hence, placed constant channels should still be considered as free.    On the other hand, an {\em unplaced} constant channel may still be useful, depending on how it interacts with the placements allowed by the theory.    This is indeed what happens when the allowed placements include the quantum {\tt SWITCH}   \cite{ebler2018enhanced}.

One might insist that operations that transform constant channels (placed or unplaced) into non-constant channels should not be allowed in a resource theory of communication.    This requirement would amount to the following:
\begin{manualtheorem}{1'} \label{nacc}
\emph{(No Activation of Constant Channels.)}
	In a resource theory of  communication, no free operation $\map S  \in \mathsf M_{\rm free}$   should be able to transform a constant channel into a non-constant channel. 
\end{manualtheorem}
  
 Note that Condition  \ref{nacc} (No Activation of Constant Channels) is stronger than Condition \ref{nsc} (No Side-Channel Generation). If a supermap violated Condition \ref{nsc},  by allowing the sender and receiver to communicate independently of the input channels, then in particular it would allow the sender and receiver to communicate with constant channels, thus violating Condition \ref{nacc}.    In fact, Condition \ref{nacc} is {\em strictly stronger} than Condition \ref{nsc}.  The quantum {\tt SWITCH} placement transforms two completely depolarising channels into a non-constant channel \cite{ebler2018enhanced}, thereby violating Condition \ref{nacc}. On the other hand,  the quantum {\tt SWITCH} placement does not permit the sender and receiver to communicate independently of the input channels: for example, if the input channels are the completely depolarising channel and the identity, the quantum {\tt SWITCH} placement  outputs the channel
  \begin{align}\label{depi}
\map S^{A,B,\omega}_{\tt SWITCH} (\map N_{\rm dep}, \map I)    =   \map N_{\rm dep}  \otimes \omega \, ,
\end{align} 
which is constant and does not permit any communication.  Hence, the quantum {\tt SWITCH} placement satisfies Condition \ref{nsc}, while it violates Condition  \ref{nacc}.

One motivation for assuming Condition \ref{nacc} would be the idea that the communication provider could ``break'' some of the available devices, by turning them into constant channels, {\em before  placing them between the sender and receiver}.  This pre-placement operation would be described by a {\em constant supermap}, of the form 
\begin{align}
\map S^{\map N_0}   :    \map N \mapsto     \map N_0  \qquad \forall \map N  \in  \Chan  (X,Y) \, , 
\end{align} 
where $\map N_0$ is a constant channel. 
If  such constant supermaps were allowed, then Conditions \ref{nsc} and \ref{nacc} would become equivalent:  a placement supermap that transforms some constant channel into a non-constant channel could be preceded by a constant supermap,  thus enabling communication independently of the input channels.  However, it is not obvious why constant supermaps should be regarded as free. Ultimately, assuming constant supermaps to be free is equivalent to assuming {\em by fiat} that constant channels are zero-resource channels, and therefore can be generated for free.

In summary, it is important to distinguish between two requirements: (a) constant channels should not be transformed into non-constant channels, and (b) it should not be possible to communicate independently of the input devices. While requirement (b) may still be too weak to guarantee that a resource theory of communication is interesting, it appears  that there are interesting resource theories of communication that violate the requirement (a) and still lead to  non-trivial Shannon-theoretic structures.

\section{Discussion of the notion of side-channel proposed in Ref.\ \cite{guerin2018shannon}}\label{app:alternativesidechannel}

The authors of Ref.\ \cite{guerin2018shannon}  write
{\em ``The abstract way to know whether a process contains a “side-channel” is to look at the reduced process''},  which they define as the quantum channel obtained by inserting completely depolarising channels into the supermap under consideration.   
 However,  it is not clear why  the reduced process should be defined in terms of completely depolarising channels, instead of arbitrary  constant channels.  The criterion for ``side-channels" proposed by the   authors of  Ref.\ \cite{guerin2018shannon} seems to be an incomplete version of Condition \ref{nacc} of the previous Appendix: instead of having all constant channels, as in Condition \ref{nacc}, they consider only the  completely depolarising channels.

 After stating their criterion for side-channels, the authors of  Ref.\ \cite{guerin2018shannon} continue by writing {\em ``We see that in the case of the quantum switch, the reduced process [...] is a quantum channel which allows for direct communication [...], no matter what noisy operations are being applied [...]''}.   While the first half of the sentence is correct, it is unclear how the second half should  be interpreted: since the reduced channel was defined by applying completely depolarising channels,  the fact that it allows for direct communication does not imply that the quantum {\tt SWITCH} allows for communication ``no matter what operations are being applied''.  In fact,  this statement is false: when a completely depolarising channel and  the identity channel are applied, the reduced process of the quantum {\tt SWITCH} is a constant channel, and does not allow for any communication, as shown by Equation (\ref{depi}) of the previous Appendix.

\section{Proof of Proposition \ref{prop:dep}}\label{app:dep_proof}

Here we provide a proof of Proposition \ref{prop:dep}. The proof follows from  the following Lemmas, proven at the end of this Appendix.

\begin{lemma}\label{lemma:homo}
Let  $\map N_1  \in  \Chan (X)$ and $\map N_2  \in  \Chan (X)$ be two quantum channels, and let  $\widetilde {\map N}_1$ and  $\widetilde {\map   N}_2$ be their vacuum extensions.   Then,    $\widetilde {\map   N}_1  \circ  \widetilde {\map   N}_2$ is a vacuum extension of $\map N_1\circ \map N_2$, and its vacuum interference operator is  $F_1  F_2$, where $F_1$  ($F_2$) is the vacuum interference operator of $\widetilde {\map N}_1$  ($\widetilde {\map N}_2$).  
\end{lemma}

\begin{lemma}\label{lemma:Fge}
Let $F  \in L  (\spc H_X)$ be the vacuum interference operator associated to a generic vacuum-extended channel  $\widetilde {\map C}  \in  \Chan  (\widetilde X)$.    Then, one has $\|  F  \|_{\infty}   \le 1$. 
\end{lemma}

\begin{lemma}\label{lemma:Fnorm}
Let $\map C  \in  \Chan  (X)$ be a quantum channel on a quantum system of dimension $d\ge 2$,   and let $\widetilde {\map C} \in \Chan (\widetilde X)$ be an arbitrary vacuum extension of $\map C$.  If the Choi operator 
\begin{align}  C:  =  \sum_{  i,j}   \,   |i\>\<j|\otimes \map C  (|i\>\<j|) \,  ,
\end{align} 
 has full rank, then the vacuum interference operator $F$ associated to $\widetilde {\map C}$  satisfies the strict inequality  $\|  F  \|_{\infty}   <1$. 
\end{lemma}
  
{\bf Proof of Proposition \ref{prop:dep}.}    Let  $\widetilde {\map N}_{\rm dep} $ be an arbitrary vacuum extension of the completely depolarising channel $\map N_{\rm dep}$.  By Lemma \ref{lemma:homo}, the relation    
\begin{align}\label{relation}
 \widetilde {\map N}_{\rm dep}  \circ \widetilde {\map N}_{\rm dep}    =  \widetilde {\map N}_{\rm dep} 
 \end{align}
 implies the relation  $F^2   =  F$, where $F$ is the vacuum interference operator of $\widetilde {\map N}_{\rm dep} $.   In turn, the relation $F^2  = F$ implies the relation  $F  =  F^n$ for every integer $n\in \N$.  In terms of the norm, this condition yields the bound 
 \begin{align}
 \nonumber \|  F\|_\infty &  =   \|   F^n  \|_\infty  \\
 &  \le   \|   F   \|_\infty^n   \qquad \forall n\in\N \, . \label{Fn}
 \end{align}
 
 Now, the Choi operator of the completely depolarising channel $\map N_{\rm dep}$ is $N  =  I  \otimes I  /d$ and has full rank. Hence, Lemma \ref{lemma:Fnorm} implies $\|  F\|_\infty  <1$.   Hence, equation (\ref{Fn}) implies $\|  F\|_\infty  =  0$, and therefore $F=  0$.  In summary, the only vacuum extension satisfying the condition (\ref{relation}) is the incoherent one. \qed 
 
 \medskip    
 
{\bf Proof  of Lemma \ref{lemma:homo}.}  Using Equation (\ref{outputstate}) for the vacuum-extended channels $\widetilde {\map N}_1$ and  $\widetilde {\map   N}_2$, we obtain the relation  
\begin{equation}
\begin{split}
 ( \widetilde {\map N}_1 \! \circ \! \widetilde {\map   N}_2)   (\rho)    =  &    (\map N_1 \! \circ \! \map N_2)( P_X \rho  P_X)   + \<{\rm vac} | \rho  |{\rm vac}\>  \,  |{\rm vac}\>\<{\rm vac} |   \\
  &  +  F_1  F_2   \rho  |{\rm vac}\>\<{\rm vac} |      +   |{\rm vac}\>\<{\rm vac} |    \rho   F_2^\dag F_1^\dag  \, ,  \label{zxcv}
\end{split}
\end{equation}
valid for every $\rho  \in  \St (\widetilde X)$.  From Equation (\ref{zxcv}) one can deduce that  $\widetilde {\map   N}_1  \circ  \widetilde {\map   N}_2$ is a vacuum extension of $\map N_1\circ \map N_2$  (Conditions  (\ref{vext1})  and (\ref{vext2}) are satisfied).  Moreover, comparison of  Equation (\ref{zxcv}) with Equation (\ref{outputstate}) shows that the vacuum interference operator of  $\widetilde {\map   N}_1  \circ  \widetilde {\map   N}_2$  is $F_1 F_2$. \qed 
 
 \medskip 
  
{\bf Proof of Lemma \ref{lemma:Fge}.}     By definition  (\ref{vacint}), the vacuum interference operator can be expressed as $F   =  \sum_i \overline \gamma_i    C_i $, where $\{\widetilde C_i  : =   C_i    \oplus  \gamma_i |{\rm vac}\>\<{\rm vac}| \}$ is an arbitrary  Kraus representation of the vacuum-extended channel $\widetilde {\map C}$.    

By definition, one has
\begin{equation}\label{key}
\norm{F}_\infty \! 
:= \!\! \max_{ \left\{  \ket{\phi}  \in  \spc H_X,~ \norm{\ket{\phi}}=1 \right\} }  \! \norm{F \ket{\phi}}  \, .
\end{equation}
Let $|\phi\>  \in \spc H_X$ be a unit vector  such that $  \| F\|_\infty   =  \|   F |\phi\> \|$.   Then, one has the  following series of (in)equalities:
\begin{align}
\nonumber  \|   F  \|_\infty^2   &   =   \|  F  |\phi\>  \|^2 \\
 \nonumber    &    = \bra{\phi} F^\dagger F \ket{\phi} \\
\nonumber  &= \sum_{ij} {\gamma_i } \bar{\gamma_j} \bra{\phi} C_i^\dagger C_j \ket{\phi} \\
\nonumber  &\leq \sum_{ij}  \abs{{\gamma_i }}  \abs{\gamma_j} \abs{\bra{\phi} C_i^\dagger C_j  \ket{\phi} }\\
\nonumber  &\leq \sum_{ij}  \abs{{\gamma_i }}  
\abs{ \gamma_j} \sqrt{\bra{\phi} C_i^\dagger C_i  \ket{\phi}  \bra{\phi} C_j^\dagger C_j  \ket{\phi} }\\
\nonumber  &= \left( \sum_i \abs{\gamma_i} \sqrt{\bra{\phi} C_i^\dagger C_i  \ket{\phi}} \right)^2\\
\nonumber  &\leq \left( \sum_i\bra{\phi} C_i^\dagger C_i  \ket{\phi} \right)^2\\
&= 1 \, , \label{qwert}
\end{align}
where the first inequality is the triangle inequality for the modulus, and the second and third inequalities are Cauchy-Schwarz inequalities.  \qed

\medskip  

{\bf Proof of Lemma \ref{lemma:Fnorm}. }    Lemma \ref{lemma:Fge} shows that the norm of $F$ is smaller than or equal to 1.  The equality  $\|  F  \|_\infty  = 1$ holds if and only if all the inequalities in Equation (\ref{qwert}) hold with the equality sign. In the following we will show that saturating the second inequality is impossible when $C$ has full rank. 

The second inequality  in Equation (\ref{qwert}) is saturated if and only if 
\begin{align}\label{poiu}
C_i \ket{\phi} &\propto C_j \ket{\phi} \qquad \forall i,j \, ,
\end{align}
that is, if and only if 
\begin{align}
 C_i \ket{\phi} &= \lambda_i   |\phi_0\>   ~\qquad \forall i  \, ,
\end{align}
where  $|\phi_0\> \in \spc H_X$ is a fixed  unit vector, and $\{\lambda_i\}$ are complex numbers.  

Let $A  = \sum_i    \alpha_i  \,   C_i$  be an arbitrary linear combination of the operators $\{C_i\}$, with complex coefficients $\{\alpha_i\}$.    Then, one has
\begin{equation}
\begin{split}
 A \ket{\phi}   & =  \sum_i   \alpha_i   \,   C_i  |\phi\>  \\
  &   =  \left(   \sum_i  \alpha_i \lambda_i \right)   \, \ket{\phi_0} \, . \label{eq:genA}
  \end{split}
\end{equation}
In other words, every linear combination of the operators $\{C_i\}$ must map $|\phi\>$ into a vector proportional to $|\phi_0\>$.

Now, the Choi operator $C$ has full rank if and only if the operators $C_i$ are a spanning set for the vector space $L(\spc H_X)$.  
  This means, in particular, that there exist coefficients $\{\alpha_i\}$  such that   $A  = \sum_i    \alpha_i  \,   C_i   =  |\phi_0^\perp\>\<\phi|$, where $|\phi_0^\perp\>$ is a unit vector orthogonal to $|\phi_0\>$  (such a vector exists because the Hilbert space $\spc H_X$ is at least two-dimensional).     
  In this case, one has $A  |\phi\>  =  |\phi_0^\perp\>$, meaning that  Equation  (\ref{eq:genA}) cannot be satisfied. This implies that Equation (\ref{poiu}) cannot be satisfied, and that  the bound $\|  F\|_\infty \le 1 $ cannot hold with the equality sign. 
  \qed

\section{Reply to the interferometric arguments of Gu\'erin {\em et al.}\ }\label{app:int}

Ref.\ \cite{guerin2018shannon}	provides an interferometric implementation of the SDPP  supermap $\map F$ defined by
	\begin{align*}
	\map F (\map N_1, \map N_2) (\rho ) = [(\map N_2 \circ \map N_1) \otimes \map I_C] \circ \map U^{\tt CNOT} (\rho \otimes \omega),
\end{align*}  
arguing that in this implementation it cannot be said that  the map $\map F$ transfers information  from the message to the control before the noisy channels are applied.  

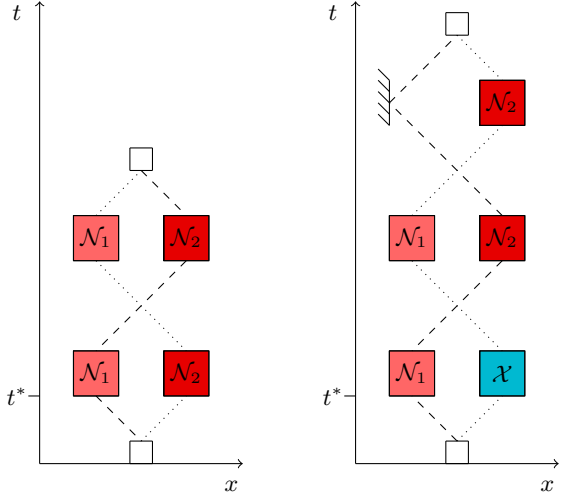
\begin{figure}
	\centering
	\begin{tikzpicture}[scale=1.2]
	\begin{pgfonlayer}{nodelayer}
	\draw[fill=red!60!white] (-4.5,0) -- (-4.5,1) -- (-3.5,1) -- (-3.5,0) -- (-4.5,0) -- (-4.5,1);
	\draw[fill=red!60!white] (-4.5,-3) -- (-4.5,-2) -- (-3.5,-2) -- (-3.5,-3) -- (-4.5,-3) -- (-4.5,-2);
	\draw[fill=red!90!black] (-2.5,0) -- (-2.5,1) -- (-1.5,1) -- (-1.5,0) -- (-2.5,0) -- (-2.5,1);
	\draw[fill=red!90!black] (-2.5,-3) -- (-2.5,-2) -- (-1.5,-2) -- (-1.5,-3) -- (-2.5,-3) -- (-2.5,-2);
	\draw[fill=red!60!white] (2.5,0) -- (2.5,1) -- (3.5,1) -- (3.5,0) -- (2.5,0) -- (2.5,1);
	\draw[fill=red!60!white] (2.5,-3) -- (2.5,-2) -- (3.5,-2) -- (3.5,-3) -- (2.5,-3) -- (2.5,-2);
	\draw[fill=red!90!black] (4.5,0) -- (4.5,1) -- (5.5,1) -- (5.5,0) -- (4.5,0) -- (4.5,1);
	\draw[fill=red!90!black] (4.5,3) -- (4.5,4) -- (5.5,4) -- (5.5,3) -- (4.5,3) -- (4.5,4);
	\draw[fill=Turquoise!90!white] (4.5,-3) -- (4.5,-2) -- (5.5,-2) -- (5.5,-3) -- (4.5,-3) -- (4.5,-2);
		\node [style=none] (0) at (-5.25, 5.75) {};
		\node [style=none] (1) at (-5.25, -4.5) {};
		\node [style=none] (2) at (-0.75, -4.5) {};
		\node [style=none] (3) at (-4.5, -2) {};
		\node [style=none] (4) at (-3.5, -2) {};
		\node [style=none] (5) at (-4.5, -3) {};
		\node [style=none] (6) at (-3.5, -3) {};
		\node [style=none] (7) at (-2.5, -2) {};
		\node [style=none] (8) at (-1.5, -2) {};
		\node [style=none] (9) at (-2.5, -3) {};
		\node [style=none] (10) at (-1.5, -3) {};
		\node [style=none] (11) at (-4.5, 0) {};
		\node [style=none] (12) at (-2.5, 1) {};
		\node [style=none] (13) at (-3.5, 1) {};
		\node [style=none] (14) at (-1.5, 1) {};
		\node [style=none] (15) at (-1.5, 0) {};
		\node [style=none] (16) at (-2.5, 0) {};
		\node [style=none] (17) at (-3.5, 0) {};
		\node [style=none] (18) at (-4.5, 1) {};
		\node [style=none] (19) at (-3.25, -4) {};
		\node [style=none] (20) at (-2.75, -4) {};
		\node [style=none] (21) at (-2.75, -4.5) {};
		\node [style=none] (22) at (-3.25, -4.5) {};
		\node [style=none] (23) at (-2.75, 2) {};
		\node [style=none] (24) at (-3.25, 2) {};
		\node [style=none] (25) at (-2.75, 2.5) {};
		\node [style=none] (26) at (-3.25, 2.5) {};
		\node [style=none] (27) at (-5.5, -3) {};
		\node [style=none] (28) at (-5.25, -3) {};
		\node [style=none] (29) at (-5.75, -3) {$t^*$};
		\node [style=none] (30) at (-4, 0.5) {$\map{N}_1$};
		\node [style=none] (31) at (-2, 0.5) {$\map{N}_2$};
		\node [style=none] (32) at (-2, -2.5) {$\map{N}_2$};
		\node [style=none] (33) at (-4, -2.5) {$\map{N}_1$};
		\node [style=none] (34) at (-4, -3) {};
		\node [style=none] (35) at (-2, -3) {};
		\node [style=none] (36) at (-4, -2) {};
		\node [style=none] (37) at (-2, -2) {};
		\node [style=none] (38) at (-2, 0) {};
		\node [style=none] (39) at (-4, 0) {};
		\node [style=none] (40) at (-4, 1) {};
		\node [style=none] (41) at (-2, 1) {};
		\node [style=none] (42) at (3.5, 0) {};
		\node [style=none] (43) at (1.75, -3) {};
		\node [style=none] (44) at (4.5, -2) {};
		\node [style=none] (45) at (5, 1) {};
		\node [style=none] (46) at (4.5, 1) {};
		\node [style=none] (47) at (1.75, 5.75) {};
		\node [style=none] (48) at (5, -2.5) {$\map{X}$};
		\node [style=none] (49) at (1.5, -3) {};
		\node [style=none] (50) at (5, 0) {};
		\node [style=none] (51) at (4.25, -4) {};
		\node [style=none] (52) at (3, -2) {};
		\node [style=none] (53) at (3, 0.5) {$\map{N}_1$};
		\node [style=none] (54) at (5.5, -3) {};
		\node [style=none] (55) at (2.5, 1) {};
		\node [style=none] (56) at (1.25, -3) {$t^*$};
		\node [style=none] (57) at (3.75, -4) {};
		\node [style=none] (58) at (5.5, 1) {};
		\node [style=none] (59) at (2.5, 0) {};
		\node [style=none] (60) at (4.5, -3) {};
		\node [style=none] (61) at (1.75, -4.5) {};
		\node [style=none] (62) at (3, -3) {};
		\node [style=none] (63) at (3.5, -2) {};
		\node [style=none] (64) at (2.5, -2) {};
		\node [style=none] (65) at (3.75, -4.5) {};
		\node [style=none] (66) at (2.5, -3) {};
		\node [style=none] (67) at (6.25, -4.5) {};
		\node [style=none] (68) at (3, 1) {};
		\node [style=none] (69) at (5, -3) {};
		\node [style=none] (70) at (3, -2.5) {$\map{N}_1$};
		\node [style=none] (71) at (5, 0.5) {$\map{N}_2$};
		\node [style=none] (72) at (4.5, 0) {};
		\node [style=none] (73) at (3.5, -3) {};
		\node [style=none] (74) at (3, 0) {};
		\node [style=none] (75) at (5.5, 0) {};
		\node [style=none] (76) at (5.5, -2) {};
		\node [style=none] (77) at (3.5, 1) {};
		\node [style=none] (78) at (4.25, -4.5) {};
		\node [style=none] (79) at (5, -2) {};
		\node [style=none] (80) at (4.5, 4) {};
		\node [style=none] (81) at (4.5, 3) {};
		\node [style=none] (82) at (5, 3) {};
		\node [style=none] (83) at (4.25, 5.5) {};
		\node [style=none] (84) at (4.25, 5) {};
		\node [style=none] (85) at (3.75, 5.5) {};
		\node [style=none] (86) at (5, 3.5) {$\map{N}_2$};
		\node [style=none] (87) at (5, 4) {};
		\node [style=none] (88) at (5.5, 4) {};
		\node [style=none] (89) at (5.5, 3) {};
		\node [style=none] (90) at (3.75, 5) {};
		\node [style=none] (91) at (2.5, 3.5) {};
		\node [style=none] (92) at (2.5, 4) {};
		\node [style=none] (93) at (2.5, 3) {};
		\node [style=none] (94) at (-3, 2) {};
		\node [style=none] (95) at (-3.25, 2) {};
		\node [style=none] (96) at (-3, 2) {};
		\node [style=none] (97) at (-3, -4) {};
		\node [style=none] (98) at (4, -4) {};
		\node [style=none] (99) at (4, 5) {};
		\node [style=none] (100) at (2.25, 4) {};
		\node [style=none] (101) at (2.25, 3.75) {};
		\node [style=none] (102) at (2.25, 3.5) {};
		\node [style=none] (103) at (2.25, 3.25) {};
		\node [style=none] (104) at (2.5, 3.75) {};
		\node [style=none] (105) at (2.5, 3.25) {};
		\node [style=none] (106) at (2.25, 4.25) {};
		\node [style=none] (107) at (2.5, 4) {};
		\node [style=none] (108) at (-5.75, 5.5) {$t$};
		\node [style=none] (109) at (1.25, 5.5) {$t$};
		\node [style=none] (110) at (-1, -5) {$x$};
		\node [style=none] (111) at (6, -5) {$x$};
	\end{pgfonlayer}
	\begin{pgfonlayer}{edgelayer}
		\draw (3.center) to (4.center);
		\draw (4.center) to (6.center);
		\draw (6.center) to (5.center);
		\draw (5.center) to (3.center);
		\draw (7.center) to (9.center);
		\draw (9.center) to (10.center);
		\draw (10.center) to (8.center);
		\draw (8.center) to (7.center);
		\draw (18.center) to (13.center);
		\draw (13.center) to (17.center);
		\draw (17.center) to (11.center);
		\draw (11.center) to (18.center);
		\draw (12.center) to (16.center);
		\draw (16.center) to (15.center);
		\draw (15.center) to (14.center);
		\draw (14.center) to (12.center);
		\draw (19.center) to (20.center);
		\draw (20.center) to (21.center);
		\draw (19.center) to (22.center);
		\draw (26.center) to (25.center);
		\draw (25.center) to (23.center);
		\draw (26.center) to (24.center);
		\draw (24.center) to (23.center);
		\draw (22.center) to (21.center);
		\draw (27.center) to (28.center);
		\draw [style=dotted] (39.center) to (37.center);
		\draw [style=dashed] (38.center) to (36.center);
		\draw (64.center) to (63.center);
		\draw (63.center) to (73.center);
		\draw (73.center) to (66.center);
		\draw (66.center) to (64.center);
		\draw (44.center) to (60.center);
		\draw (60.center) to (54.center);
		\draw (54.center) to (76.center);
		\draw (76.center) to (44.center);
		\draw (55.center) to (77.center);
		\draw (77.center) to (42.center);
		\draw (42.center) to (59.center);
		\draw (59.center) to (55.center);
		\draw (46.center) to (72.center);
		\draw (72.center) to (75.center);
		\draw (75.center) to (58.center);
		\draw (58.center) to (46.center);
		\draw (57.center) to (51.center);
		\draw (51.center) to (78.center);
		\draw (57.center) to (65.center);
		\draw (65.center) to (78.center);
		\draw (49.center) to (43.center);
		\draw [style=dotted] (74.center) to (79.center);
		\draw [style=dashed] (50.center) to (52.center);
		\draw (80.center) to (81.center);
		\draw (81.center) to (89.center);
		\draw (89.center) to (88.center);
		\draw (88.center) to (80.center);
		\draw (85.center) to (83.center);
		\draw (83.center) to (84.center);
		\draw (85.center) to (90.center);
		\draw (90.center) to (84.center);
		\draw [style=dotted] (68.center) to (82.center);
		\draw [style=dashed] (45.center) to (91.center);
		\draw (92.center) to (93.center);
		\draw [->] (61.center) to (47.center);
		\draw [->] (1.center) to (0.center);
		\draw [->] (1.center) to (2.center);
		\draw [->] (61.center) to (67.center);
		\draw [style=dotted] (40.center) to (94.center);
		\draw [style=dashed] (94.center) to (41.center);
		\draw [style=dashed] (34.center) to (97.center);
		\draw [style=dotted] (97.center) to (35.center);
		\draw [style=dashed] (98.center) to (62.center);
		\draw [style=dotted] (98.center) to (69.center);
		\draw [style=dashed] (91.center) to (99.center);
		\draw [style=dotted] (99.center) to (87.center);
		\draw (100.center) to (104.center);
		\draw (101.center) to (91.center);
		\draw (102.center) to (105.center);
		\draw (103.center) to (93.center);
		\draw (106.center) to (107.center);
	\end{pgfonlayer}
\end{tikzpicture}
	\caption{\label{fig:interferometer} {\em Spacetime diagrams of an interferometric simulation of the quantum {\tt SWITCH} (left) and an interferometric implementation of the process $\map F$ of Equation \eqref{eq:sig_enc} (right).} In both diagrams, the small white squares are beam splitters. The combed line in the diagram on the right is a mirror. The dashed and dotted lines represent the alternative paths taken by the photon in a superposition. Note that this is not a formal circuit diagram, such that the two applications of each channel $\map N_1$ and $\map N_2$ are not independent. 	
	}
\end{figure}

 The proposal of Ref.\ \cite{guerin2018shannon}  is shown  on the right-hand side of Figure \ref{fig:interferometer}. It  consists of an interferometer with two spatial models 0 and 1.   On mode 0, the noisy channel $\map N_2 \circ \map N_1$  is applied right away, while in mode 1 it is applied after a  {\tt NOT} gate.   
 
 Note that the time of application of the channel $\map N_2 \circ \map N_1$ depends on the mode: the channel is applied at an earlier time on mode 0, and at a later time on mode 1.  Since the control is in a superposition state, the message is sent through both modes in a coherent superposition, and therefore the time of application of the channel $\map N_2 \circ \map N_1$ ends up in a coherent superposition.  
 
In this particular implementation, the $\tt NOT$ gate on mode 1 takes place at the same time as the noisy channel $\map N_2\circ \map N_1$ on mode 0.  One could also arrange the setup in such a way that the $\tt NOT$ gate on mode 1 takes place before or after the noisy channel $\map N_2\circ \map N_1$ on mode 0.   The authors of    Ref.\ \cite{guerin2018shannon} argue  that {\em ``this already shows an ambiguity regarding whether [the {\tt NOT} gate] should be considered as part of the `encoding' or not''}.  

We point out, however, that  it is misleading to compare the time when the ${\tt NOT}$ gate takes place {\em on mode 1} with the time when the noisy channel $\map N_2\circ \map N_1$ takes place {\em on mode 0}.  
Instead, one should compare  the times on the same mode: on all possible implementations,  the ${\tt NOT}$ gate on mode 1  takes place before the noisy channel  $\map N_2\circ \map N_1$.   

The authors  of  Ref.\ \cite{guerin2018shannon} appear to have missed  the fact that a quantum particle can be sent through a noisy channel at a superposition of different times, and therefore,  the encoding operations performed before the transmission can also take place at a superposition of different times. The times of application of  the encoding operations and  of the noisy channel can be different  in different branches of the superposition, but the fact that the encoding causally precedes the noisy channel  is true in all branches, and is independent of the specific implementation of the SDPP supermap $\map F$.

The authors of Ref.\ \cite{guerin2018shannon} insist that the ``encoding'' should be defined as the set of operations that take place before a given time, {\em in all branches of the superposition} (specifically, they choose the time denoted as $t^*$ in Figure \ref{fig:interferometer}).  
Starting from this  premise, they compare the interferometric implementation of the SDPP supermap $\map F$  with an interferometric simulation  of the quantum {\tt SWITCH}, shown on the left-hand side of Figure \ref{fig:interferometer},   and     claim that {\em ``the encoding  [for the supermap $\map F$] is the same as for the switch.''}  Instead, if one takes into account that the transmission through the noisy channel $\map N_2\circ \map N_1$ happens at a superposition of two different times, depending on the modes, then  the encoding operations are completely different: 
\begin{itemize}
\item for the quantum {\tt SWITCH}, one has the encoding 
\begin{align}
\map E(\rho)   =  \rho   \otimes \omega  \, ,
\end{align}  
which does not transfer any information from the message to the control, 
\item for the supermap  $\map F$, one has the encoding 
\begin{align}
\map E (\rho)   =  U^{\tt CNOT}  (\rho \otimes \omega) U^{\tt CNOT} \,,  
\end{align}
which transfers  information from the message to the control, thereby exploiting the control as a side-channel that completely bypasses the noisy channel $\map N_2\circ \map N_1$.  
\end{itemize}  

\end{document}